\documentclass{article}
\usepackage{graphicx}
\usepackage{subfig}
\usepackage{algorithm}
\usepackage{lscape}
\usepackage[modulo]{lineno}
\usepackage{pgfplots}
\usepackage{booktabs}
\usepackage[affil-it]{authblk}
\usepackage{geometry}
\usepackage[round]{natbib}
\usepackage{hyperref}
\hypersetup{
    colorlinks=true,
    citecolor=black,
    urlcolor=blue,
    linkcolor=black,
    pdfborder={0 0 0},
}

\begin{document}
\title{Dwell time modulation restrictions do not necessarily improve treatment plan quality for prostate HDR brachytherapy}
\author{Marleen Balvert$^1$, Bram L. Gorissen$^1$, Dick den Hertog$^1$ and Aswin L. Hoffmann$^2$}
\affil{$^1$ Department of Econometrics and Operations Research/Center for Economic Research (CentER), Tilburg University, PO Box 90153, 5000 LE Tilburg, The Netherlands}
\affil{$^2$ Department of Radiation Oncology (MAASTRO), GROW School for Oncology and Developmental Biology, Maastricht University Medical Center, 6201 BN Maastricht, The Netherlands}
\date{}

\maketitle

\begin{abstract}
Inverse planning algorithms for dwell time optimisation in interstitial high-dose-rate (HDR) brachytherapy may produce solutions with large dwell time variations within catheters, which may result in undesirable selective high-dose subvolumes. Extending the dwell time optimisation model with a dwell time modulation restriction (DTMR) that limits dwell time differences between neighboring dwell positions has been suggested to eliminate this problem. DTMRs may additionally reduce the sensitivity for uncertainties in dwell positions that inevitably result from catheter reconstruction errors and afterloader source positioning inaccuracies. This study quantifies the reduction of high-dose subvolumes and the robustness against these uncertainties by applying a DTMR to template-based prostate HDR brachytherapy implants.

Three different DTMRs were consecutively applied to a linear dose-based penalty model (LD) and a dose-volume based model (LDV), both obtained from literature. The models were solved with DTMR levels ranging from no restriction to uniform dwell times within catheters in discrete steps. Uncertainties were simulated on clinical cases using in-house developed software, and dose-volume metrics were calculated in each simulation. For the assessment of high-dose subvolumes, the dose homogeneity index (DHI) and the contiguous dose volume histogram were analysed. Robustness was measured by the improvement of the lowest $D_{90\%}$ of the planning target volume (PTV) observed in the simulations.

For (LD), a DTMR yields an increase in DHI of approximately 30\% and reduces the size of the largest high-dose volume by 2 to 5 cc. However, this comes at a cost of a reduction in $D_{90\%}$ of the PTV of 10\%, which often implies that it drops below the desired minimum of 100\%. For (LDV), none of the DTMRs were able to improve high-dose volume measures. DTMRs were not capable of improving robustness of PTV $D_{90\%}$ against uncertainty in dwell positions for both models.
\end{abstract}

\begin{tikzpicture}[remember picture,overlay]
\node[anchor=south,yshift=10pt] at (current page.south) {\fbox{\parbox{\dimexpr\textwidth-\fboxsep-\fboxrule\relax}{\footnotesize This is an author-created, un-copyedited version of an article published in Physics in Medicine and Biology \href{http://dx.doi.org/10.1088/0031-9155/60/2/537}{DOI:10.1088/0031-9155/60/2/537}.}}};
\end{tikzpicture} 

\section{Introduction}
Interstitial high-dose-rate (HDR) brachytherapy with a high activity $^{192}$Ir stepping source using remote afterloading has shown to be an excellent treatment option for localised prostate cancer in any risk category. Its high tumour control and low toxicity rates result from the precision and control with which this highly conformal treatment can be delivered \citep{Yamada:2012}.

In modern treatment planning systems, automated techniques for anatomy-based inverse treatment planning enable a fast adjustment of the source dwell time distribution within implanted catheters. The fundament of these automated techniques is a mathematical optimisation model that uses dose penalty functions for the planning target volume (PTV) and all relevant organs at risk (OARs) to achieve pre-set dose requirements. Several optimisation algorithms, like the inverse planning simulated annealing (IPSA, \citeauthor{Lessard:2001} 2001) and hybrid inverse planning optimisation (HIPO, \citeauthor{Karabis:2005} 2005) algorithms have been described in the literature to solve this task.

Often these algorithms produce solutions with large dwell time variations within catheters \citep{Holm:2012}. This may give rise to the following problems related to treatment plan robustness against uncertainties in dose delivery. Firstly, large dwell times may produce selective high-dose subvolumes around dominant dwell positions. As stated by \citet{Baltas:2009}, these high-dose subvolumes should be avoided unless dose inhomogeneities in the target volume are biologically motivated. Secondly, catheters with large dwell time variations are expected to yield a heterogeneous dose distribution. Since a longitudinal displacement of the catheter implies that a dwell position may shift to the location of its neighbor, heterogeneous dose distributions are expected to be more susceptible to uncertainties in dwell positions. Such uncertainties can be caused by catheter reconstruction errors and (mechanical) source positioning inaccuracies of the afterloader as well as inter- and intra-fraction catheter displacement. 

\subsection{Clinical procedure and workflow}
This study is based on a local clinical procedure for HDR brachytherapy delivered in two fractions of 8.5 Gy with a one-week interfraction interval. The brachytherapy fractions were a boost to external radiotherapy delivered in 13 fractions of 2.75 Gy, based on the work by \citet{Hoskin:2007}. Prior to treatment, the patient was placed in dorsal lithotomy position, and was not moved until the treatment fraction had been completed. First, transrectal ultrasound (TRUS) images were acquired using an endocavity biplane transducer (type: 8848, BK Medical, Herlev, Denmark) with an image resolution of 0.1 mm/pixel at 6 MHz, and relevant tissue structures were contoured. Second, a so-called pre-plan was made, where the catheter configuration as well as the dwell times were optimised via inverse planning. For each fraction, a new implant was made using rigid stainless steel trocar needles (hereafter referred to as `catheters'). In the third step, catheters were inserted under TRUS guidance, using a transperineal template with a hole spacing of 5 mm. Fourth, new TRUS images were acquired, on which catheter locations were reconstructed and structure delineations were adapted when necessary. Since we made use of rigid needles, catheter reconstruction was performed by localising two points: the first point is the location of the catheter at the template, which was known, and the second point is the catheter tip, which was identified on the TRUS scan. In the fifth phase, a treatment plan was developed based on the delineations and catheter reconstructions using inverse planning (HDRplus version 3.0, Eckert and Ziegler BEBIG GmbH, Berlin, Germany). This work only considers the dwell time optimisation in this fifth step, where a fixed catheter configuration was given. Finally, after plan evaluation and approval the treatment was delivered using a $^{192}$Ir source.

\subsection{High-dose subvolumes}
Small high-dose subvolumes inevitably occur around each of the dwell positions, where the sizes of the regions depend on the respective dwell times. When large dwell times occur, the high-dose subvolumes around two neighboring dwell positions may get connected, resulting in a large high-dose subvolume. Without (radio)biological information about intra-tumour heterogeneity, such large high-dose subvolumes may cause irreversible damage to the stromal tissue causing necrosis, and hence are considered an undesirable property of a dose distribution. Therefore, it is reasonable to avoid the formation of such high-dose subvolumes.

The dose-volume characteristics of the largest high-dose subvolume for a range of dose levels expressed as a percentage of the prescription dose is reflected in the contiguous dose-volume histogram (DVH$^c$). \citet{Thomas:2007} used the DVH$^c$ to quantify dose heterogeneity and perform a contiguous volume analysis in post-implant dosimetry of permanent prostate brachytherapy.

Besides DVH$^c$, which is a local measure as it concerns only a single high-dose volume, we consider the global dose homogeneity index (DHI), which is calculated as \citep{Wu:1988}:
\[
	\textrm{DHI} = \frac{V_{100\%}-V_{150\%}}{V_{100\%}},
\]
where $V_{x\%}$ denotes the volume fraction receiving at least $x\%$ of the prescribed dose. DHI can be interpreted as the volume fraction receiving a dose between 100\% and 150\% of the prescribed dose relative to the volume receiving at least 100\% of the prescribed dose. The higher the DHI, the more uniform is the dose distribution within the target volume.

While DVH$^c$ applies to the largest volume receiving at least a certain dose, interest may also be in the second or third largest volume. We only take these into account indirectly via the DHI for a dose of 150\% of the prescribed dose. In order to see this, note that $1- $ DHI $= V_{150\%}/V_{100\%}$, which is the total volume of all regions receiving at least 150\% of the prescribed dose as a fraction of the total volume receiving at least the prescribed dose.

\subsection{Uncertainty in dwell locations}
A generally acknowledged problem encountered in HDR brachytherapy is the uncertainty in the location of the dwell positions, resulting in a difference between the intended dose distribution and the one actually delivered \citep{Kirisits:2014}. Here, we identify three causes. The first cause is a catheter positioning error, which occurs when the catheter is not exactly positioned during implantation as it was planned in the pre-plan \citep{Abolhassani:2007}. We do not consider this type of error here, since we focus on cases where the catheters have already been inserted. The second cause is a reconstruction error: it is difficult to determine the exact location of the catheter tip from TRUS imaging data \citep{Siebert:2009}. This uncertainty translates to an uncertainty in the location of each of the dwell positions within that catheter. The third cause is a source positioning error caused by mechanical inaccuracies of the remote afterloading system, resulting from the accuracy of the afterloader as well as uncertainty in the pathlength of the source guide transit tube. The latter may be bent, causing an error in the distance from the afterloader to the source. These inaccuracies result in all dwell positions within one catheter to shift in the same direction with the same magnitude.

When the expected locations of the dwell positions differ from the true locations, the delivered dose distribution differs from the expected dose distribution. Treatment plans that are more robust, \textit{i.e.}, for which the dose distribution is less susceptible to uncertainties, are preferred in clinical practice.

\subsection{Dwell time modulation restrictions}
In the literature it has been proposed to regularise the dwell time distribution per catheter by adding a dwell time modulation restriction (DTMR) to the optimisation model \citep{Baltas:2009}. This is a constraint that puts a restriction on the difference between dwell times of adjacent dwell positions. In \citet{Gorissen:2012}, a DTMR restricting the relative dwell time difference was used. \citet{VanderLaarse:1994} considered the sum of squared differences between dwell times of adjacent dwell locations.

DTMRs are available in commercial treatment planning systems, as for example in the real-time intra-operative planning system Oncentra Prostate (Nucletron B.V., Veenendaal, The Netherlands) that employs the HIPO \citep{Karabis:2005} and the IPSA algorithm \citep{Lessard:2001,Alterovitz:2006}. Here, a user-selectable level between 0 and 1 allows for different degrees of dwell time modulation. To the best of our knowledge, no mathematical definition of the DTMR has been published for these particular dwell-time optimisation algorithms, and the quantitative interpretation of the various restriction levels hence remains unclear. \citet{Baltas:2009} and \citet{Mavroidis:2010} have studied the effects of a DTMR in HIPO, and stated that including a DTMR results in treatment plans with fewer high-dose subvolumes and lower total dwell time. Unfortunately, neither of these articles substantiates the statement regarding the reduction in high-dose subvolumes.

\subsection{Aim of the paper}
The aim of this study is to quantify the assumed improvement in treatment plan quality caused by DTMRs in HDR brachytherapy of the prostate. We measure the reduction in high-dose subvolumes caused by three different DTMRs in existing dose- or dose-volume based inverse treatment planning models and are the first to do so in a quantitative manner. Furthermore, this study is the first to assess robustness against uncertainties in dwell locations as a result of these DTMRs. It is our aim to investigate the effect of an increasing DTMR level, not to consider treatment quality for a single DTMR level.

\section{Methods and materials}
\label{sec:Methods}
\subsection{Dwell time optimisation models}
For our analysis, we used two different dwell time optimisation models. The first model is the linear dose (LD) model by \citet{Alterovitz:2006}, which forms the basis for the IPSA algorithm. This model assigns a penalty to each dose calculation point receiving a dose below the preset lower bound or above the preset upper bound, where the penalty is linear in the difference between the dose and the corresponding bound. The objective is to minimise the total penalty. Recently, new optimisation models have been developed that directly optimise dose-volume histogram (DVH) parameters \citep{Siauw:2011,Gorissen:2012}. The second model we use is the linear dose volume (LDV) model decribed by \citet{Gorissen:2012}. This model maximises the number of dose calculation points in the PTV that receive at least the prescribed dose, while restricting the dose received by the hottest ten percent of the rectum and urethra, denoted by $D_{10\%}$(rectum) and $D_{10\%}$(urethra), respectively. 

\subsection{Modulation restrictions}
Both the (LD) and (LDV) models contain the variable $t_j$, which denotes the dwell time of dwell position $j$ in seconds. A restriction can be placed on the dwell time gradient of neighboring dwell positions using these dwell time variables. \citet{Gorissen:2012} have introduced DTMR-R:

\begin{equation}
	t_{j_1} \leq (1+\gamma) t_{j_2} \qquad \forall j_1 \in J, \ \forall j_2 \in \Gamma(j_1), 	 
	\label{relative diff mr}
\end{equation}
where $\gamma$ denotes the pre-set maximum \emph{relative} difference between two adjacent dwell positions, $j_1$ and $j_2$ denote two dwell positions, $J$ denotes the set of dwell positions and $\Gamma(j)$ denotes the set of all dwell positions adjacent to dwell position $j$. Thus, constraint (\ref{relative diff mr}) holds for all dwell positions and all his neighbours. Note that this also implies the reversed constraint, where the dwell time of dwell position $j_2$ cannot exceed $(1+\gamma) t_{j_1}$.

Instead of restricting the relative differences as in constraint (\ref{relative diff mr}) we can restrict the absolute difference between dwell times of two adjacent dwell positions. We introduce DTMR-A, formulated as:

\begin{equation}
	t_{j_1}-t_{j_2} \leq \theta \qquad \forall j_1 \in J, \ \forall j_2 \in \Gamma(j_1), \label{absolute diff mr}
\end{equation}
where $\theta$ is the pre-set maximum \emph{absolute} difference between two adjacent dwell positions. Just as constraint (\ref{relative diff mr}), constraint (\ref{absolute diff mr}) works two ways.

Note that the behavior of DTMR-R and DTMR-A is very different. For DTMR-A, the allowed dwell time difference is independent of the dwell times of neighbouring dwell positions. On the other hand, when using DTMR-R, a larger dwell time allows a larger absolute difference between the dwell times of neighbouring dwell positions.

Finally, DTMR-Q is introduced as a modification of the quadratic penalty first described by \citet{VanderLaarse:1994}:

\begin{equation}
	\frac{1}{2} \sum_{j_1 \in J} \sum_{j_2 \in \Gamma(j_1)} (t_{j_1}-t_{j_2})^2 \leq \rho, 
\end{equation}
where $\rho$ is some pre-set maximum on the sum of squared differences between dwell times of adjacent dwell positions.

\subsection{Patient data and simulations}
\label{sec:MethMat}
In order to investigate the effects of the different DTMRs on the quality of the dose distribution, the three restrictions were applied to the (LD) and (LDV) dwell time optimisation models.

For the numerical evaluation clinical data from three prostate cancer patients were used, where the rectum and urethra are the delineated OARs. These three patients cover various prostate sizes: 32, 55 and 48 cc, respectively. The PTV was defined as the clinical target volume (\textit{i.e.}, the prostate) extended with a 2 mm margin. The catheter positions used had been chosen by an experienced treatment planner. For patients 1 and 2, 16 catheters were used, and 14 were used for patient 3.

Patient data were obtained from the treatment planning system (HDRplus, version 3.0, Eckert and Ziegler BEBIG GmbH, Berlin, Germany), comprising the sets of dwell positions, catheter positions and dose calculation points, the parameters necessary for optimisation, and the dose rate kernel matrix containing the dose rate contribution from each dwell position to each calculation point. We used the same data as \citet{Gorissen:2012}. Dose calculation points had been hexagonally distributed over the delineated structures. The dose rates were determined according to the TG-43 formalism \citep{Nath:1995}, with source specific parameters according to \citet{Granero:2006}.

For  all three patients different treatment plans were obtained by solving the (LD) and (LDV) models extended with each DTMR for different values for $\gamma$, $\theta$ and $\rho$. In the models extended with DTMR-R, $1+\gamma$ ranges from 1 to 4.6 with an incremental step size of 10\%, implying that the relative difference between dwell times of adjacent dwell positions is restricted to be 0 up to 360\%. Note that $\ 1 + \gamma=4.6$ is closest to the unrestricted case, and for $1+\gamma=1$ all dwell times within the same catheter are forced to be equal. In the models extended with DTMR-A, $\theta$ varies from 0 to 5 in steps of 0.05, where $\theta = 5$ corresponds to the unrestricted case. In the models extended with DTMR-Q, $\rho$ varies from 0 to the value implying free modulation, taking 50 steps. The optimisation models extended with DTMR-R or DMTR-A were solved using CPLEX 12.2 Optimiser (IBM Corporation, Somers, USA), which is one of the strongest solvers for linear optimisation problems. For (LD) extended with DTMR-Q we obtained more accurate results using the MOSEK 6.0 solver (Mosek ApS, Copenhagen, Denmark) due to its strongly developed interior point method. For model (LDV) we stopped the solver as soon as 95\% of the PTV received at least the prescribed dose, or after 30 minutes (on a computer with an Intel Q8400 processor). The (LDV) model extended with DTMR-Q could not be solved within reasonable time, and is thus not included in the analysis.

After the treatment plans had been generated, the actual locations of the catheters were perturbed by means of simulation, resulting in different locations of the dwell positions. Due to the large number of possible scenarios, at least 10,000 simulations were calculated per patient. Simulated locations are based on deviations from the nominal (measured) scenario. The accuracies used for simulation are consistent with values reported in the literature \cite[page 62]{Pantelis:2004}. Since an error in locating the catheter tip can be in any direction, for each simulation, the location of the catheter tip was uniformly sampled from a sphere of 2 mm around the measured position. Together with the fixed and known location of the catheter at the template, this results in changes both in the angle between the catheter and the template and in the insertion depth of the catheter in cranial direction. Consequently, all dwell positions in that catheter were moved with the catheter. A single longitudinal shift was uniformly sampled on the interval [-1,1] mm, which applies to all dwell positions within the catheter simultaneously. Dislocations were sampled for each catheter separately. For each treatment plan obtained, the resulting objective values and DVH evaluation criteria were calculated for every simulation. In order to assess whether the models provide good and robust treatment plans, the objective value and the DVH criteria were compared for different DTMR parameter values by simulation.

\subsection{Plan quality indicators}
We consider the following performance indicators to assess treatment plan quality. In Table \ref{tbl:DVH} the DVH criteria are shown, based on the 8.5 Gy per fraction boost according to the local treatment protocol. Here, $D_{90\%}$(PTV) reflects the minimum dose received by the hottest 90\% of the PTV as a fraction of the prescribed dose.
\begin{table}[h]
	\footnotesize
	\caption{Dose-volume criteria, based on the protocol by \protect\citet{Hoskin:2007}.}\label{tbl:DVH}
	\begin{center}
		\begin{tabular}{@{}lll}
			\toprule[0.13em]
			\multicolumn{1}{c}{PTV}                 & \multicolumn{1}{c}{Rectum}   & \multicolumn{1}{c}{Urethra} \cr
			\midrule
			$D_{90\%} \geq 100\%$	& $D_{10\%} \leq 7.2$ Gy		& $D_{10\%} \leq 10$ Gy \cr
			$V_{100\%}\geq 95\%$	& $D_{2\textnormal{cc}} \leq 6.7$ Gy	& $D_{0.1\textnormal{cc}} \leq 10$ Gy \cr
			$V_{150\%}\leq 55\%$	&  	&  \cr
			$V_{200\%}\leq 20\%$ \cr
			\bottomrule[0.13em]
		\end{tabular}
	\end{center}
\end{table}

For the assessment of robustness against uncertainty in catheter positions, the objective value is a useful indicator. Ideally, a robust model does not deteriorate the objective value, while it improves the worst-case value in the simulations. Furthermore, it decreases the standard deviation among simulations.

The objective of (LD) is only a surrogate for the actual goal, which is to satisfy the preset DVH criteria as well as possible \citep{Holm:2011,Gorissen:2012}. Therefore, we also consider the results for $D_{90\%}$(PTV), where robustness in $D_{90\%}$ is more important than robustness in the objective value. For the (LDV) model, we consider the objective (\textit{i.e.}, $V_{100\%}$) as well as $D_{90\%}$ of the PTV.

\section{Results}
\label{sec:Results}
The objective values, DVH metrics and high-dose volume measures resulting from the different treatment plans applied to the simulations are summarised in graphs. Since the number of graphs is huge, we only present the graphs for the (LD) model using DTMR-R showing the total penalty, $D_{90\%}$(PTV), DHI ad DVH$^c$ (Figures \ref{LPreldevLOVpt11}, \ref{LPreldevPTVD90pt11}, \ref{LPrelDHIpt11} and \ref{LPreldevDVHcpt11}, respectively). The remaining graphs have been included in appendices of the supplementary data.

\subsection{Interpretation of the figures}
In all graphs except for those concerning DVH$^c$, the modulation parameter is shown on the horizontal axis. The smallest value implies a strong DTMR, while the largest value implies free or almost free modulation. Consequently, the two extreme observations are the same for all modulation restrictions. For example, it does not matter whether DTMR-A, DTMR-R or DTMR-Q is used when the best plan is the one with the strongest DTMR, because for all DTMRs a parameter value exists that forces all dwell times to be equal within the same catheter. In contrast, a specific DTMR is favourable if the best plan is found for non-extreme DTMR parameter values, {\it i.e.}, in the middle of the graph. Note that, since (LDV) could not always be solved to optimality, the results for the extreme modulation restrictions are not exactly the same.

In the figures displaying the total penalty or dose-volume parameters, the three solid lines represent the average, the maximum and the minimum value over all simulations, and the dotted curve represents the value when all locations of the dwell positions are as derived from the imaging data ({\it i.e.}, nominal value). The grey area denotes the values within a distance of one standard deviation from the mean.

By way of example, we briefly describe the interpretation of Figure \ref{LPreldevPTVD90pt11}. From this figure we see that for the strongest modulation restriction, {\it i.e.}, when $1+\gamma=1$, the minimum, mean and maximum $D_{90\%}$(PTV) over all simulations are approximately 99.9\%, 105\% and 109\%, respectively. This implies that for patient 1 $D_{90\%}$(PTV) is 99.9\% in the worst case scenario. The $D_{90\%}$(PTV) is 106\% if the dwell locations are exactly as in the nominal scenario, which can be concluded from the dotted line representing the nominal value. When the treatment plan developed with (LD) and DTMR-R with $1+\gamma=1$ is used, the values for $D_{90\%}$(PTV) that deviate at most one standard deviation from the mean lie between 104\% and 106\%.

As opposed to DHI and dose-volume metrics, DVH$^c$ is not a single value, but a 2D graph that depends on the DTMR level. Therefore, a 3D graph was included to show the DVH$^c$ as a function of the DTMR levels. Since we do not consider robustness of DVH$^c$, only the nominal case is shown.

\subsection{Relative dwell time difference restricted in (LD)}
The results obtained from (LD) extended with the relative dwell time difference constraint (\ref{relative diff mr}) are discussed in this subsection. First, the results concerning robustness  are discussed, followed by high-dose volume measures. Note that we are not interested in the absolute values of the quality indicators, but merely in the improvements as a result of applying a DTMR. Hence, we will only consider deviations in parameter values caused by applying a stronger restriction.

\begin{figure}[!htb]
	\centering
   	\subfloat[]{\label{LPreldevLOVpt11}\includegraphics[width=0.4\textwidth]{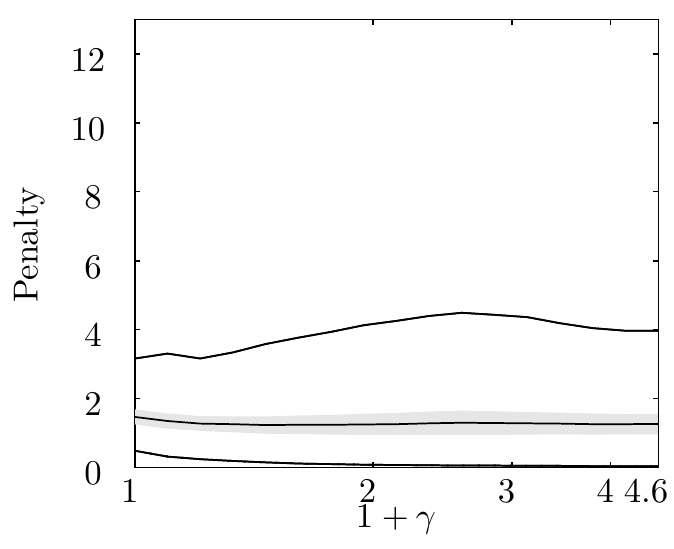}}
   	\qquad\qquad\qquad
   	 \subfloat[]{\label{LPreldevPTVD90pt11}\includegraphics[width=0.4\textwidth]{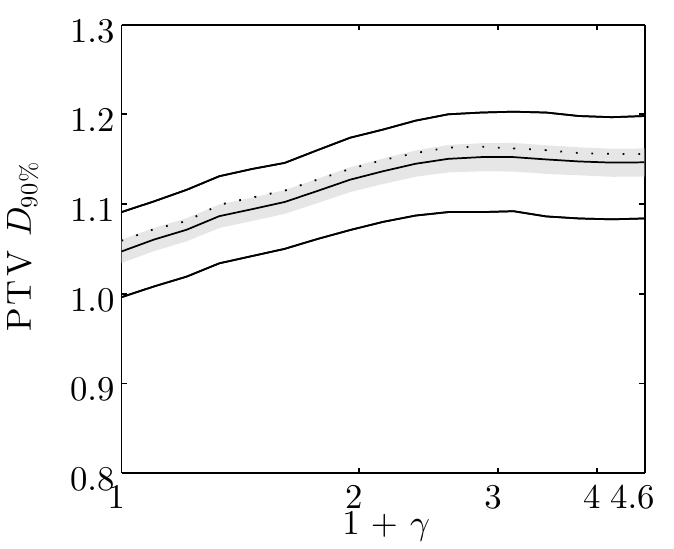}}
   	\caption{Graphs for patient 1 showing: (a) penalty and (b) $D_{90\%}$(PTV) generated by (LD) extended with DTMR-R. The solid lines represent the minimum, mean and maximum values, and the dotted line is the pre-plan value. The grey area denotes values at most one standard deviation from the mean. \label{LPreldevLOV1}}
\end{figure}

From Figure \ref{LPreldevLOVpt11}, it is evident that the range of penalties among the simulations becomes smaller when the DTMR gets stronger. Furthermore, the penalty for the worst-case scenario becomes smaller. From these two observations, we conclude that adding a strong DTMR-R to the (LD) model yields solutions with robust penalty values. However, the results in Figure \ref{LPreldevPTVD90pt11} show that a strong modulation restriction yields a decrease in the $D_{90\%}$(PTV). For patient 1 this decrease does not result in insufficient target coverage, but for patients 2 and 3 it drops below the minimum desired level of 100\% (supplementary data: Figures A5b and A5c). The standard deviation decreases slightly, but the $D_{90\%}$(PTV) for the worst-case scenario strongly decreases. We thus conclude that DTMR-R does not yield robustness of the $D_{90\%}$ for the PTV, and leads to compromised target coverage.

The results for the remaining DVH parameters for the PTV as well as those for the OARs can be found in Appendix A of the supplementary data. From these figures, we conclude that for all patients, a strong restriction on the relative dwell time differences results in a large reduction of $V_{100\%}$, and a slight reduction of $V_{150\%}$ and $V_{200\%}$. Also rectum $D_{2\textnormal{cc}}$ and $D_{10\%}$ slightly decrease, whereas urethra $D_{0.1\textnormal{cc}}$ and $D_{10\%}$ stay at the same level.

\begin{figure}[!htb]
   \centering
   \subfloat[]{\label{LPrelDHIpt11}\includegraphics[width=0.35\textwidth]{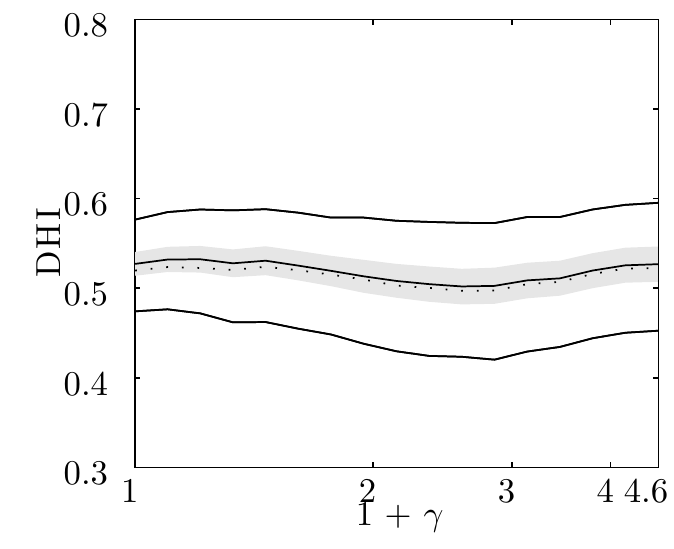}}
   \qquad
   \subfloat[]{\label{LPreldevDVHcpt11}\includegraphics[width=0.35\textwidth]{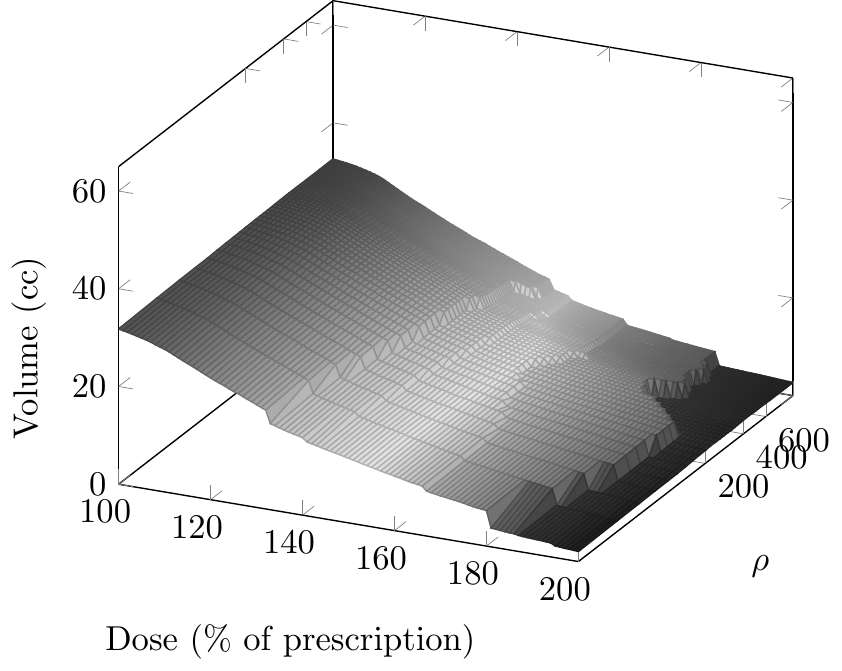}}
%   	\resizebox{0.45\textwidth}{!}{\begin{tikzpicture}
%   \selectcolormodel{gray}
%   \begin{axis}[
%   zmin=0,
%   zmax=65,
%   ytick={0,1,2,3,4,5},
%   y tick label style={anchor=north west},
%   z tick label style={anchor=east},
%   xlabel={Percentage of prescription dose},
%   ylabel={1+$\gamma$},
%   zlabel={Volume (cc)}]
%   \addplot3[surf,mesh/ordering=y varies] table{LDreldevpt1.dat};
%   \end{axis}
%   \end{tikzpicture}}}
   \caption{Graphs for patient 1 showing: high-dose volume performance indicators (a) DHI and (b) DVH$^c$ generated by (LD) extended with DTMR-R. The solid lines in graph (a) represent the minimum, mean and maximum values, and the dotted line is the pre-plan value. The grey area denotes values at most one standard deviation from the mean. \label{LPreldevHSPI1}}
\end{figure}

High-dose subvolumes are the second feature under consideration. Figure \ref{LPrelDHIpt11} shows that for patient 1, DHI is optimal for the weakest as well as the strongest DTMR. For patients 2 and 3, the highest DHI is obtained for a strong modulation restriction (supplementary data: Figures A2b and A2c). From Figure \ref{LPreldevDVHcpt11}, one can see that the strongest DTMR yields the best DVH$^c$ for patient 1. The same conclusion can be drawn for patients 2 and 3 (supplementary data: Figures A3b and A3c). Note that this is a trivial consequence from the decrease in $D_{90\%}$.

From the above we conclude that including DMTR-R in (LD) can result in undesirable treatment plans when considering the $D_{90\%}$ for the PTV. Furthermore, it does result in robustness of the penalty, but not in robustness of the $D_{90\%}$. A strong DMTR-R slightly improves high-dose subvolumes according to the DHI and DVH$^c$ and slightly reduces the dose to the rectum, but this is a trivial consequence of an undesirably low $D_{90\%}$.

\subsection{General results}
The results for the (LD) model extended with DTMR-A and DTMR-Q are similar to the results for the (LD) model extended with DTMR-R and do not need a separate discussion. The (LDV) curves do show a large variation, of which the main cause is that we did not solve these models to optimality. Therefore, we should consider the trend, rather than the individual values. For the (LDV) model extended with DTMR-R and DTMR-A the DTMRs did not provide treatment plans with improved high-dose volume indicator values, and sometimes even worse high-dose volume indicator values were found. Furthermore, for some patients our DTMRs yield lower values for the $D_{90\%}$ of the PTV.

\section{Discussion}
\label{sec:Discussion}
For prostate HDR brachytherapy dose distributions, the effects of DTMRs were assessed in several other papers as well. \citet{Baltas:2009} and \citet{Mavroidis:2010} used data from 12 clinical implants in combination with the HIPO algorithm. The dosimetric quality of these implants was assessed for plans optimised with and without DTMR, such that DVH parameters of their protocol were completely fulfilled. This resulted in a slight reduction ($<$2\%) in DVHs for the prostate and a more pronounced sparing of the OARs, especially urethra and bladder. However, our work does not support this conclusion. We observed a decrease in target coverage of approximately 10\% when using a DTMR.

\citet{Baltas:2009} and \citet{Mavroidis:2010} have found a lower mean dwell time per implant and a mean total dwell time reduction of 1.4\%, that both were proven to be statistically significant. In our opinion, part of the reduction in total dwell time for the plan with DTMR can be explained by the average observed reduction in $D_{90\%}$ for the PTV. Using (LD), we observed that each of the three DTMRs resulted in a mean total dwell time reduction of 3$-$5$\%$. For DTMR-R and DTMR-A the total dwell time decreased gradually as the restriction became stronger, whereas for DTMR-Q the reduction takes place only when dwell times are forced to be equal within the same catheter. For (LDV) there were no differences in mean total dwell time between the plans with and without DTMR.

High-dose volume parameters do not show improvements as a result of the DTMR for the (LDV) model. For the (LD) model, the DTMR can improve high-dose volume measures only while compromising target coverage. A trade-off between these two goals thus needs to be made. For patient 1, the decrease in coverage is not problematic, but for patients 2 and 3 it drops below the desired minimum. The effects of a DTMR do not only differ among patients, but also among models, which indicates that the different effects may not depend on the implant quality and patient characteristics.

\citet{Holm:2013} suggested another method to improve the dose homogeneity by incorporating the DHI into the objective function. When comparing their optimisation model to the original linear penalty model by \citet{Alterovitz:2006}, they found no significant difference in treatment plan quality when comparing the DHI. Thus, an effective optimisation method to improve dose homogeneity in prostate HDR brachytherapy still needs to be established.

For the assessment of a DTMR's capabilities to improve robustness against uncertainties in dwell locations, we took catheter reconstruction errors and the source positioning inaccuracy of the afterloader into account. The simulated errors are of the same order of magnitude as those reported in literature \citep{Siebert:2009,Pantelis:2004,Peikari:2012}, though we neglected the difference in error magnitude between the various directions. The magnitude of variation in $D_{90\%}$ and $V_{100\%}$ for the PTV we found as a result of these errors corresponds to values reported by \citet{Kirisits:2014}, though \citet{Pantelis:2004} found variabilities of approximately half the magnitude of the errors we found. This difference may arise from the fact that \citet{Pantelis:2004} derived DVH metrics from the average dose to a calculation point, and as such ignored the risks occurring in individual scenarios.

The results of the current study show that the DTMR is not suitable for the development of robust treatment plans. Therefore, these uncertainties need to be included in the optimisation model in a different way. Examples of methods to take uncertainties into account are robust optimisation \citep{Ben-Tal:2009} and stochastic programming \citep{Kall:1994}. Future research is required to assess the potential benefit for robust brachytherapy treatment planning.

Additional uncertainties arise in various stages of the treatment process. Examples are delineation variabilities, catheter deflection, organ deformations, dislocation of the implant between fractions \citep{Kirisits:2014} and transit dose contribution \citep{Fonseca:2014}. The effects of those uncertainties as well as methods to overcome them are interesting topics for future research, especially for different catheter configurations.

Our study has some limitations. Firstly, the use of rigid needles allowed us to determine catheter locations by localising two points on each catheter. However, flexible catheters are used in clinics where the implant remains \textit{in situ}\ between fractions. When performing a similar study with flexible catheters, indicating two points on a catheter is insufficient for identification of the complete catheter location, and uncertainties should be simulated in a different way.

Secondly, the number of patients is small. This allowed us to perform the highly detailed study that was necessary to address the effects of a DTMR on high-dose subvolumes and robustness. The results for all three patients were negative, and DTMRs sometimes even showed to result in deterioration of the plan quality. Including more patients in this study would not have changed the negative results for these three patients. Therefore, despite the small number of patients, we conclude that DTMRs do not in general improve treatment plan quality, as they were expected to. This does not imply that a DTMR has negative effects for every patient: patients for whom a DTMR has positive effects may exist. Investigating the effects of a DTMR for each patient prior to treatment would take up too much time. However, we do recommend other institutions to perform a similar study in order to quantitatively assess the influence of a DTMR on robustness, homogeneity and DVH parameters.

\section{Conclusion}
\label{sec:Conclusion}
Although robustness in the penalty of the (LD) model is obtained from all three DTMRs, no improvement in robustness of the $D_{90\%}$(PTV) was obtained by applying any of the three DTMRs for the (LD) and (LDV) models. Furthermore, these DTMRs do not reduce the high-dose subvolumes without simultaneously deteriorating $D_{90\%}$(PTV). Finally, no significant sparing of the OARs is achieved, unless the dose delivered at the PTV is decreased as well. 

\section*{Acknowledgement}
We thank Ulrich Wimmert$^\dagger$ from SonoTECH GmbH (Neu-Ulm, Germany) for providing a research version of the HDRplus software that has the ability to export the dose rate kernel matrix and the coordinates of surface points, dose calculation points and dwell positions.

%\section*{References}
\bibliographystyle{abbrvnatnew}
\bibliography{Balv20140930}

\begin{thebibliography}{25}
\providecommand{\natexlab}[1]{#1}
\providecommand{\url}[1]{\texttt{#1}}
\expandafter\ifx\csname urlstyle\endcsname\relax
  \providecommand{\doi}[1]{doi: #1}\else
  \providecommand{\doi}{doi: \begingroup \urlstyle{rm}\Url}\fi

\bibitem[Abolhassani et~al.(2007)Abolhassani, Patel, and
  Moallem]{Abolhassani:2007}
N.~Abolhassani, R.~Patel, and M.~Moallem.
\newblock Needle insertion into soft tissue: a survey.
\newblock \emph{Medical Engineering and Physics},
  \href{http://dx.doi.org/10.1016/j.medengphy.2006.07.003}{29\penalty0
  (4):\penalty0 413--431}, 2007.

\bibitem[Alterovitz et~al.({2006})Alterovitz, Lessard, Pouliot, Hsu, O'Brien,
  and Goldberg]{Alterovitz:2006}
R.~Alterovitz, E.~Lessard, J.~Pouliot, I.~J. Hsu, J.~F. O'Brien, and
  K.~Goldberg.
\newblock {Optimization of HDR brachytherapy dose distributions using linear
  programming with penalty costs}.
\newblock \emph{{Medical Physics}},
  \href{{http://dx.doi.org/10.1118/1.2349685}}{{33}\penalty0 ({11}):\penalty0
  {4012--4019}}, {2006}.

\bibitem[Baltas et~al.(2009)Baltas, Katsilieri, Kefala, Papaioannou, Karabis,
  Mavroidis, and Zamboglou]{Baltas:2009}
D.~Baltas, Z.~Katsilieri, V.~Kefala, S.~Papaioannou, A.~Karabis, P.~Mavroidis,
  and N.~Zamboglou.
\newblock Influence of modulation restriction in inverse optimization with
  {HIPO} of prostate implants on plan quality: Analysis using dosimetric and
  radiobiological indices.
\newblock \emph{IFMBE Proceedings},
  \href{http://dx.doi.org/10.1007/978-3-642-03474-9\_81}{25\penalty0
  (1):\penalty0 283--286}, 2009.

\bibitem[Ben-Tal et~al.(2009)Ben-Tal, {El Ghaoui}, and
  Nemirovski]{Ben-Tal:2009}
A.~Ben-Tal, L.~{El Ghaoui}, and A.~Nemirovski.
\newblock \emph{Robust Optimization}.
\newblock Princeton Series in Applied Mathematics. Princeton University Press,
  \href{http://sites.google.com/site/robustoptimization/}{2009}.

\bibitem[Fonseca et~al.(2014)Fonseca, Landry, Reniers, Hoffmann, Rubo, Antunes,
  Yoriyaz, and Verhaegen]{Fonseca:2014}
G.~Fonseca, G.~Landry, B.~Reniers, A.~Hoffmann, R.~Rubo, P.~Antunes,
  H.~Yoriyaz, and F.~Verhaegen.
\newblock The contribution from transit dose for $^{192}${Ir} {HDR}
  brachytherapy treatments.
\newblock \emph{Physics in Medicine and Biology},
  \href{http://dx.doi.org/10.1088/0031-9155/59/7/1831}{59\penalty0
  (7):\penalty0 1831--1844}, 2014.

\bibitem[Gorissen et~al.(2013)Gorissen, den Hertog, and
  Hoffmann]{Gorissen:2012}
B.~Gorissen, D.~den Hertog, and A.~Hoffmann.
\newblock Mixed integer programming improves comprehensibility and plan quality
  in inverse optimization of prostate {HDR} brachytherapy.
\newblock \emph{Physics in Medicine and Biology},
  \href{http://dx.doi.org/10.1088/0031-9155/58/4/1041}{58\penalty0
  (4):\penalty0 1041--1057}, 2013.

\bibitem[Granero et~al.(2006)Granero, Perez-Calatayud, Casal, Ballester, and
  Venselaar]{Granero:2006}
D.~Granero, J.~Perez-Calatayud, E.~Casal, F.~Ballester, and J.~Venselaar.
\newblock A dosimetric study on the ir-192 high dose rate flexisource.
\newblock \emph{Medical Physics},
  \href{http://dx.doi.org/10.1118/1.2388154}{33\penalty0 (12):\penalty0
  4578--4582}, 2006.

\bibitem[Holm(2011)]{Holm:2011}
{\AA}.~Holm.
\newblock \emph{Dose plan optimization in HDR brachytherapy using penalties
  properties and extensions}.
\newblock PhD thesis, Link\"{o}ping University, Link\"{o}ping, Sweden,
  \href{http://urn.kb.se/resolve?urn=urn:nbn:se:liu:diva-67790}{2011}.

\bibitem[Holm(2013)]{Holm:2013}
{\AA}.~Holm.
\newblock \emph{Mathematical Optimization of HDR Brachytherapy}.
\newblock PhD thesis, Link\"{o}ping University, Link\"{o}ping, Sweden,
  \href{http://dx.doi.org/10.3384/diss.diva-99795}{2013}.

\bibitem[Holm et~al.(2012)Holm, Larsson, and Tedgren]{Holm:2012}
{\AA}.~Holm, T.~Larsson, and {\AA}.~C. Tedgren.
\newblock Impact of using linear optimization models in dose planning for {HDR}
  brachytherapy.
\newblock \emph{Medical Physics},
  \href{http://dx.doi.org/10.1118/1.3676179}{39\penalty0 (2):\penalty0
  1021--1028}, 2012.

\bibitem[Hoskin et~al.(2007)Hoskin, Motohashi, Bownes, Bryant, and
  Ostler]{Hoskin:2007}
P.~J. Hoskin, K.~Motohashi, P.~Bownes, L.~Bryant, and P.~Ostler.
\newblock High dose rate brachytherapy in combination with external beam
  radiotherapy in the radical treatment of prostate cancer: initial results of
  a randomised phase three trial.
\newblock \emph{Radiotherapy and Oncology},
  \href{http://dx.doi.org/10.1016/j.radonc.2007.04.011}{84\penalty0
  (2):\penalty0 114--120}, 2007.

\bibitem[Kall and Wallace(1994)]{Kall:1994}
P.~Kall and S.~W. Wallace.
\newblock \emph{Stochastic programming}.
\newblock Wiley Interscience Series in Systems and Optimization. Wiley, 1994.

\bibitem[Karabis et~al.(2005)Karabis, Giannouli, and Baltas]{Karabis:2005}
A.~Karabis, S.~Giannouli, and D.~Baltas.
\newblock {HIPO}: A hybrid inverse treatment planning optimization algorithm in
  {HDR} brachytherapy.
\newblock \emph{Radiotherapy and Oncolology},
  \href{http://dx.doi.org/10.1016/S0167-8140(05)81018-7}{76:\penalty0 S29},
  2005.

\bibitem[Kirisits et~al.(2014)Kirisits, Rivard, Baltas, Ballester, Brabandere,
  van~der Laarse, Niatsetski, Papagiannis, Hellebust, Perez-Calatayud,
  Tanderup, Venselaar, and Siebert]{Kirisits:2014}
C.~Kirisits, M.~J. Rivard, D.~Baltas, F.~Ballester, M.~D. Brabandere,
  R.~van~der Laarse, Y.~Niatsetski, P.~Papagiannis, T.~P. Hellebust,
  J.~Perez-Calatayud, K.~Tanderup, J.~L. Venselaar, and F.-A. Siebert.
\newblock Review of clinical brachytherapy uncertainties: Analysis guidelines
  of {GEC-ESTRO} and the {AAPM}.
\newblock \emph{Radiotherapy and Oncology},
  \href{http://dx.doi.org/10.1016/j.radonc.2013.11.002}{110\penalty0
  (1):\penalty0 199--212}, 2014.

\bibitem[Lessard and Pouliot({2001})]{Lessard:2001}
E.~Lessard and J.~Pouliot.
\newblock Inverse planning anatomy-based dose optimization for
  {HDR}-brachytherapy of the prostate using fast simulated annealing algorithm
  and dedicated objective function.
\newblock \emph{Medical Physics},
  \href{http://dx.doi.org/10.1118/1.1368127}{{28}\penalty0 ({5}):\penalty0
  {773--779}}, {2001}.

\bibitem[Mavroidis et~al.(2010)Mavroidis, Katsilieri, Kefala, Milickovic,
  Papanikolaou, Karabis, Zamboglou, and Baltas]{Mavroidis:2010}
P.~Mavroidis, Z.~Katsilieri, V.~Kefala, N.~Milickovic, N.~Papanikolaou,
  A.~Karabis, N.~Zamboglou, and D.~Baltas.
\newblock Radiobiological evaluation of the influence of dwell time modulation
  restriction in {HIPO} optimized {HDR} prostate brachytherapy implants.
\newblock \emph{Journal of Contemporary Brachytherapy},
  \href{http://dx.doi.org/10.5114/jcb.2010.16923}{2\penalty0 (3):\penalty0
  117--128}, 2010.

\bibitem[Nath et~al.(1995)Nath, Anderson, Luxton, Weaver, Williamson, and
  Meigooni]{Nath:1995}
R.~Nath, L.~Anderson, G.~Luxton, K.~Weaver, J.~Williamson, and A.~Meigooni.
\newblock Dosimetry of interstitial brachytherapy sources: recommendations of
  the {AAPM} radiation therapy committee task group no. 43.
\newblock \emph{Medical Physics},
  \href{http://dx.doi.org/10.1118/1.597458}{22\penalty0 (2):\penalty0
  209--334}, 1995.

\bibitem[Pantelis et~al.(2004)Pantelis, Papagiannis, Anagnostopoulos, Baltas,
  Karaiskos, Sandilos, and Sakelliou]{Pantelis:2004}
E.~Pantelis, P.~Papagiannis, G.~Anagnostopoulos, D.~Baltas, P.~Karaiskos,
  P.~Sandilos, and L.~Sakelliou.
\newblock Evaluation of a {TG-43} compliant analytical dosimetry model in
  clinical $^{192}${Ir} {HDR} brachytherapy treatment planning and assessment
  of the significance of source position and catheter reconstruction
  uncertainties.
\newblock \emph{Physics in Medicine and Biology},
  \href{http://dx.doi.org/10.1088/0031-9155/49/1/004}{49\penalty0
  (55):\penalty0 55--67}, 2004.

\bibitem[Peikari et~al.(2012)Peikari, Chen, Lasso, Heffter, and
  Fichtinger]{Peikari:2012}
M.~Peikari, T.~Chen, A.~Lasso, T.~Heffter, and G.~Fichtinger.
\newblock Characterization of ultrasound elevation beam artifacts for prostate
  brachytherapy needle insertion.
\newblock \emph{Medical Physics},
  \href{http://dx.doi.org/10.1118/1.3669488}{39\penalty0 (1):\penalty0
  246--256}, 2012.

\bibitem[Siauw et~al.(2011)Siauw, Cunha, Atamt{\"u}rk, Hsu, Pouliot, and
  Goldberg]{Siauw:2011}
T.~Siauw, A.~Cunha, A.~Atamt{\"u}rk, I.-C. Hsu, J.~Pouliot, and K.~Goldberg.
\newblock {IPIP}: A new approach to inverse planning for {HDR} brachytherapy by
  directly optimizing dosimetric indices.
\newblock \emph{Medical Physics},
  \href{http://dx.doi.org/10.1118/1.3598437}{38\penalty0 (7):\penalty0
  4045--4051}, 2011.

\bibitem[Siebert et~al.(2009)Siebert, Hirt, Niehoff, and
  Kov{\'a}cs]{Siebert:2009}
F.~Siebert, M.~Hirt, P.~Niehoff, and G.~Kov{\'a}cs.
\newblock Imaging of implant needles for real-time {HDR}-brachytherapy prostate
  treatment using biplane ultrasound transducers.
\newblock \emph{Medical Physics},
  \href{http://dx.doi.org/10.1118/1.3157107}{39\penalty0 (8):\penalty0
  3406--3412}, 2009.

\bibitem[Thomas et~al.(2007)Thomas, Kruk, McGahan, Spadinger, and
  Morris]{Thomas:2007}
C.~Thomas, A.~Kruk, C.~McGahan, I.~Spadinger, and W.~Morris.
\newblock Prostate brachytherapy post-implant dosimetry: a comparison between
  higher and lower source density.
\newblock \emph{Radiotherapy \& Oncology},
  \href{http://dx.doi.org/10.1016/j.radonc.2007.02.004}{83\penalty0
  (1):\penalty0 18--24}, 2007.

\bibitem[Van~der Laarse and Prins(1994)]{VanderLaarse:1994}
R.~Van~der Laarse and T.~Prins.
\newblock \emph{Introduction to HDR brachytherapy optimization. In:
  Brachytherapy from Radium to Optimization, R. Mould, J. Battermann, A.
  Martinez and B. Speiser (Eds)}.
\newblock Nucletron International B.V., 1994.

\bibitem[Wu et~al.(1988)Wu, Ulin, and Sternick]{Wu:1988}
A.~Wu, K.~Ulin, and E.~Sternick.
\newblock A dose homogeneity index for evaluating $^{192}${Ir} interstitial
  breast implants.
\newblock \emph{Medical Physics},
  \href{http://dx.doi.org/10.1118/1.596152}{15\penalty0 (1):\penalty0
  104--107}, 1988.

\bibitem[Yamada et~al.(2012)Yamada, Rogers, Demanes, Morton, Prestidge,
  Pouliot, Cohen, Zaider, Ghilezan, and Hsu]{Yamada:2012}
Y.~Yamada, L.~Rogers, D.~J. Demanes, G.~Morton, B.~R. Prestidge, J.~Pouliot,
  G.~N. Cohen, M.~Zaider, M.~Ghilezan, and I.-C. Hsu.
\newblock {American} {Brachytherapy} {Society} consensus guidelines for
  high-dose-rate prostate brachytherapy.
\newblock \emph{Brachytherapy},
  \href{http://dx.doi.org/10.1016/j.brachy.2011.09.008}{11\penalty0
  (1):\penalty0 20--32}, 2012.

\end{thebibliography}
\appendix
%% filename : SupplData_20120421[AH].tex

%\addtolength{\hoffset}{-3cm}
%\addtolength{\topmargin}{-2cm}
%\addtolength{\textheight}{4cm}
%\addtolength{\textwidth}{6cm}

\begin{landscape}

\section{Relative dwell time difference restricted}\label{appmrrel}

\subsection{(LD) model}
\begin{figure}[!h]
   \centering
   \subfloat[]{\label{LPreldevLOVpt1}\includegraphics[width=0.35\textwidth]{figures/LPreldevLOVpt1.pdf}}
   \qquad\qquad
   \subfloat[]{\label{LPreldevLOVpt2}\includegraphics[width=0.35\textwidth]{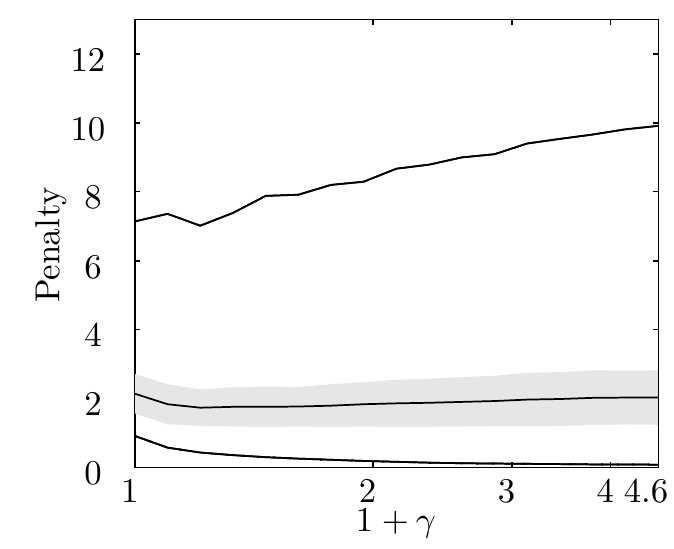}}
   \qquad\qquad
   \subfloat[]{\label{LPreldevLOVpt3}\includegraphics[width=0.35\textwidth]{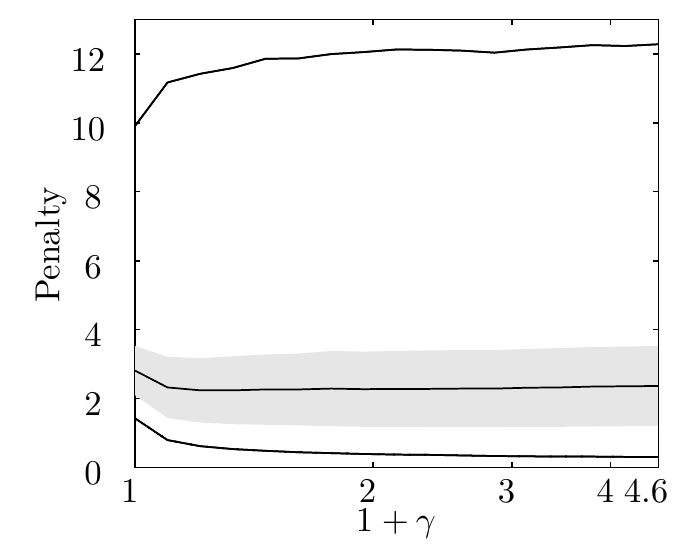}}
   \caption{Objective function value for all patients as function of the maximum relative dwell time difference $\gamma$. The solid lines represent the minimum, mean and maximum values, and the dotted line is the pre-plan value. The grey area denotes values at most one standard deviation from the mean.}\label{LPreldevLOV}
\end{figure}
\end{landscape}
\begin{landscape}
%DHI
\begin{figure}
   \centering
   \subfloat[]{\label{LPreldevDHIpt1}\includegraphics[width=0.35\textwidth]{figures/LPreldevDHIpt1.pdf}}
   \qquad\qquad
   \subfloat[]{\label{LPreldevDHIpt2}\includegraphics[width=0.35\textwidth]{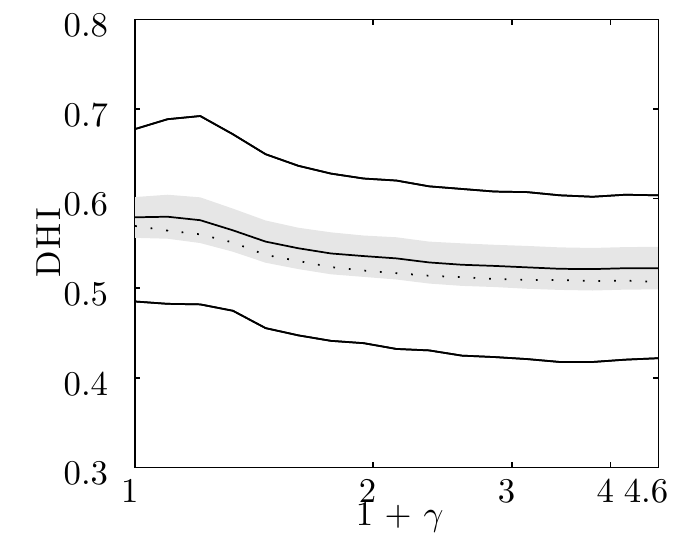}}
   \qquad\qquad
   \subfloat[]{\label{LPreldevDHIpt3}\includegraphics[width=0.35\textwidth]{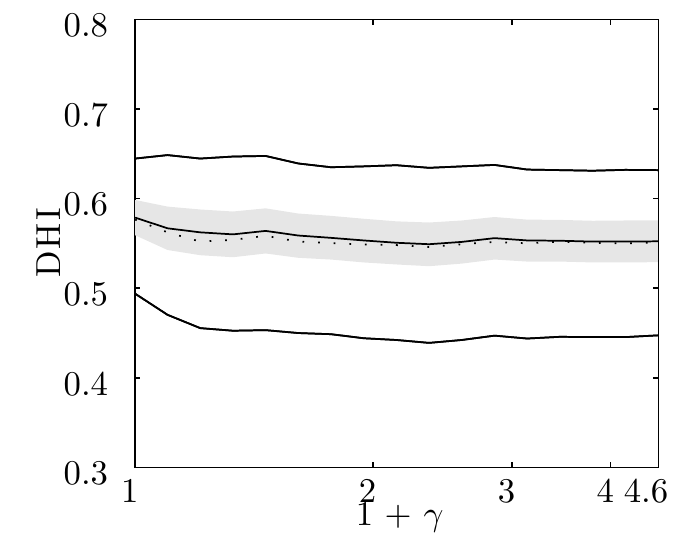}}
   \caption{DHI for all patients as function of the maximum relative dwell time difference $\gamma$. The solid lines represent the minimum, mean and maximum values, and the dotted line is the pre-plan value. The grey area denotes values at most one standard deviation from the mean.}\label{LPreldevDHI}
\end{figure}
%DVHc
\begin{figure}
   \centering
   \subfloat[]{\label{LPreldevDVHcpt1}\includegraphics[width=0.35\textwidth]{figures/LPreldevDVHcpt1.pdf}}
   \qquad
   \subfloat[]{\label{LPreldevDVHcpt2}\includegraphics[width=0.35\textwidth]{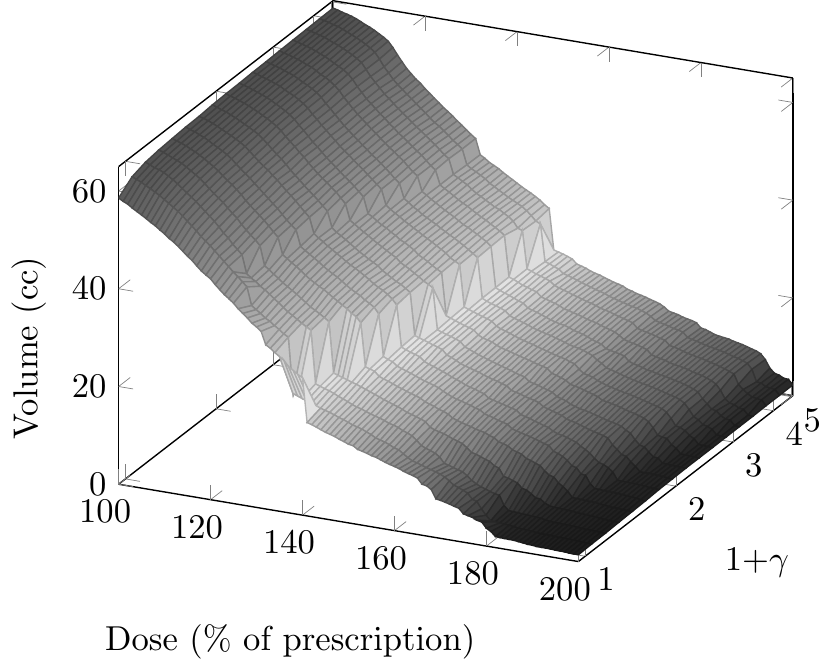}}
   \qquad
   \subfloat[]{\label{LPreldevDVHcpt3}\includegraphics[width=0.35\textwidth]{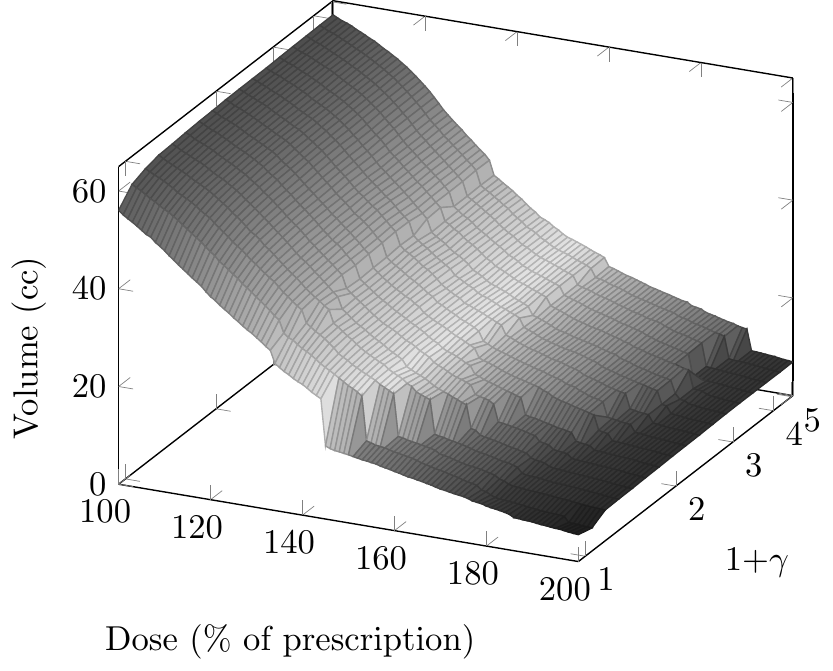}}
   \caption{PTV DVH$^c$ for all patients.}\label{LPreldevDVHc}
\end{figure}
\end{landscape}
\begin{landscape}
%PTV D90
\begin{figure}
   \centering
   \subfloat[]{\label{LPreldevPTVD90pt1}\includegraphics[width=0.35\textwidth]{figures/LPreldevPTVD90pt1.pdf}}
   \qquad\qquad
   \subfloat[]{\label{LPreldevPTVD90pt2}\includegraphics[width=0.35\textwidth]{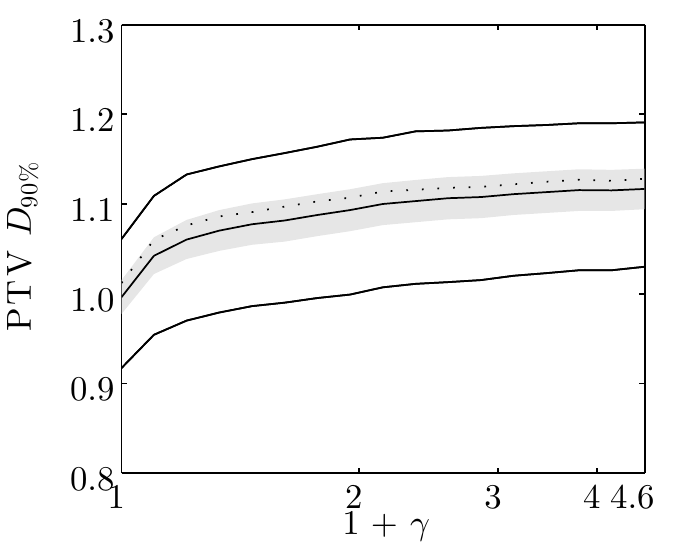}}
   \qquad\qquad
   \subfloat[]{\label{LPreldevPTVD90pt3}\includegraphics[width=0.35\textwidth]{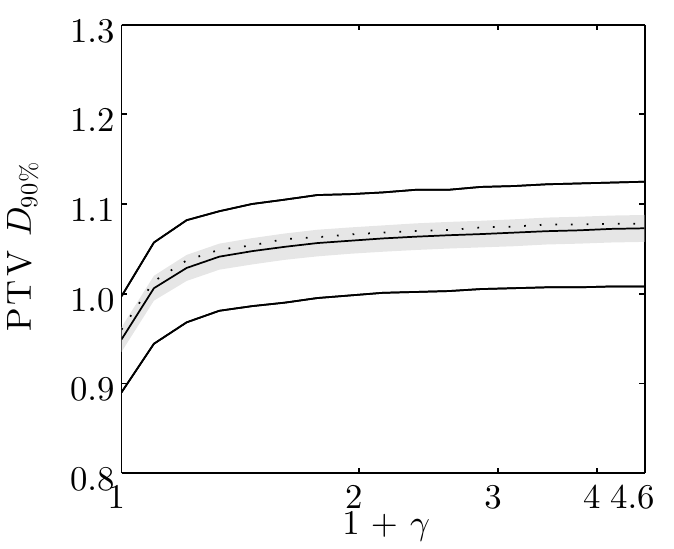}}
   \caption{$D_{90\%}$(PTV) for all patients as function of the maximum relative dwell time difference $\gamma$. The solid lines represent the minimum, mean and maximum values, and the dotted line is the pre-plan value. The grey area denotes values at most one standard deviation from the mean.}\label{LPreldevPTVD90}
\end{figure}
%PTV V100
\begin{figure}
   \centering
   \subfloat[]{\label{LPreldevPTVV100pt1}\includegraphics[width=0.35\textwidth]{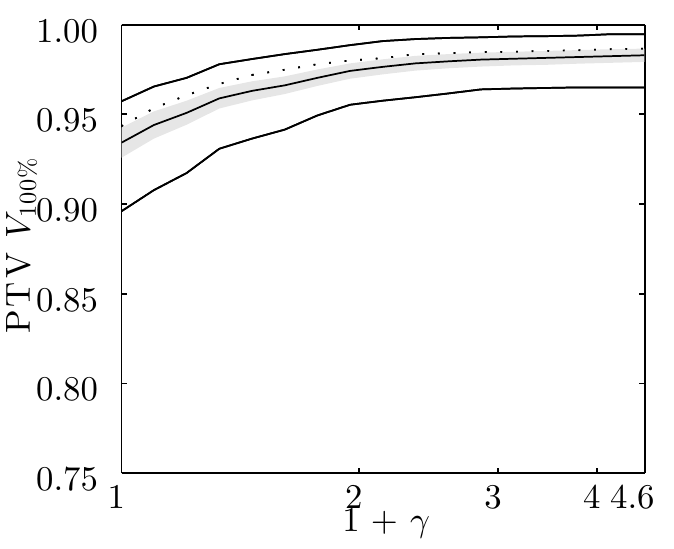}}
   \qquad\qquad
   \subfloat[]{\label{LPreldevPTVV100pt2}\includegraphics[width=0.35\textwidth]{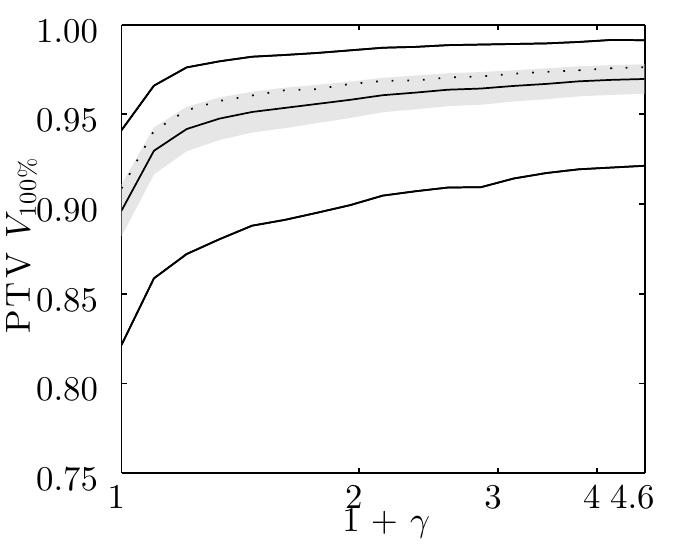}}
   \qquad\qquad
   \subfloat[]{\label{LPreldevPTVV100pt3}\includegraphics[width=0.35\textwidth]{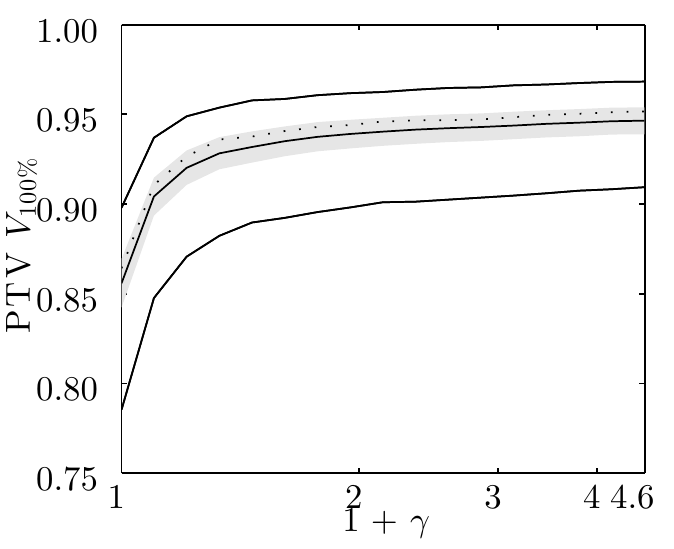}}
   \caption{$V_{100\%}$(PTV) for all patients as function of the maximum relative dwell time difference $\gamma$. The solid lines represent the minimum, mean and maximum values, and the dotted line is the pre-plan value. The grey area denotes values at most one standard deviation from the mean.}\label{LPreldevPTVV100}
\end{figure}
%PTV V150
\begin{figure}
   \centering
   \subfloat[]{\label{LPreldevPTVV150pt1}\includegraphics[width=0.35\textwidth]{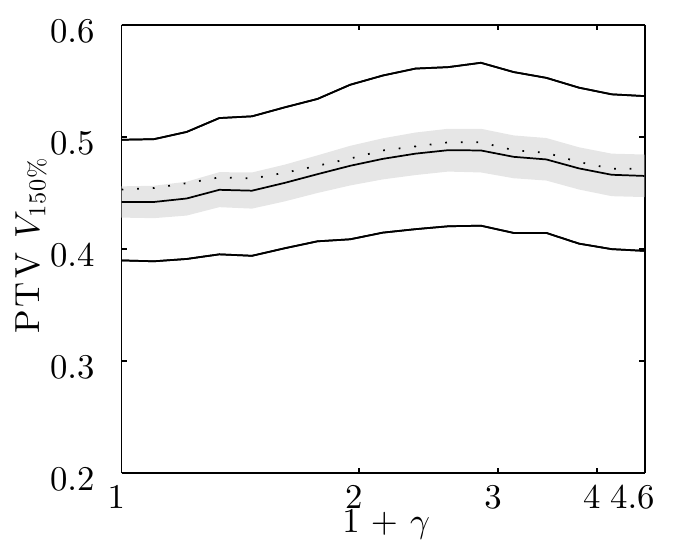}}
   \qquad\qquad
   \subfloat[]{\label{LPreldevPTVV150pt2}\includegraphics[width=0.35\textwidth]{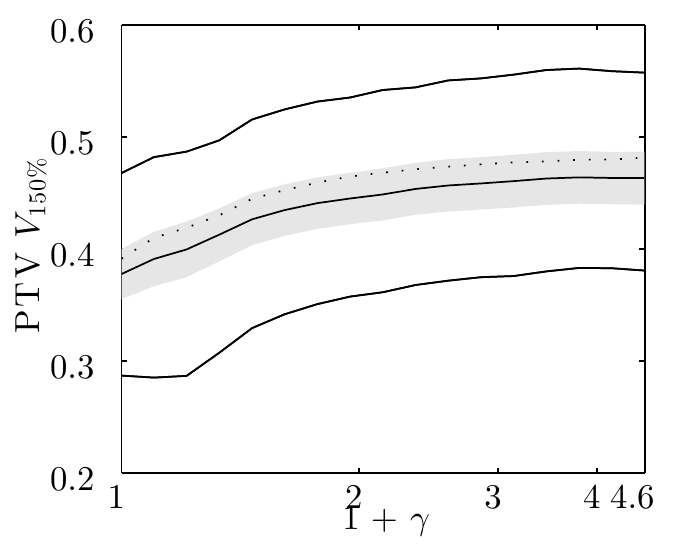}}
   \qquad\qquad
   \subfloat[]{\label{LPreldevPTVV150pt3}\includegraphics[width=0.35\textwidth]{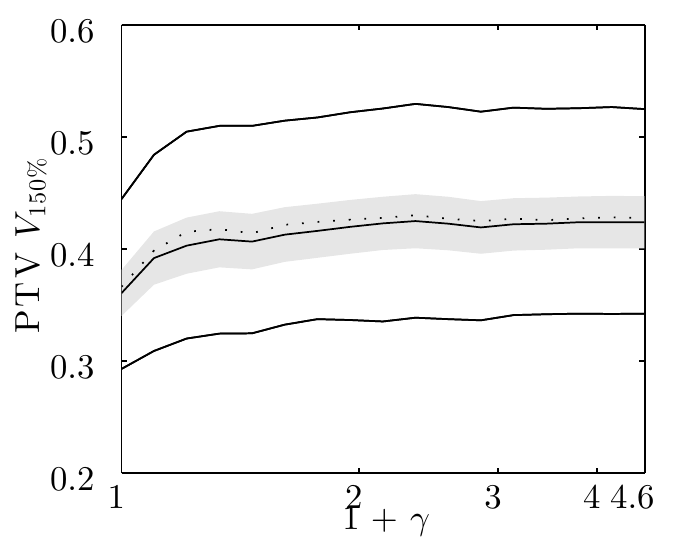}}
   \caption{$V_{150\%}$(PTV) for all patients as function of the maximum relative dwell time difference $\gamma$. The solid lines represent the minimum, mean and maximum values, and the dotted line is the pre-plan value. The grey area denotes values at most one standard deviation from the mean.}\label{LPreldevPTVV150}
\end{figure}
%PTV V200
\begin{figure}
   \centering
   \subfloat[]{\label{LPreldevPTVV200pt1}\includegraphics[width=0.35\textwidth]{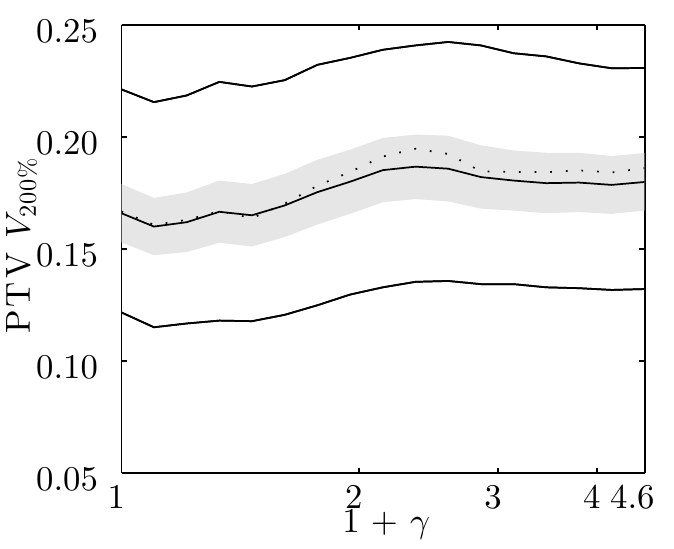}}
   \qquad\qquad
   \subfloat[]{\label{LPreldevPTVV200pt2}\includegraphics[width=0.35\textwidth]{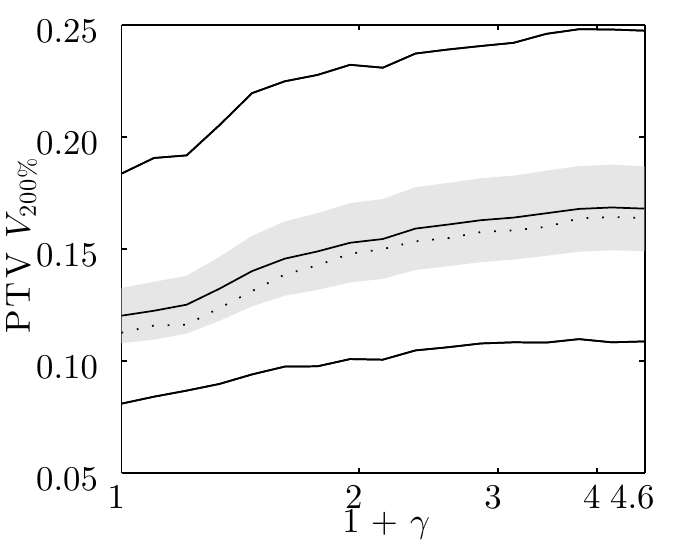}}
   \qquad\qquad
   \subfloat[]{\label{LPreldevPTVV200pt3}\includegraphics[width=0.35\textwidth]{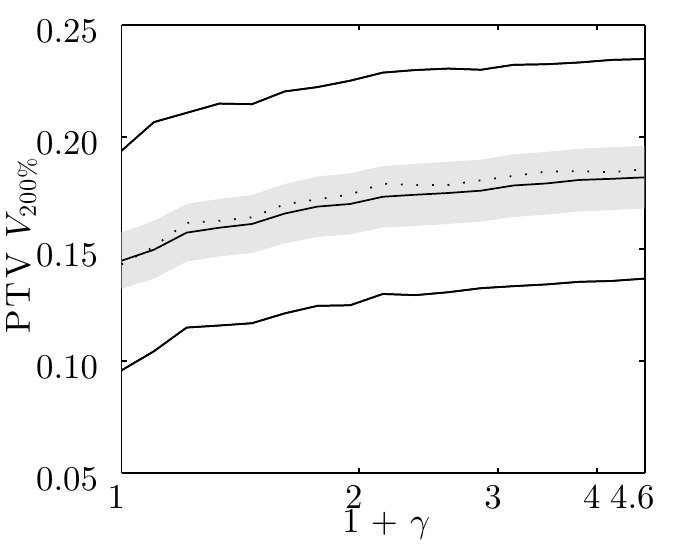}}
   \caption{$V_{200\%}$(PTV) for all patients as function of the maximum relative dwell time difference $\gamma$. The solid lines represent the minimum, mean and maximum values, and the dotted line is the pre-plan value. The grey area denotes values at most one standard deviation from the mean.}\label{LPreldevPTVV200}
\end{figure}
\end{landscape}
\begin{landscape}
% Rectum D10
\begin{figure}
   \centering
   \subfloat[]{\label{LPreldevRD10pt1}\includegraphics[width=0.35\textwidth]{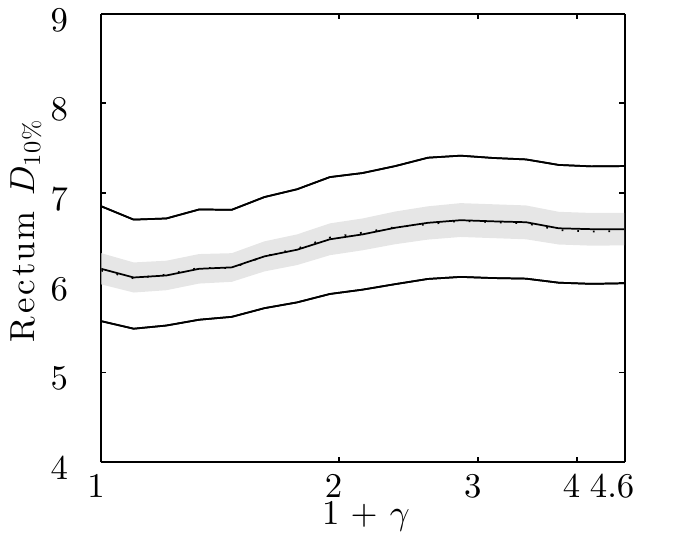}}
   \qquad\qquad
   \subfloat[]{\label{LPreldevRD10pt2}\includegraphics[width=0.35\textwidth]{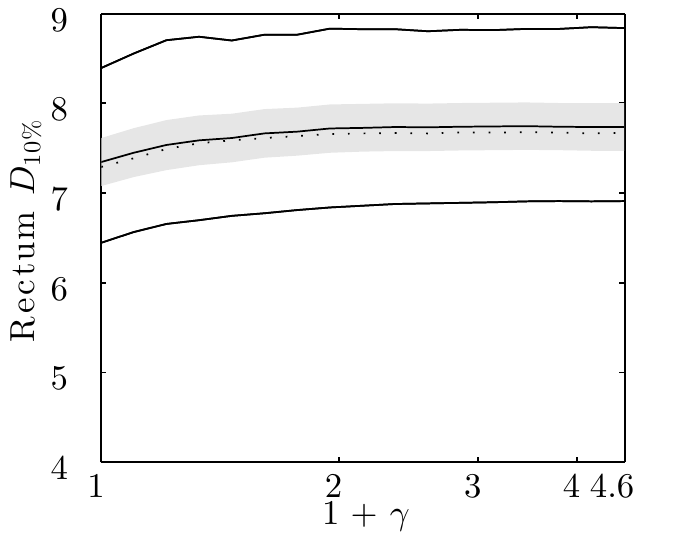}}
   \qquad\qquad
   \subfloat[]{\label{LPreldevRD10pt3}\includegraphics[width=0.35\textwidth]{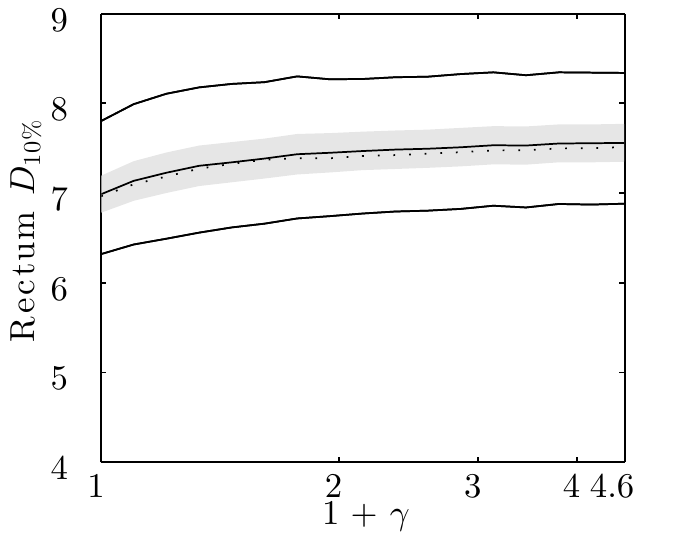}}
   \caption{$D_{10\%}$(rectum) for all patients as function of the maximum relative dwell time difference $\gamma$. The solid lines represent the minimum, mean and maximum values, and the dotted line is the pre-plan value. The grey area denotes values at most one standard deviation from the mean.}\label{LPreldevRD10}
\end{figure}
% Rectum D2cc
\begin{figure}
   \centering
   \subfloat[]{\label{LPreldevRD2ccpt1}\includegraphics[width=0.35\textwidth]{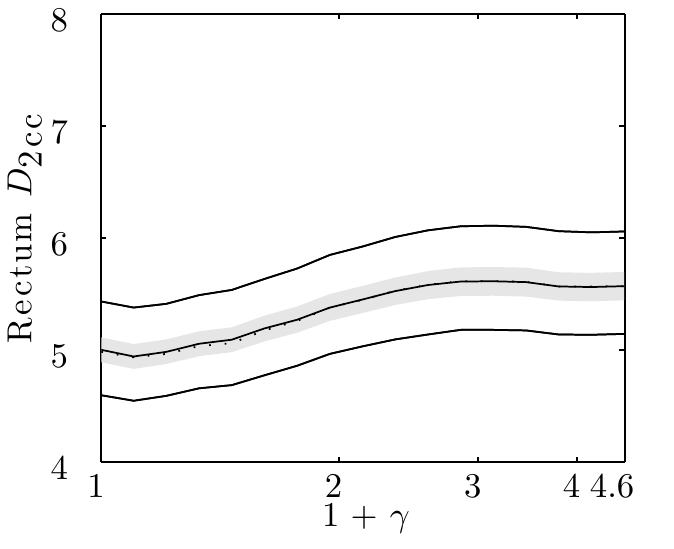}}
   \qquad\qquad
   \subfloat[]{\label{LPreldevRD2ccpt2}\includegraphics[width=0.35\textwidth]{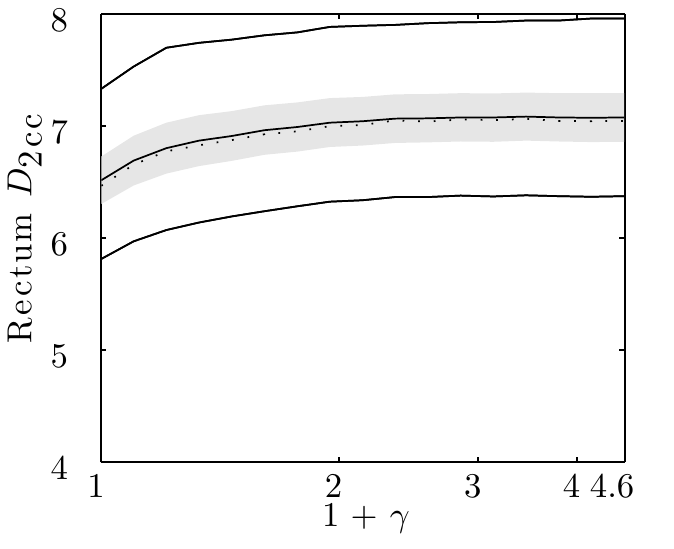}}
   \qquad\qquad
   \subfloat[]{\label{LPreldevRD2ccpt3}\includegraphics[width=0.35\textwidth]{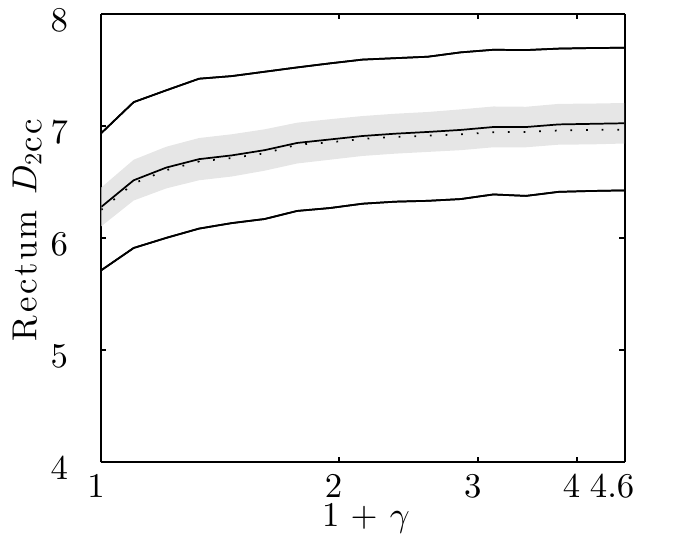}}
   \caption{$D_{2\textnormal{cc}}$(rectum) for all patients as function of the maximum relative dwell time difference $\gamma$. The solid lines represent the minimum, mean and maximum values, and the dotted line is the pre-plan value. The grey area denotes values at most one standard deviation from the mean.}\label{LPreldevRD2cc}
\end{figure}
\end{landscape}
\begin{landscape}
% Urethra D10
\begin{figure}
   \centering
   \subfloat[]{\label{LPreldevUD10pt1}\includegraphics[width=0.35\textwidth]{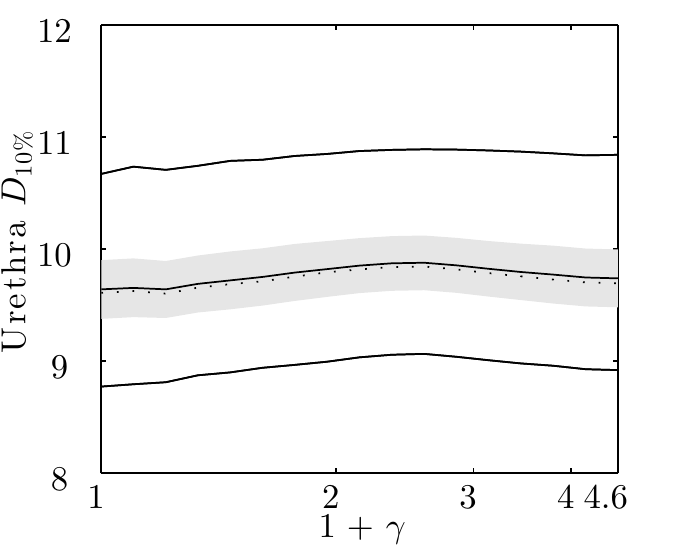}}
   \qquad\qquad
   \subfloat[]{\label{LPreldevUD10pt2}\includegraphics[width=0.35\textwidth]{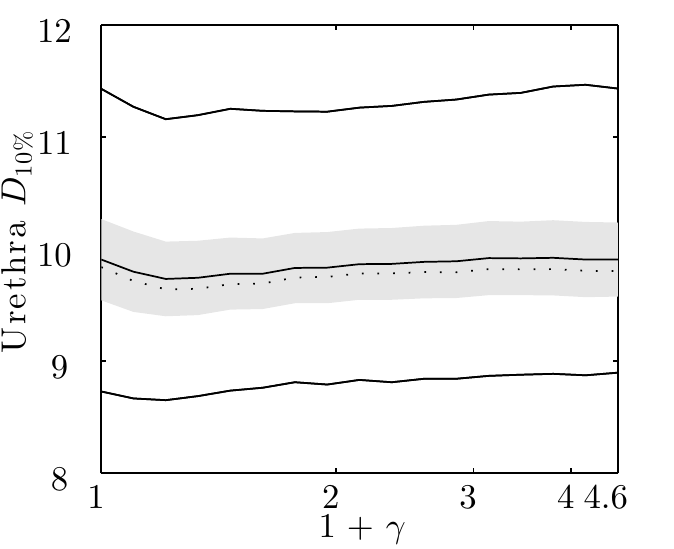}}
   \qquad\qquad
   \subfloat[]{\label{LPreldevUD10pt3}\includegraphics[width=0.35\textwidth]{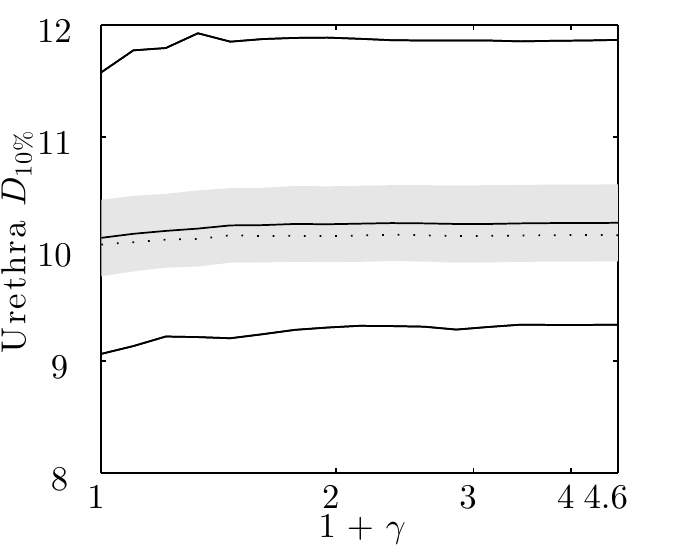}}
   \caption{$D_{10\%}$(urethra) for all patients as function of the maximum relative dwell time difference $\gamma$. The solid lines represent the minimum, mean and maximum values, and the dotted line is the pre-plan value. The grey area denotes values at most one standard deviation from the mean.}\label{LPreldevUD10}
\end{figure}
\nopagebreak[4]
% Urethra D01cc
\begin{figure}[t]
   \centering
   \subfloat[]{\label{LPreldevUD0.1ccpt1}\includegraphics[width=0.35\textwidth]{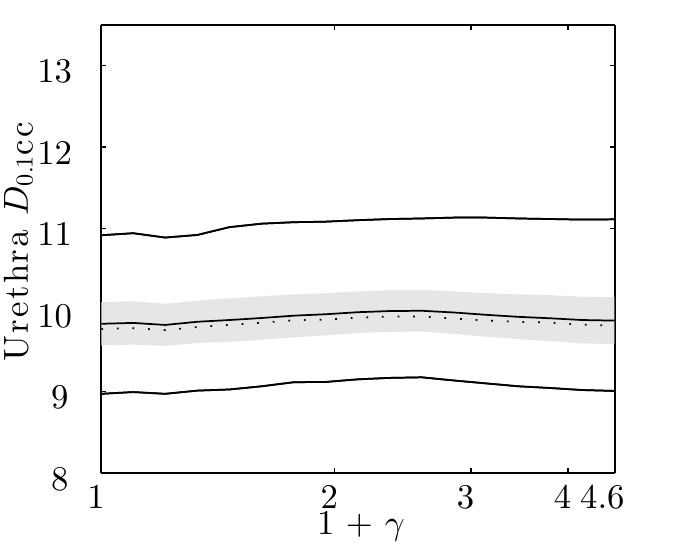}}
   \qquad\qquad
   \subfloat[]{\label{LPreldevUD0.1ccpt2}\includegraphics[width=0.35\textwidth]{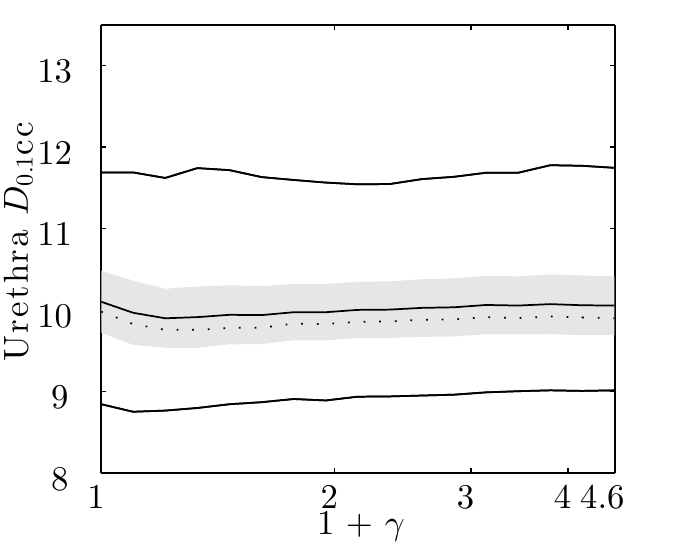}}
   \qquad\qquad
   \subfloat[]{\label{LPreldevUD0.1ccpt3}\includegraphics[width=0.35\textwidth]{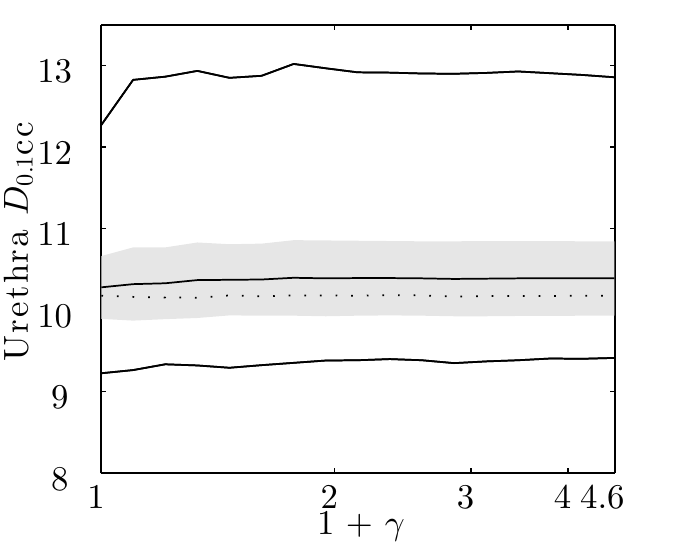}}
   \caption{$D_{0.1\textnormal{cc}}$(urethra) for all patients as function of the maximum relative dwell time difference $\gamma$. The solid lines represent the minimum, mean and maximum values, and the dotted line is the pre-plan value. The grey area denotes values at most one standard deviation from the mean.}\label{LPreldevUD0.1cc}
\end{figure}
%\qquad
%\clearpage
%\pagebreak[4]
%\qquad
\end{landscape}
\begin{landscape}
\subsection{(LDV) model}
%PTV V100
\begin{figure}[!h]
   \centering
   \subfloat[]{\label{DVHreldevPTVV100pt1}\includegraphics[width=0.35\textwidth]{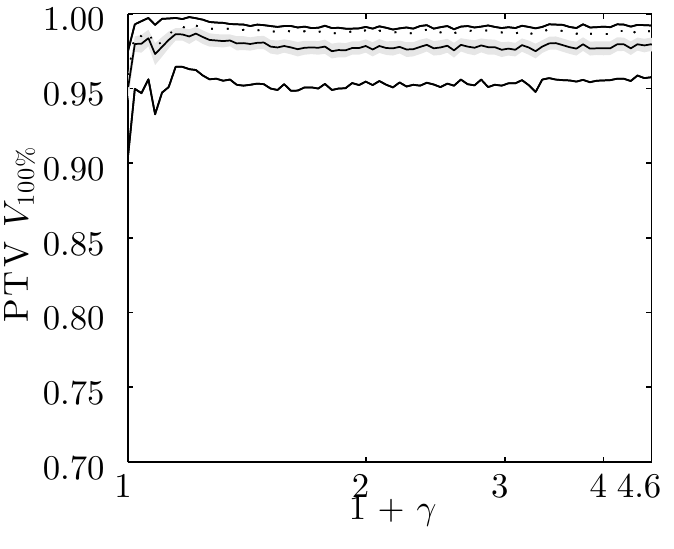}}
   \qquad\qquad
   \subfloat[]{\label{DVHreldevPTVV100pt2}\includegraphics[width=0.35\textwidth]{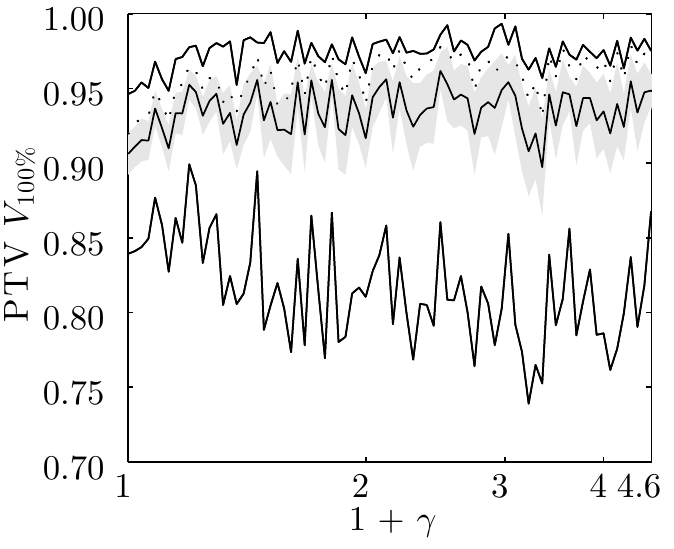}}
   \qquad\qquad
   \subfloat[]{\label{DVHreldevPTVV100pt3}\includegraphics[width=0.35\textwidth]{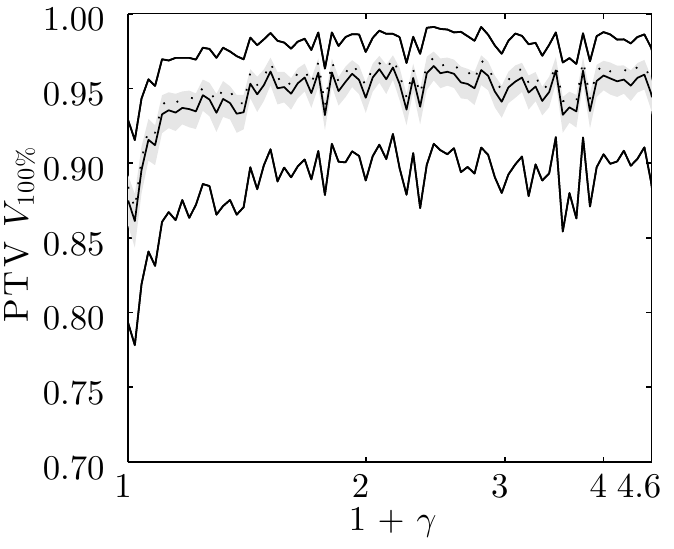}}
   \caption{$V_{100\%}$(PTV) for all patients as function of the maximum relative dwell time difference $\gamma$. The solid lines represent the minimum, mean and maximum values, and the dotted line is the pre-plan value. The grey area denotes values at most one standard deviation from the mean.}\label{DVHreldevPTVV100}
\end{figure}
%DHI
\begin{figure}[h]
   \centering
   \subfloat[.]{\label{DVHreldevDHIpt1}\includegraphics[width=0.35\textwidth]{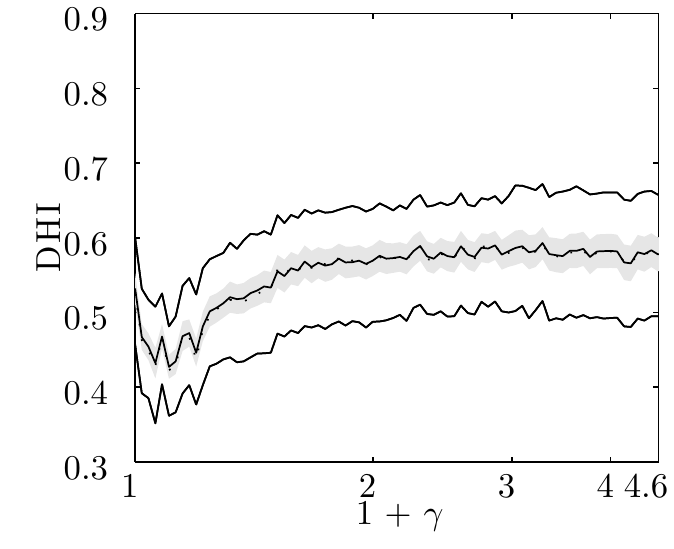}}
   \qquad\qquad
   \subfloat[.]{\label{DVHreldevDHIpt2}\includegraphics[width=0.35\textwidth]{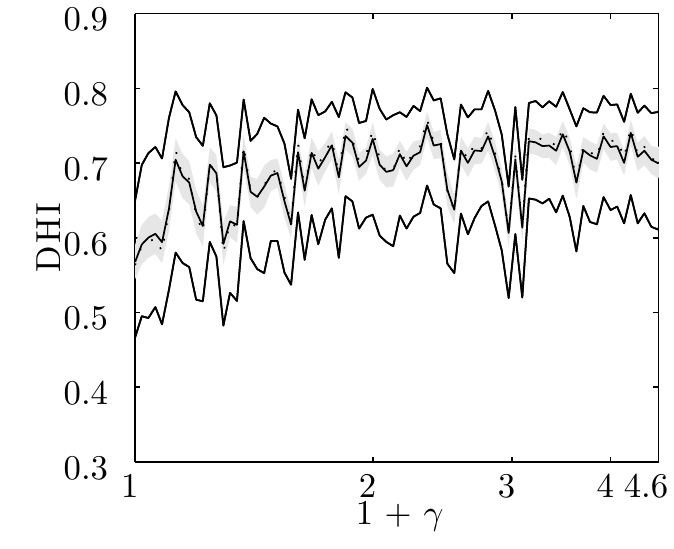}}
   \qquad\qquad
   \subfloat[]{\label{DVHreldevDHIpt3}\includegraphics[width=0.35\textwidth]{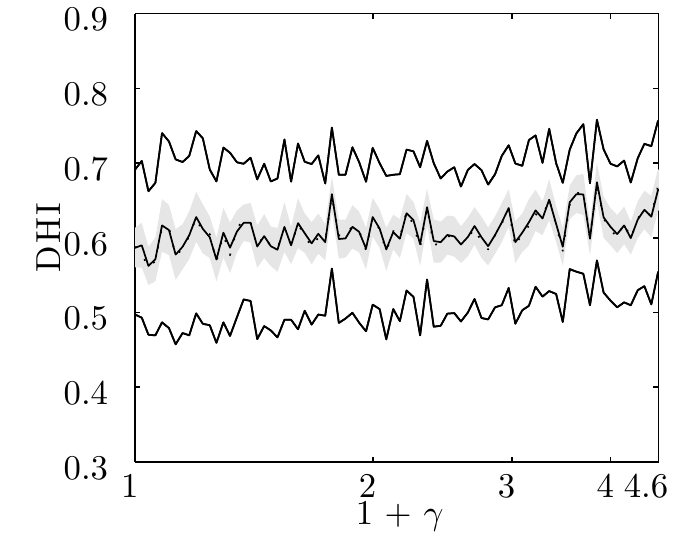}}
   \caption{DHI for all patients as function of the maximum relative dwell time difference $\gamma$. The solid lines represent the minimum, mean and maximum values, and the dotted line is the pre-plan value. The grey area denotes values at most one standard deviation from the mean.}\label{DVHreldevDHI}
\end{figure}
%DVHc
\begin{figure}[h]
   \centering
   \subfloat[]{\label{DVHreldevDVHcpt1}\includegraphics[width=0.35\textwidth]{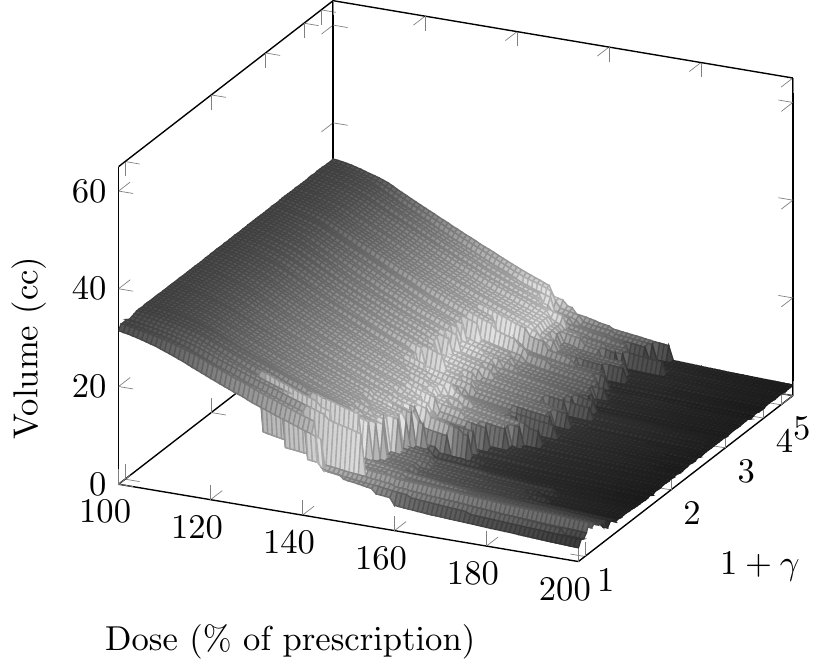}}
      \qquad
      \subfloat[]{\label{DVHreldevDVHcpt2}\includegraphics[width=0.35\textwidth]{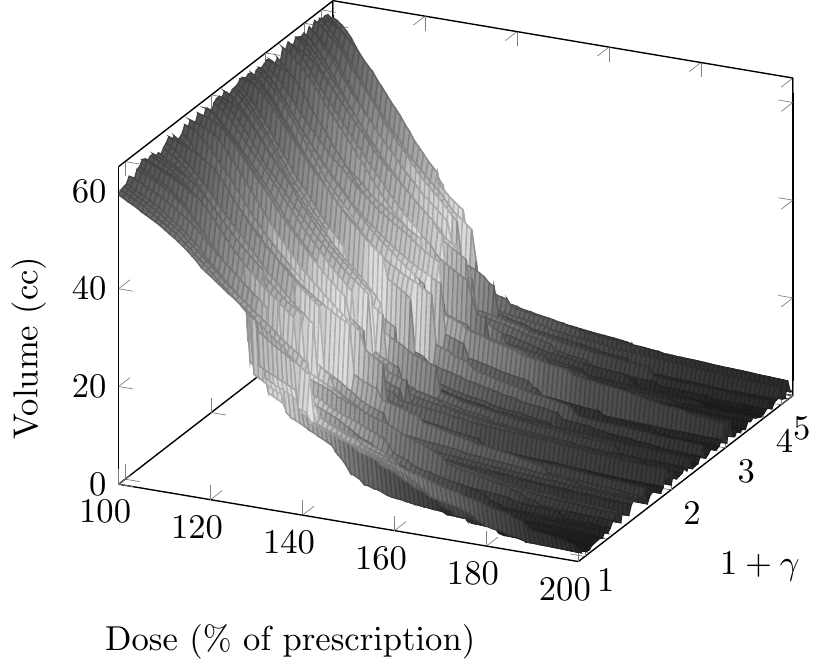}}
      \qquad
      \subfloat[]{\label{DVHreldevDVHcpt3}\includegraphics[width=0.35\textwidth]{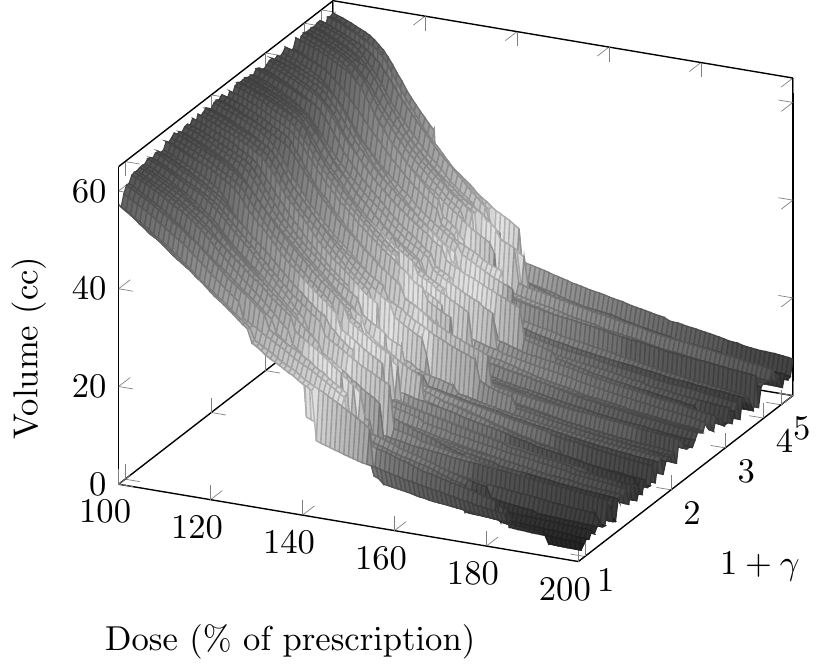}}
   \caption{PTV DVH$^c$ for all patients.}\label{DVHreldevDVHc}
\end{figure}
\pagebreak[4]
%PTV D90
\begin{figure}[!h]
   \centering
   \subfloat[]{\label{DVHreldevPTVD90pt1}\includegraphics[width=0.35\textwidth]{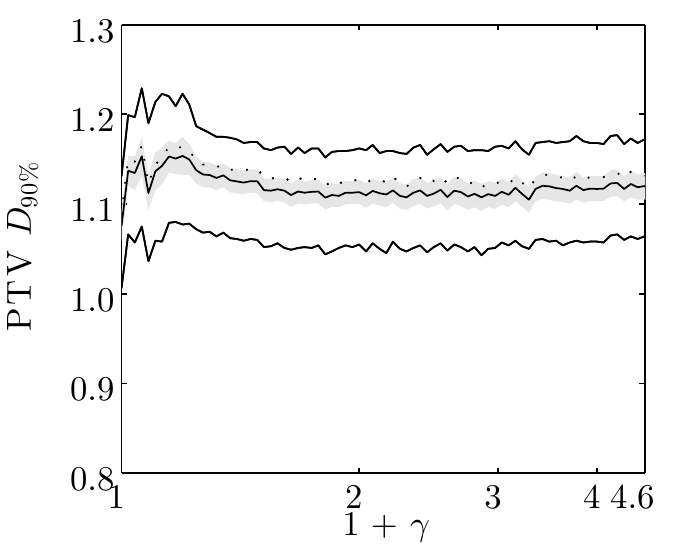}}
   \qquad\qquad
   \subfloat[]{\label{DVHreldevPTVD90pt2}\includegraphics[width=0.35\textwidth]{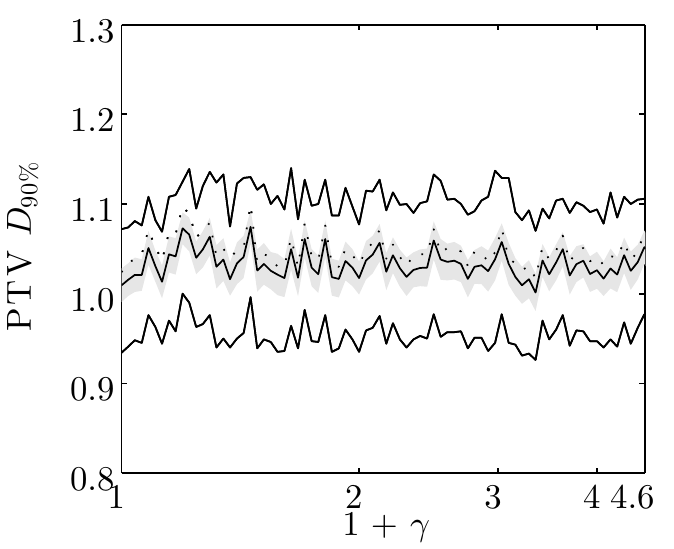}}
   \qquad\qquad
   \subfloat[]{\label{DVHreldevPTVD90pt3}\includegraphics[width=0.35\textwidth]{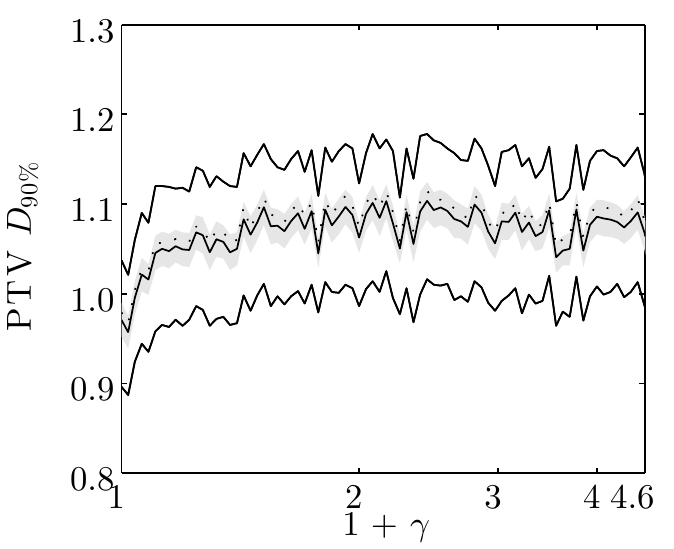}}
   \caption{$D_{90\%}$(PTV) for all patients as function of the maximum relative dwell time difference $\gamma$. The solid lines represent the minimum, mean and maximum values, and the dotted line is the pre-plan value. The grey area denotes values at most one standard deviation from the mean.}\label{DVHreldevPTVD90}
\end{figure}
%PTV V150
\begin{figure}[h]
   \centering
   \subfloat[]{\label{DVHreldevPTVV150pt1}\includegraphics[width=0.35\textwidth]{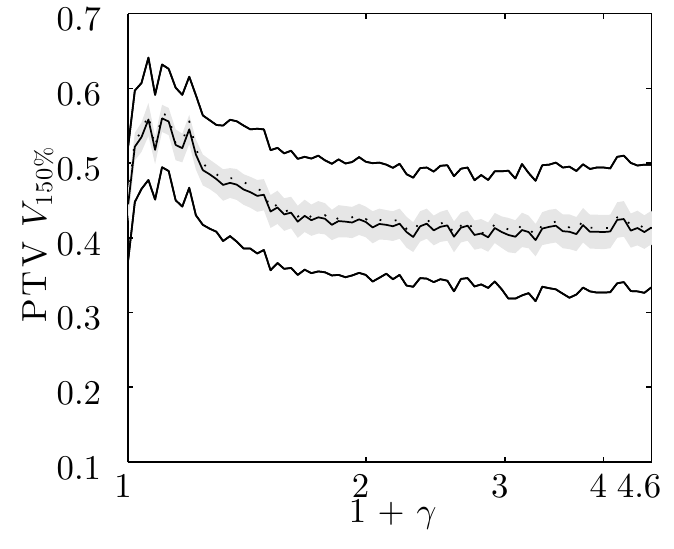}}
   \qquad\qquad
   \subfloat[]{\label{DVHreldevPTVV150pt2}\includegraphics[width=0.35\textwidth]{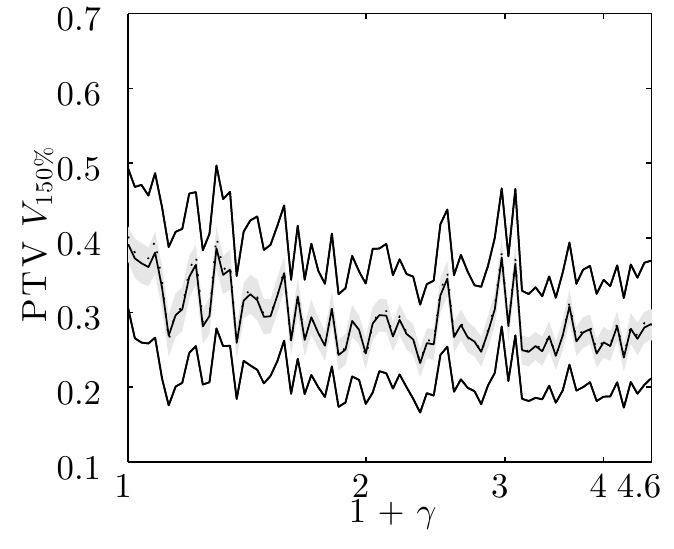}}
   \qquad\qquad
   \subfloat[]{\label{DVHreldevPTVV150pt3}\includegraphics[width=0.35\textwidth]{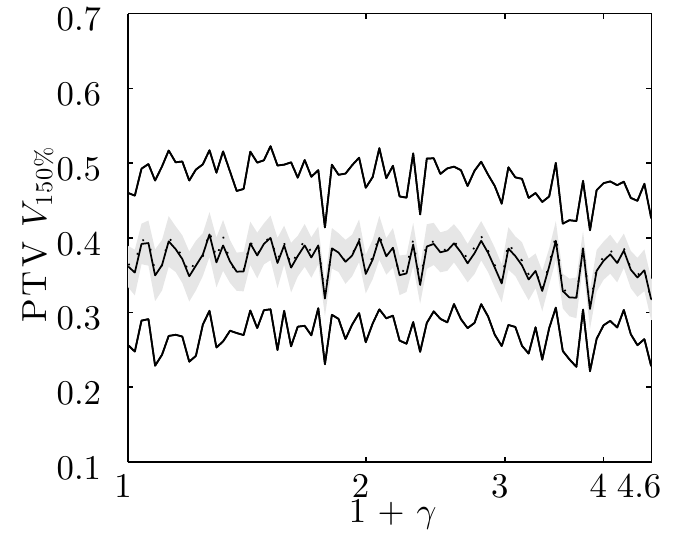}}
   \caption{$V_{150\%}$(PTV) for all patients as function of the maximum relative dwell time difference $\gamma$. The solid lines represent the minimum, mean and maximum values, and the dotted line is the pre-plan value. The grey area denotes values at most one standard deviation from the mean.}\label{DVHreldevPTVV150}
\end{figure}
\nopagebreak[4]
%PTV V200
\begin{figure}[h]
   \centering
   \subfloat[]{\label{DVHreldevPTVV200pt1}\includegraphics[width=0.35\textwidth]{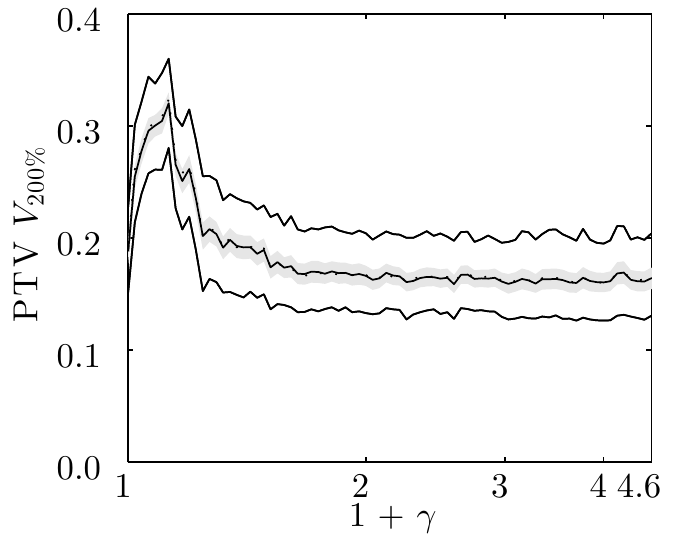}}
   \qquad\qquad
   \subfloat[]{\label{DVHreldevPTVV200pt2}\includegraphics[width=0.35\textwidth]{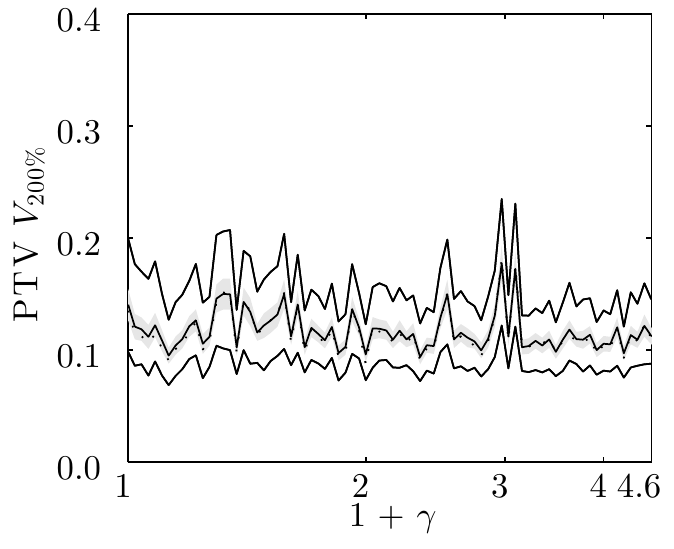}}
   \qquad\qquad
   \subfloat[]{\label{DVHreldevPTVV200pt3}\includegraphics[width=0.35\textwidth]{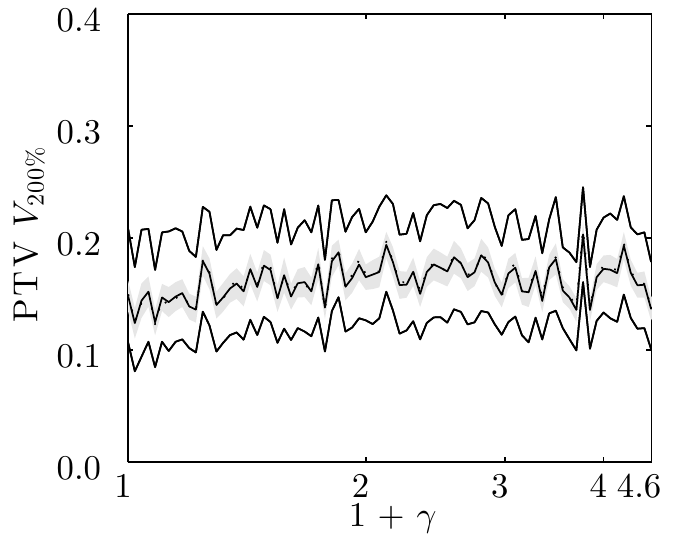}}
   \caption{$V_{200\%}$(PTV) for all patients as function of the maximum relative dwell time difference $\gamma$. The solid lines represent the minimum, mean and maximum values, and the dotted line is the pre-plan value. The grey area denotes values at most one standard deviation from the mean.}\label{DVHreldevPTVV200}
\end{figure}
\pagebreak[4]
% Rectum D10
\begin{figure}[h]
   \centering
   \subfloat[]{\label{DVHreldevRD10pt1}\includegraphics[width=0.35\textwidth]{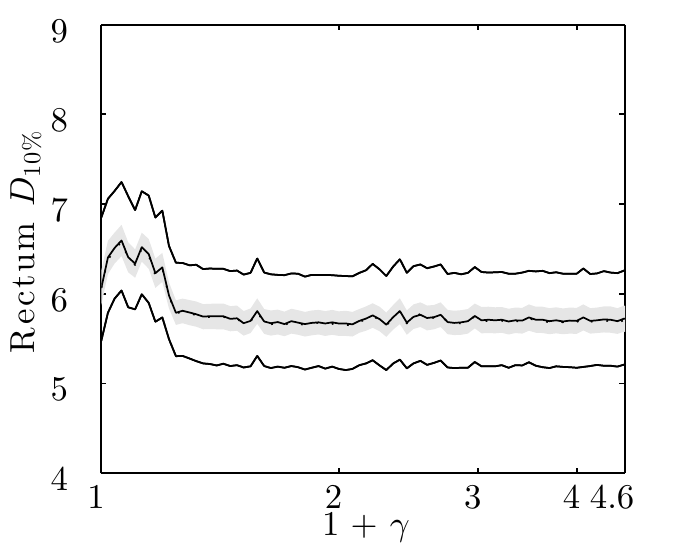}}
   \qquad\qquad
   \subfloat[]{\label{DVHreldevRD10pt2}\includegraphics[width=0.35\textwidth]{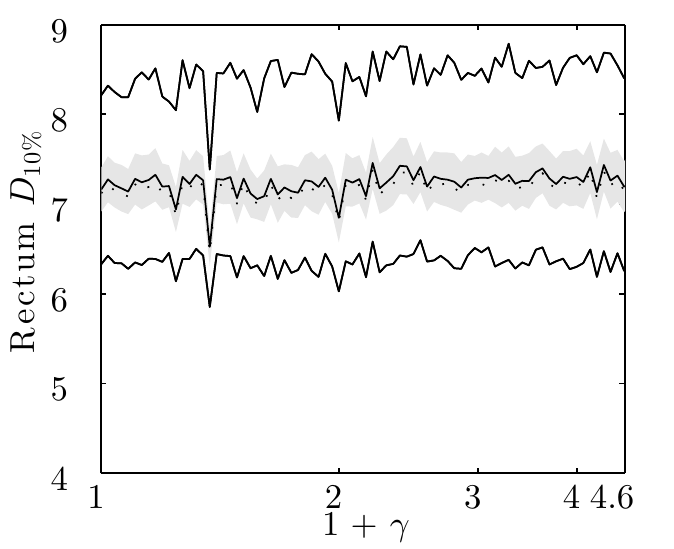}}
   \qquad\qquad
   \subfloat[]{\label{DVHreldevRD10pt3}\includegraphics[width=0.35\textwidth]{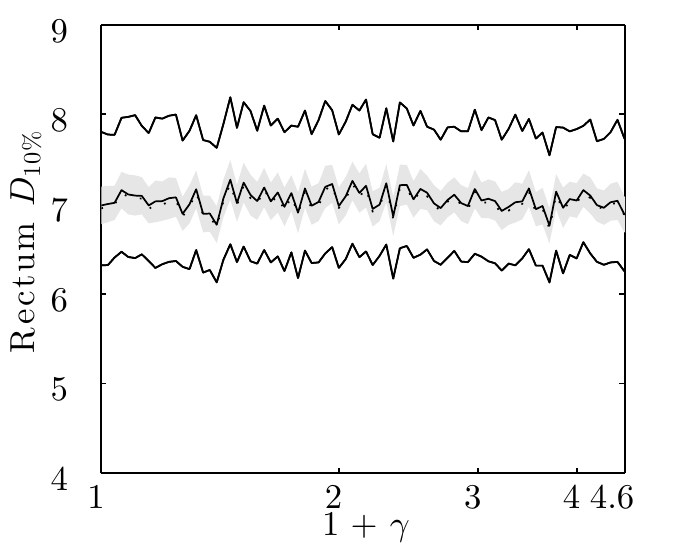}}
   \caption{$D_{10\%}$(rectum) for all patients as function of the maximum relative dwell time difference $\gamma$. The solid lines represent the minimum, mean and maximum values, and the dotted line is the pre-plan value. The grey area denotes values at most one standard deviation from the mean.}\label{DVHreldevRD10}
\end{figure}
\nopagebreak[4]
% Rectum D2cc
\begin{figure}[h]
   \centering
   \subfloat[]{\label{DVHreldevRD2ccpt1}\includegraphics[width=0.35\textwidth]{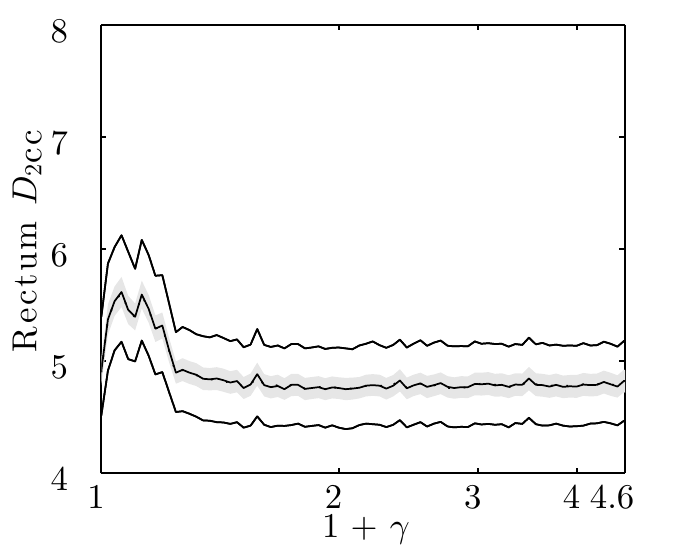}}
   \qquad\qquad
   \subfloat[]{\label{DVHreldevRD2ccpt2}\includegraphics[width=0.35\textwidth]{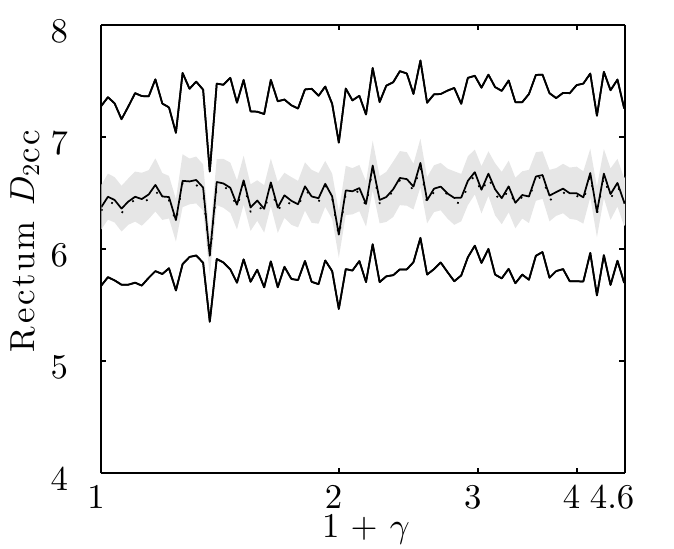}}
   \qquad\qquad
   \subfloat[]{\label{DVHreldevRD2ccpt3}\includegraphics[width=0.35\textwidth]{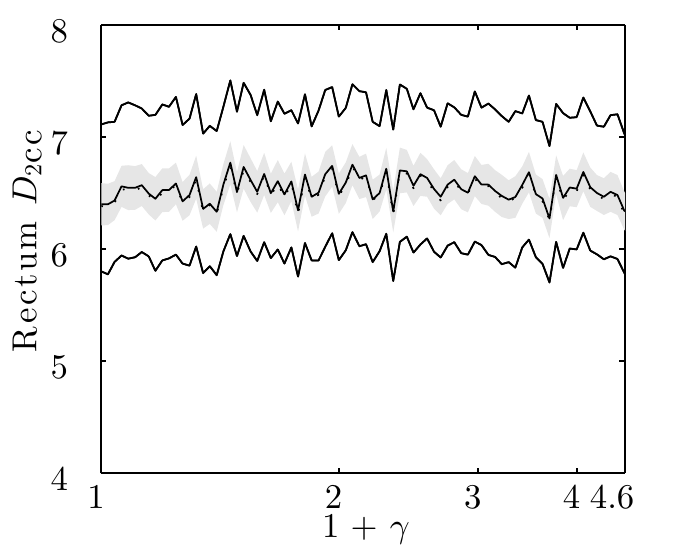}}
   \caption{$D_{2\textnormal{cc}}$(rectum) for all patients as function of the maximum relative dwell time difference $\gamma$. The solid lines represent the minimum, mean and maximum values, and the dotted line is the pre-plan value. The grey area denotes values at most one standard deviation from the mean.}\label{DVHreldevRD2cc}
\end{figure}
\pagebreak[4]
% Urethra D10
\begin{figure}[h]
   \centering
   \subfloat[]{\label{DVHreldevUD10pt1}\includegraphics[width=0.35\textwidth]{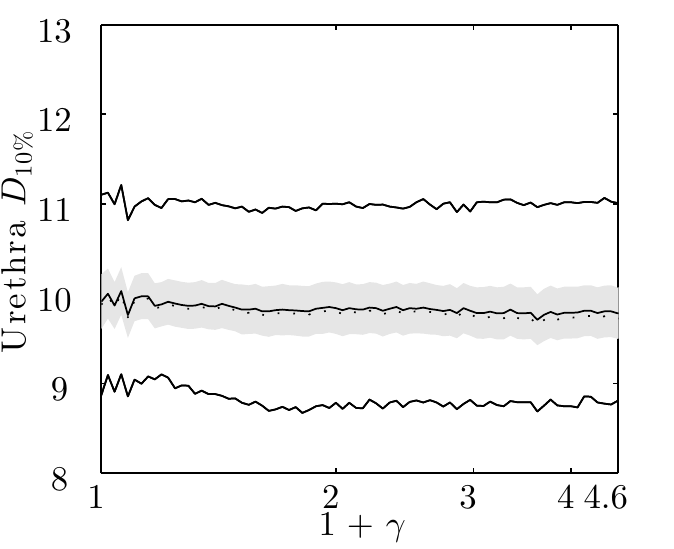}}
   \qquad\qquad
   \subfloat[]{\label{DVHreldevUD10pt2}\includegraphics[width=0.35\textwidth]{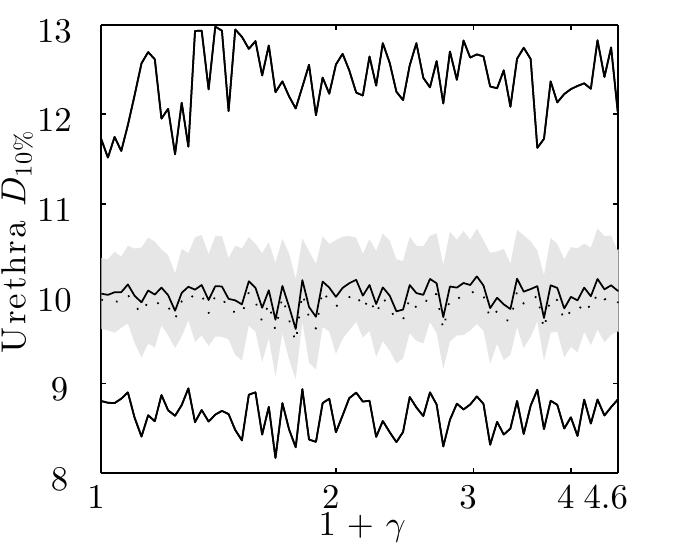}}
   \qquad\qquad
   \subfloat[]{\label{DVHreldevUD10pt3}\includegraphics[width=0.35\textwidth]{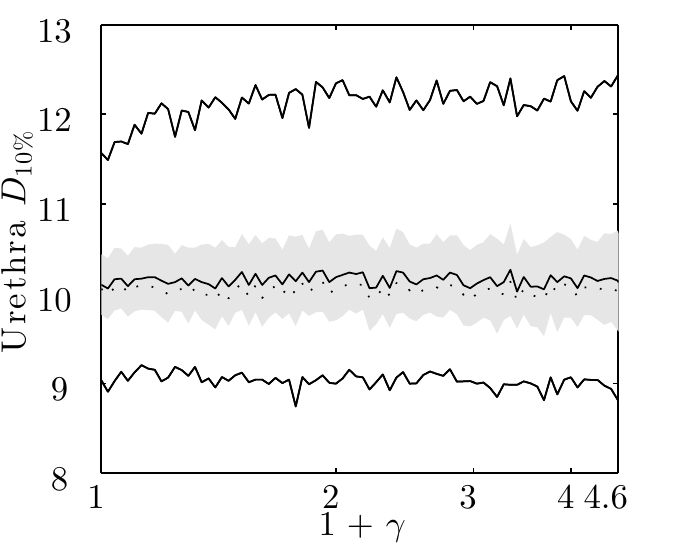}}
   \caption{$D_{10\%}$(urethra) for all patients as function of the maximum relative dwell time difference $\gamma$. The solid lines represent the minimum, mean and maximum values, and the dotted line is the pre-plan value. The grey area denotes values at most one standard deviation from the mean.}\label{DVHreldevUD10}
\end{figure}
\nopagebreak[4]
% Urethra D01cc
\begin{figure}[h]
   \centering
   \subfloat[]{\label{DVHreldevUD0.1ccpt1}\includegraphics[width=0.35\textwidth]{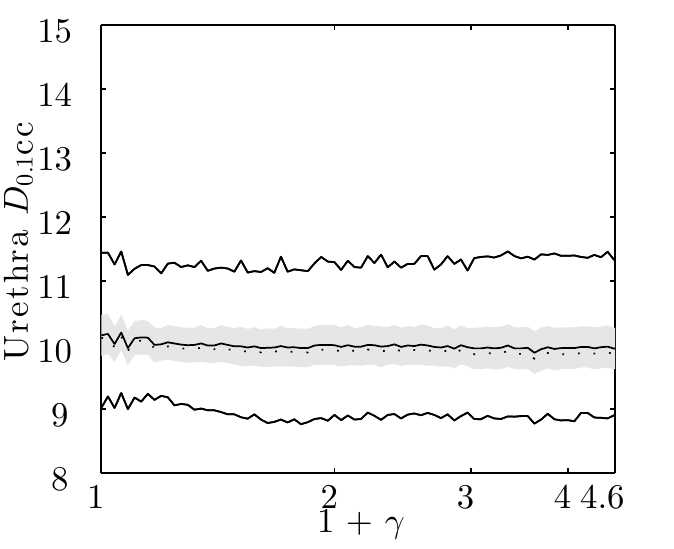}}
   \qquad\qquad
   \subfloat[]{\label{DVHreldevUD0.1ccpt2}\includegraphics[width=0.35\textwidth]{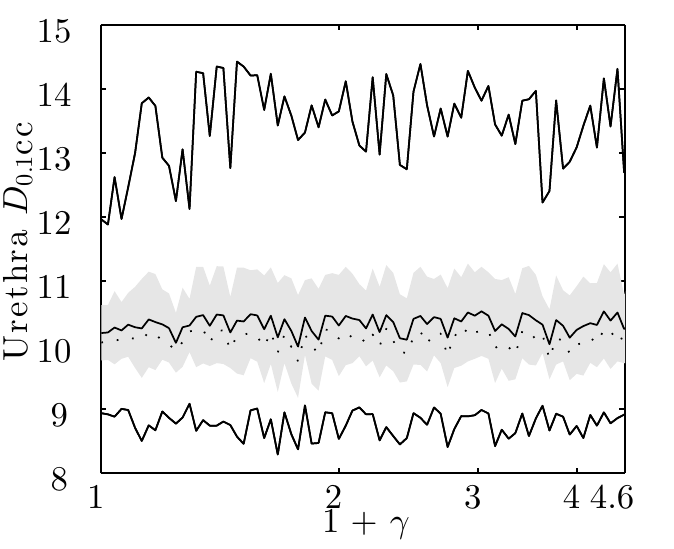}}
   \qquad\qquad
   \subfloat[]{\label{DVHreldevUD0.1ccpt3}\includegraphics[width=0.35\textwidth]{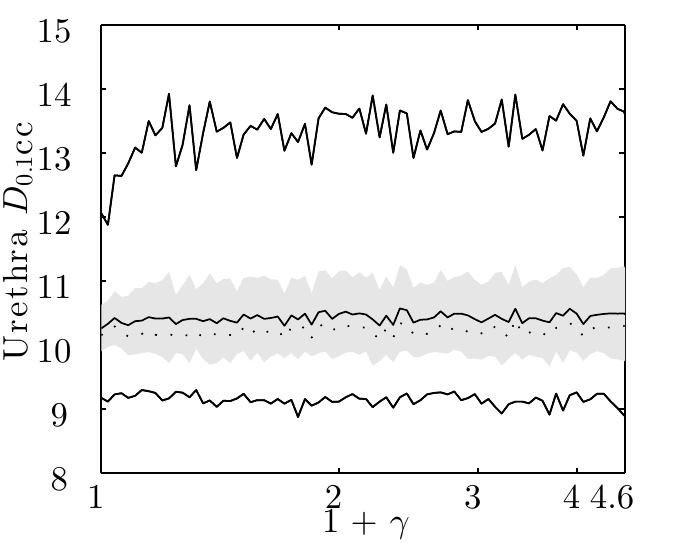}}
   \caption{$D_{0.1\textnormal{cc}}$(urethra) for all patients as function of the maximum relative dwell time difference $\gamma$. The solid lines represent the minimum, mean and maximum values, and the dotted line is the pre-plan value. The grey area denotes values at most one standard deviation from the mean.}\label{DVHreldevUD0.1cc}
\end{figure}

\end{landscape}
\begin{landscape}
\section{Absolute dwell time difference restricted}\label{appmrabs}

\subsection{(LD) model}
\begin{figure}[h]
   \centering
   \subfloat[]{\label{LPabsdevLOVpt1}\includegraphics[width=0.35\textwidth]{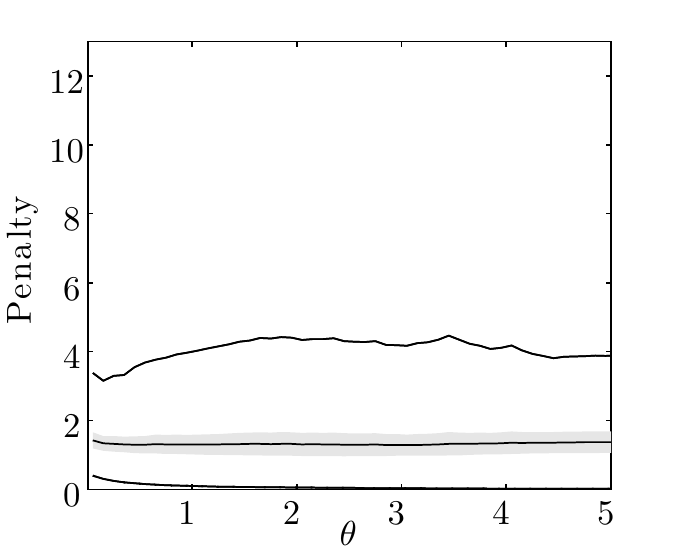}}
   \qquad\qquad
   \subfloat[]{\label{LPabsdevLOVpt2}\includegraphics[width=0.35\textwidth]{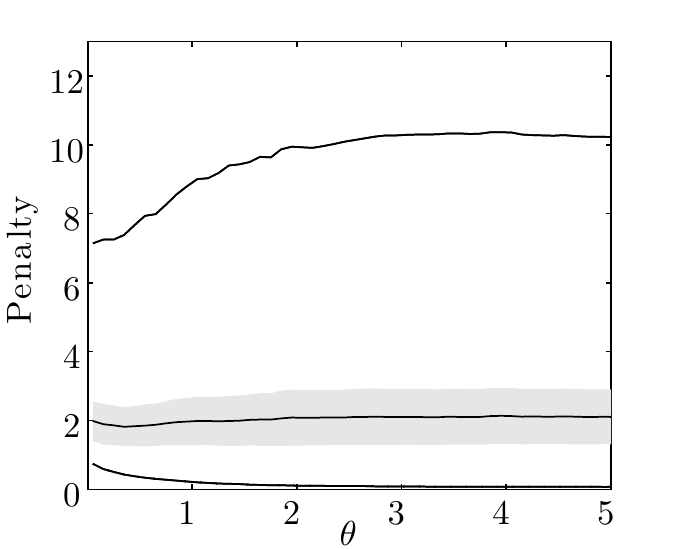}}
   \qquad\qquad
   \subfloat[]{\label{LPabsdevLOVpt3}\includegraphics[width=0.35\textwidth]{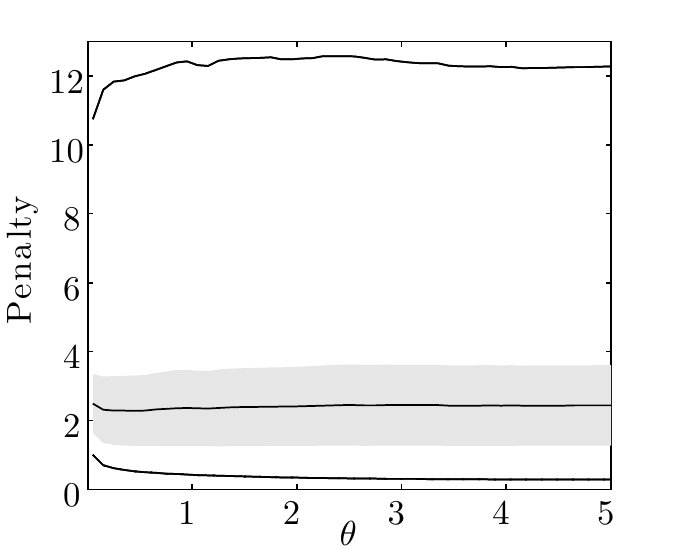}}
   \caption{Objective function value for all patients as function of the absolute dwell time difference $\theta$. The solid lines represent the minimum, mean and maximum values, and the dotted line is the pre-plan value. The grey area denotes values at most one standard deviation from the mean.}\label{LPabsdevLOV}
\end{figure}
%DHI
\begin{figure}[h]
   \centering
   \subfloat[]{\label{LPabsdevDHIpt1}\includegraphics[width=0.35\textwidth]{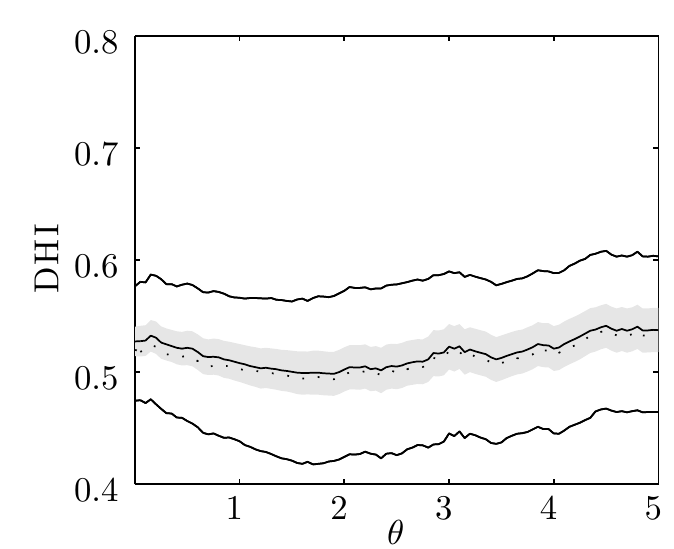}}
   \qquad\qquad
   \subfloat[]{\label{LPabsdevDHIpt2}\includegraphics[width=0.35\textwidth]{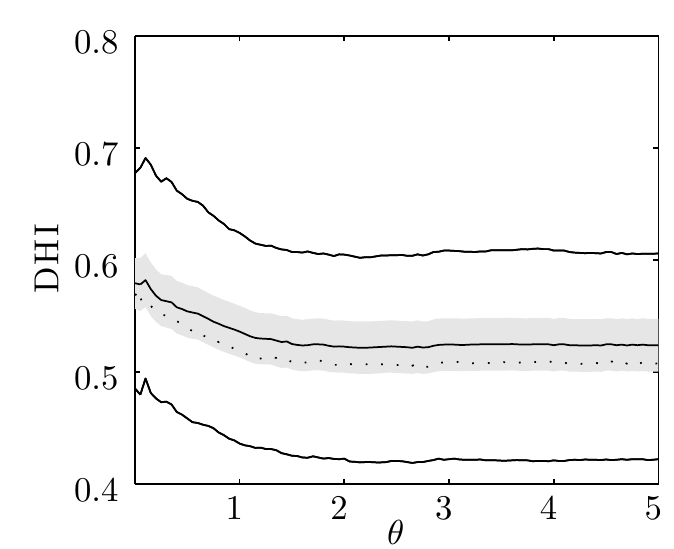}}
   \qquad\qquad
   \subfloat[]{\label{LPabsdevDHIpt3}\includegraphics[width=0.35\textwidth]{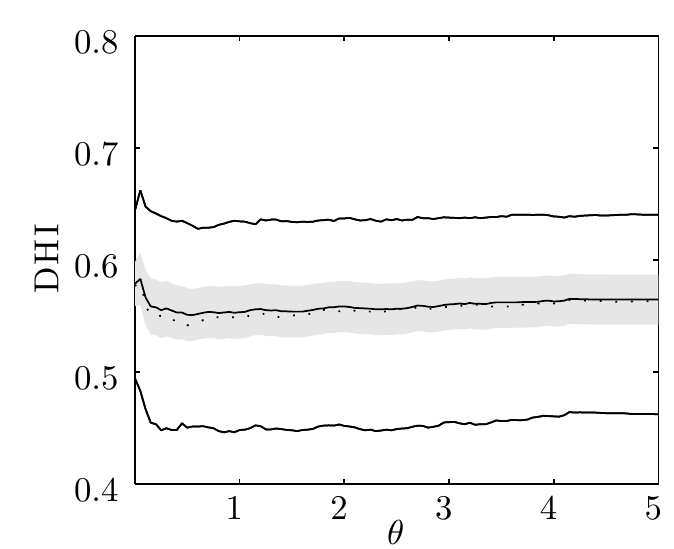}}
   \caption{DHI for all patients as function of the absolute dwell time difference $\theta$. The solid lines represent the minimum, mean and maximum values, and the dotted line is the pre-plan value. The grey area denotes values at most one standard deviation from the mean.}\label{LPabsdevDHI}
\end{figure}
%DVHc
\begin{figure}[h]
   \centering
   \subfloat[]{\label{LPabsdevDVHcpt1}\includegraphics[width=0.35\textwidth]{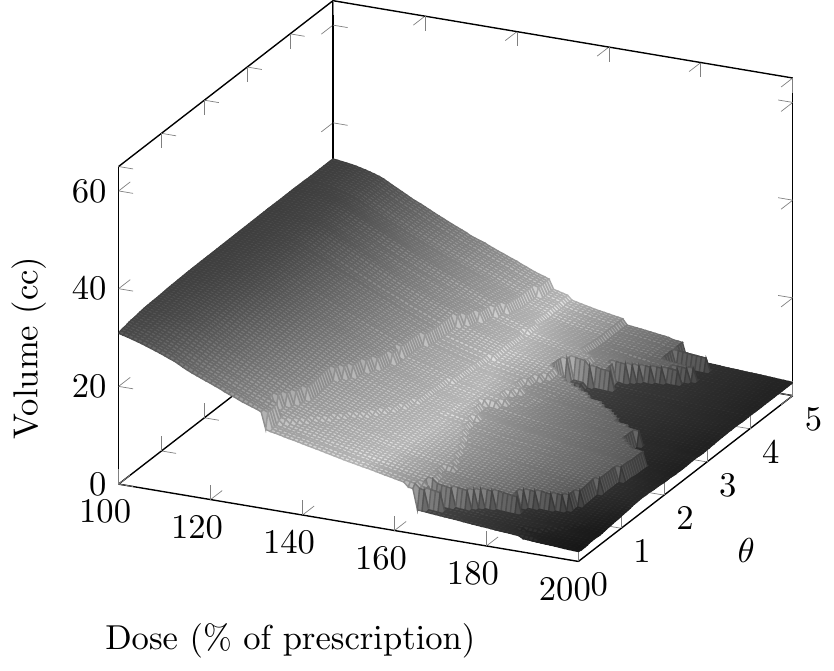}}
      \qquad
      \subfloat[]{\label{LPabsdevDVHcpt2}\includegraphics[width=0.35\textwidth]{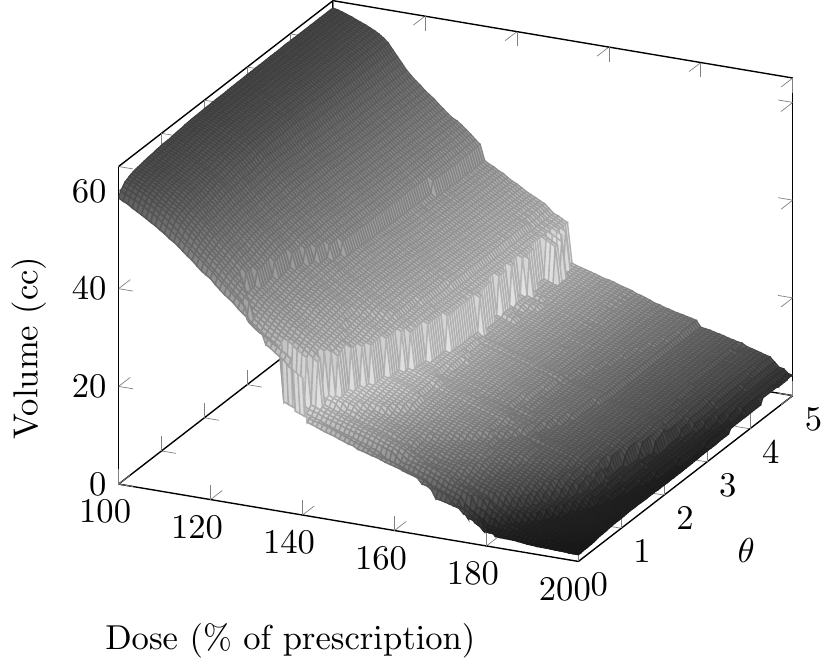}}
      \qquad
      \subfloat[]{\label{LPabsdevDVHcpt3}\includegraphics[width=0.35\textwidth]{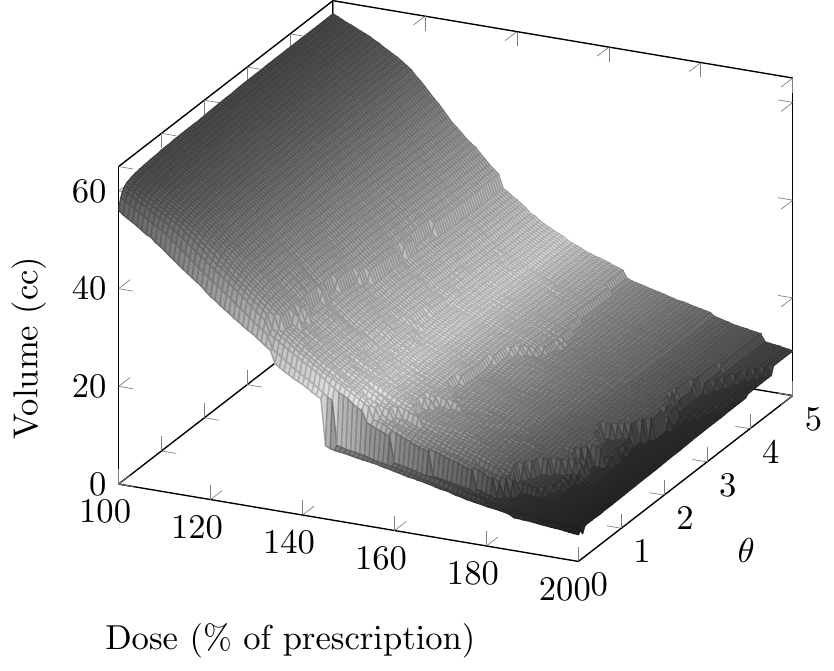}}
   \caption{PTV DVH$^c$ for all patients.}\label{LPabsdevDVHc}
\end{figure}
%PTV D90
\begin{figure}[h]
   \centering
   \subfloat[]{\label{LPabsdevPTVD90pt1}\includegraphics[width=0.35\textwidth]{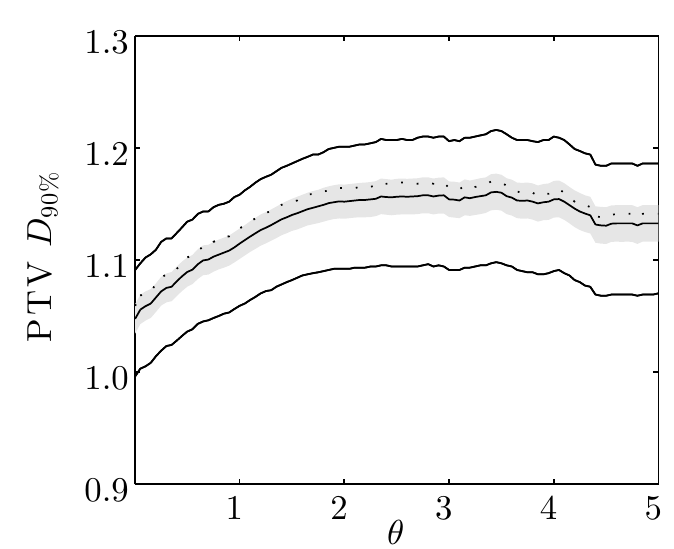}}
   \qquad\qquad
   \subfloat[]{\label{LPabsdevPTVD90pt2}\includegraphics[width=0.35\textwidth]{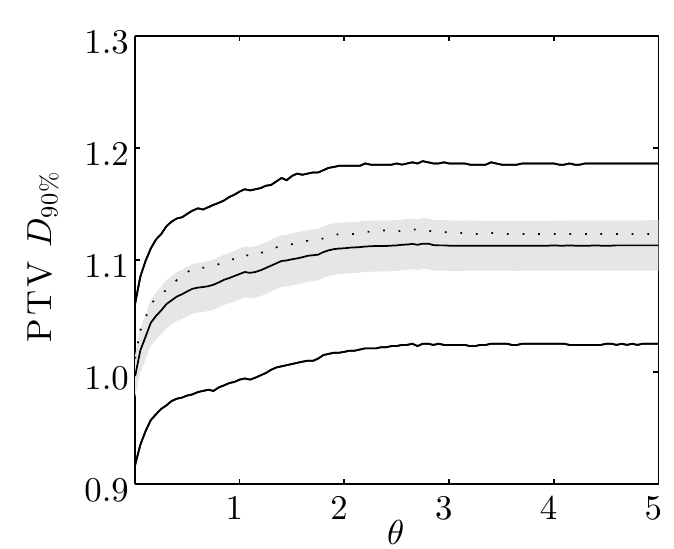}}
   \qquad\qquad
   \subfloat[]{\label{LPabsdevPTVD90pt3}\includegraphics[width=0.35\textwidth]{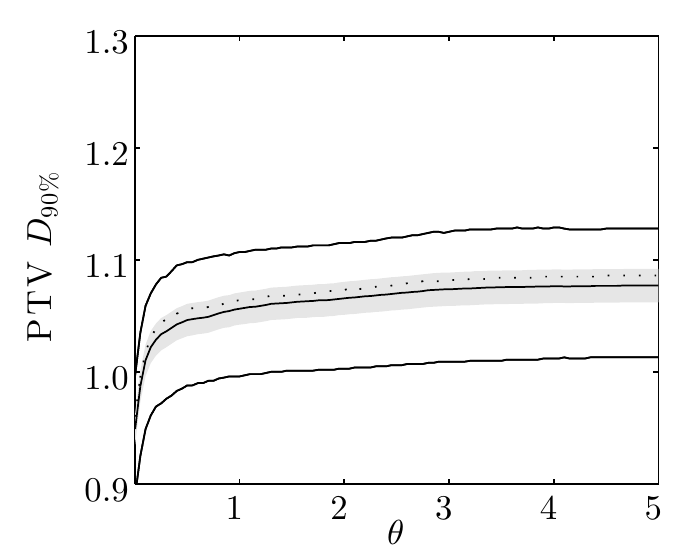}}
   \caption{$D_{90\%}$(PTV) for all patients as function of the absolute dwell time difference $\theta$. The solid lines represent the minimum, mean and maximum values, and the dotted line is the pre-plan value. The grey area denotes values at most one standard deviation from the mean.}\label{LPabsdevPTVD90}
\end{figure}
%PTV V100
\begin{figure}[h]
   \centering
   \subfloat[]{\label{LPabsdevPTVV100pt1}\includegraphics[width=0.35\textwidth]{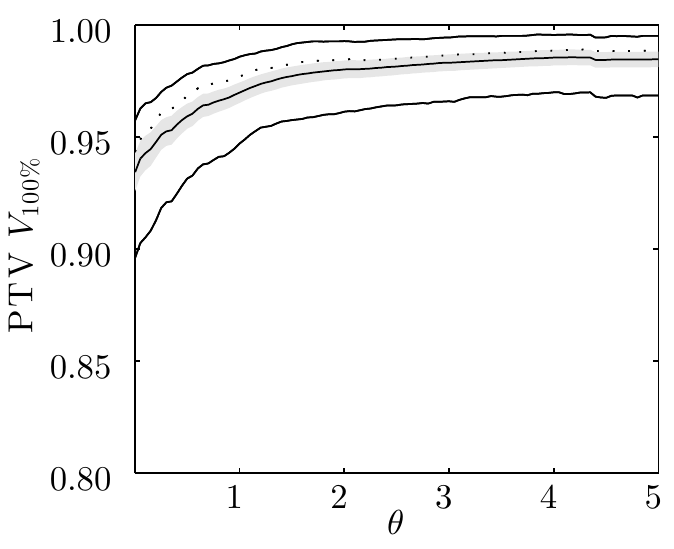}}
   \qquad\qquad
   \subfloat[]{\label{LPabsdevPTVV100pt2}\includegraphics[width=0.35\textwidth]{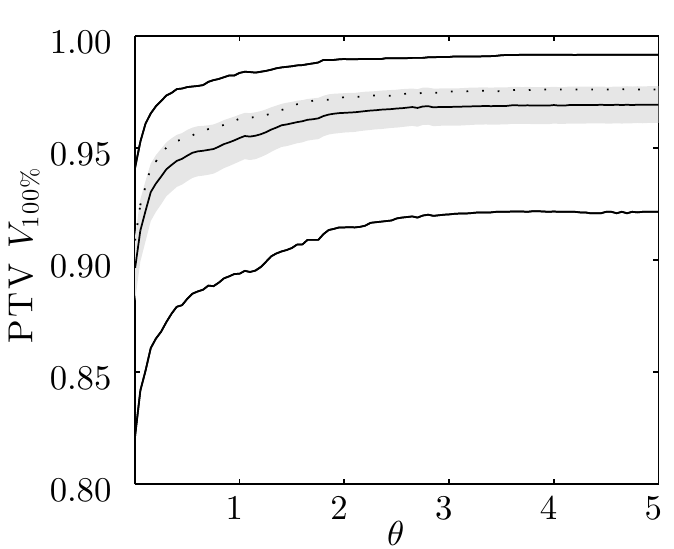}}
   \qquad\qquad
   \subfloat[]{\label{LPabsdevPTVV100pt3}\includegraphics[width=0.35\textwidth]{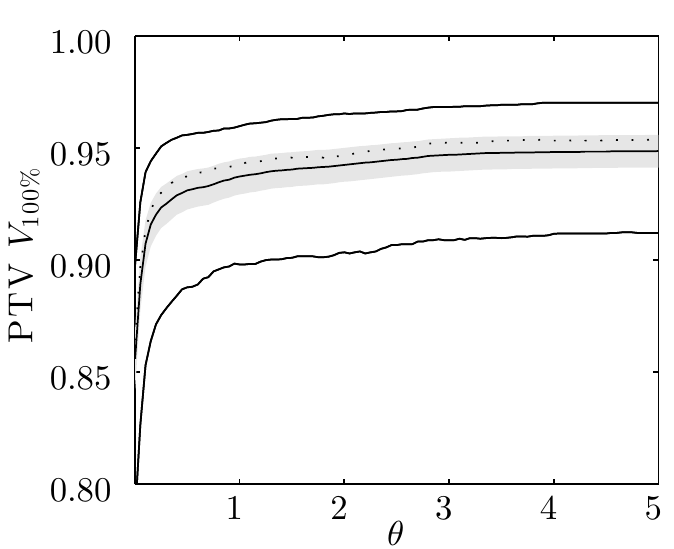}}
   \caption{$V_{100\%}$(PTV) for all patients as function of the absolute dwell time difference $\theta$. The solid lines represent the minimum, mean and maximum values, and the dotted line is the pre-plan value. The grey area denotes values at most one standard deviation from the mean.}\label{LPabsdevPTVV100}
\end{figure}
%PTV V150
\begin{figure}[h]
   \centering
   \subfloat[]{\label{LPabsdevPTVV150pt1}\includegraphics[width=0.35\textwidth]{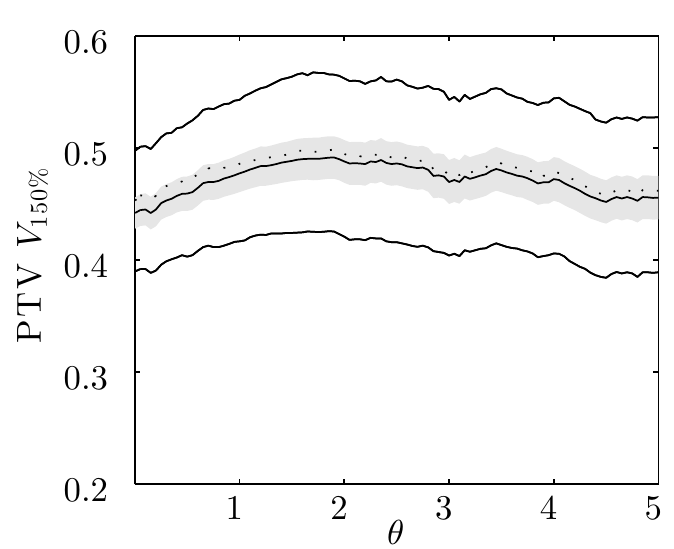}}
   \qquad\qquad
   \subfloat[]{\label{LPabsdevPTVV150pt2}\includegraphics[width=0.35\textwidth]{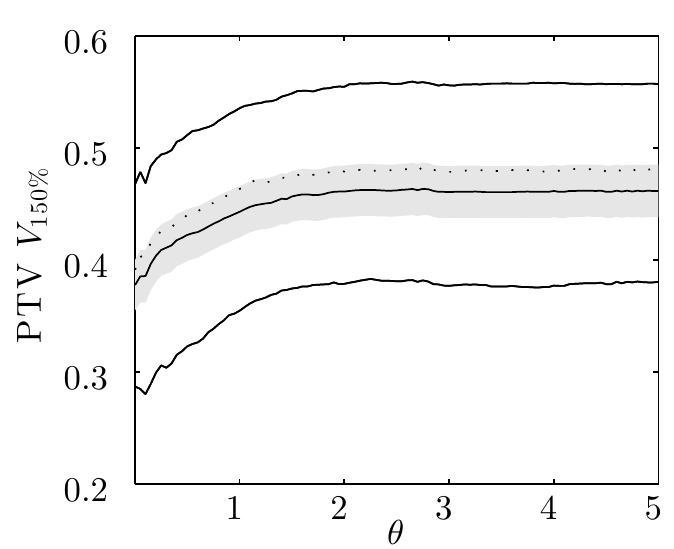}}
   \qquad\qquad
   \subfloat[]{\label{LPabsdevPTVV150pt3}\includegraphics[width=0.35\textwidth]{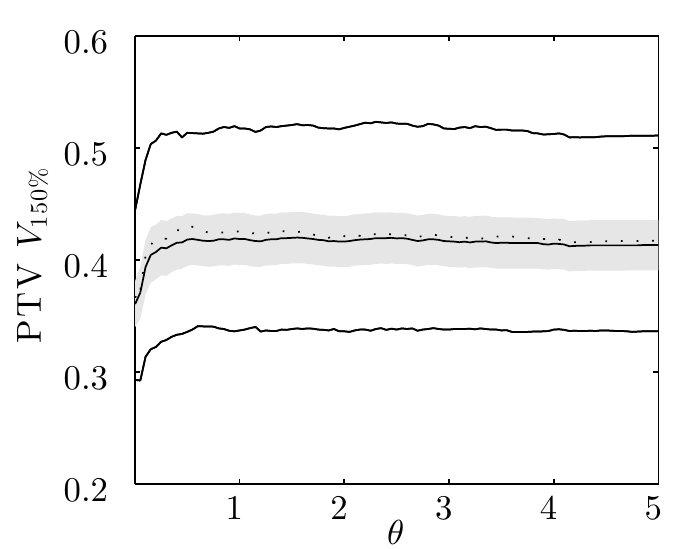}}
   \caption{$V_{150\%}$(PTV) for all patients as function of the absolute dwell time difference $\theta$. The solid lines represent the minimum, mean and maximum values, and the dotted line is the pre-plan value. The grey area denotes values at most one standard deviation from the mean.}\label{LPabsdevPTVV150}
\end{figure}
%PTV V200
\begin{figure}[h]
   \centering
   \subfloat[]{\label{LPabsdevPTVV200pt1}\includegraphics[width=0.35\textwidth]{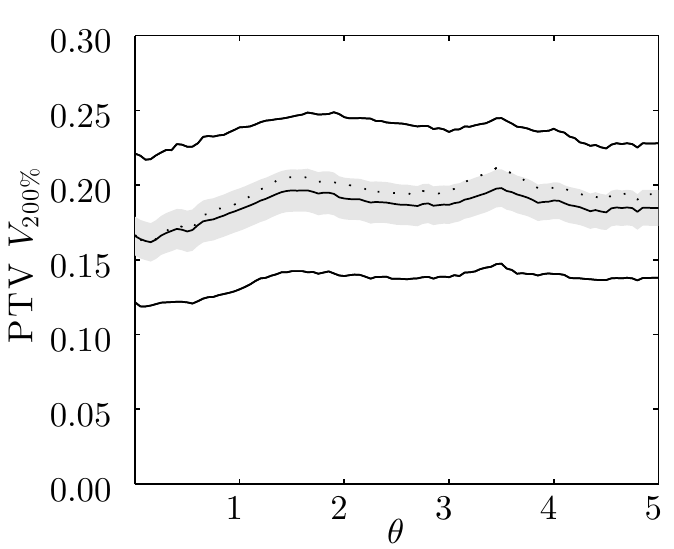}}
   \qquad\qquad
   \subfloat[]{\label{LPabsdevPTVV200pt2}\includegraphics[width=0.35\textwidth]{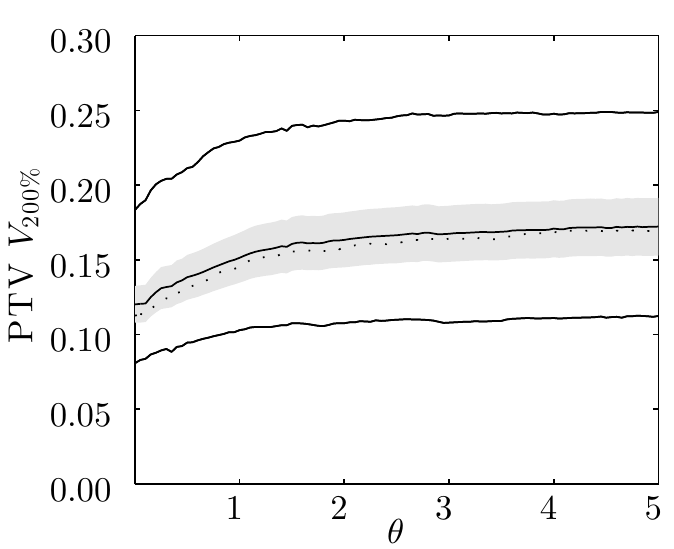}}
   \qquad\qquad
   \subfloat[]{\label{LPabsdevPTVV200pt3}\includegraphics[width=0.35\textwidth]{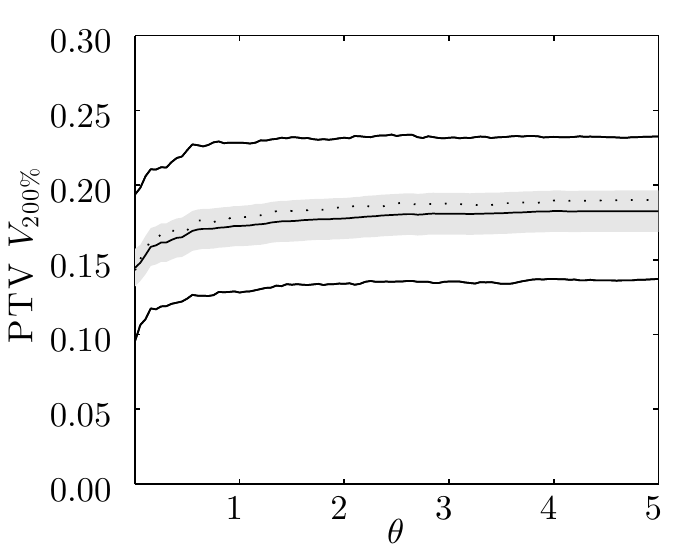}}
   \caption{$V_{200\%}$(PTV) for all patients as function of the absolute dwell time difference $\theta$. The solid lines represent the minimum, mean and maximum values, and the dotted line is the pre-plan value. The grey area denotes values at most one standard deviation from the mean.}\label{LPabsdevPTVV200}
\end{figure}
\pagebreak
% Rectum D10
\begin{figure}[h]
   \centering
   \subfloat[]{\label{LPabsdevRD10pt1}\includegraphics[width=0.35\textwidth]{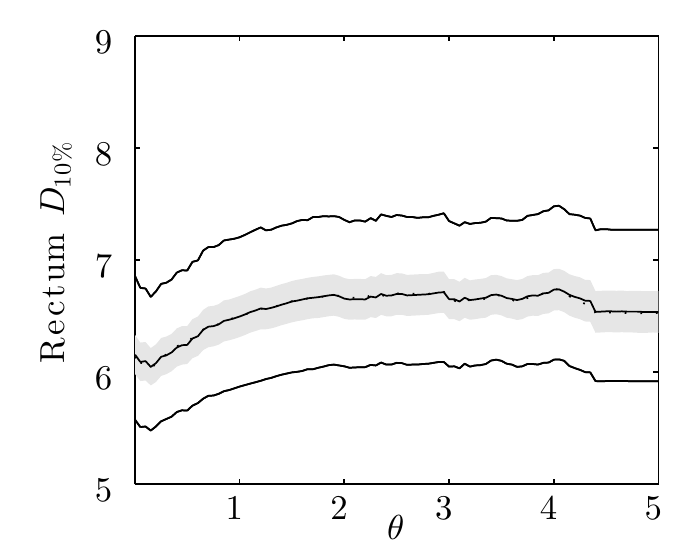}}
   \qquad\qquad
   \subfloat[]{\label{LPabsdevRD10pt2}\includegraphics[width=0.35\textwidth]{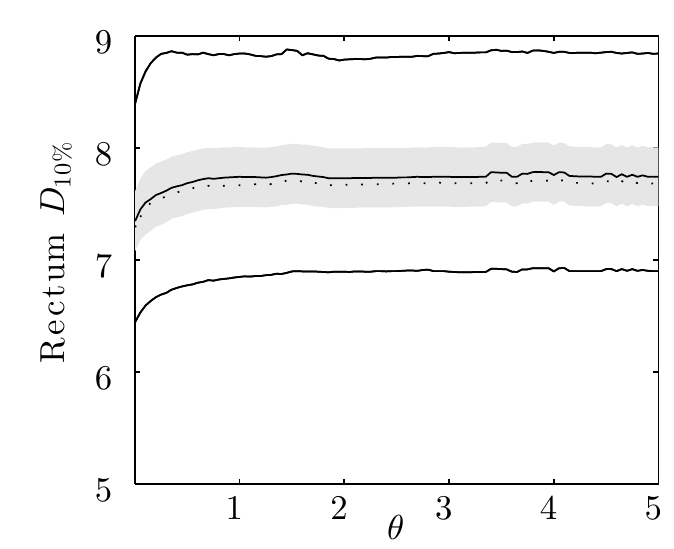}}
   \qquad\qquad
   \subfloat[]{\label{LPabsdevRD10pt3}\includegraphics[width=0.35\textwidth]{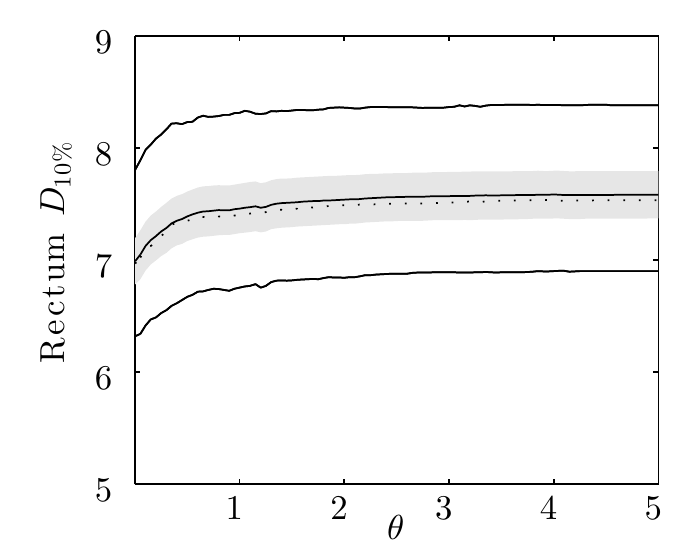}}
   \caption{$D_{10\%}$(rectum) for all patients as function of the absolute dwell time difference $\theta$. The solid lines represent the minimum, mean and maximum values, and the dotted line is the pre-plan value. The grey area denotes values at most one standard deviation from the mean.}\label{LPabsdevRD10}
\end{figure}
\nopagebreak[4]
% Rectum D2cc
\begin{figure}[h]
   \centering
   \subfloat[]{\label{LPabsdevRD2ccpt1}\includegraphics[width=0.35\textwidth]{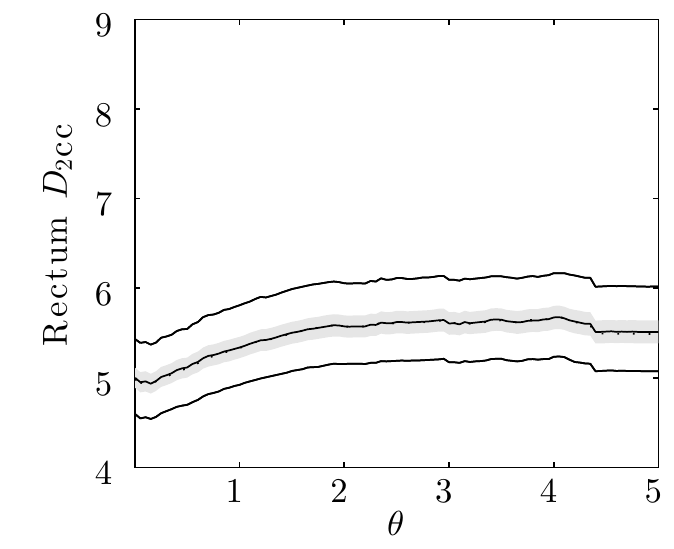}}
   \qquad\qquad
   \subfloat[]{\label{LPabsdevRD2ccpt2}\includegraphics[width=0.35\textwidth]{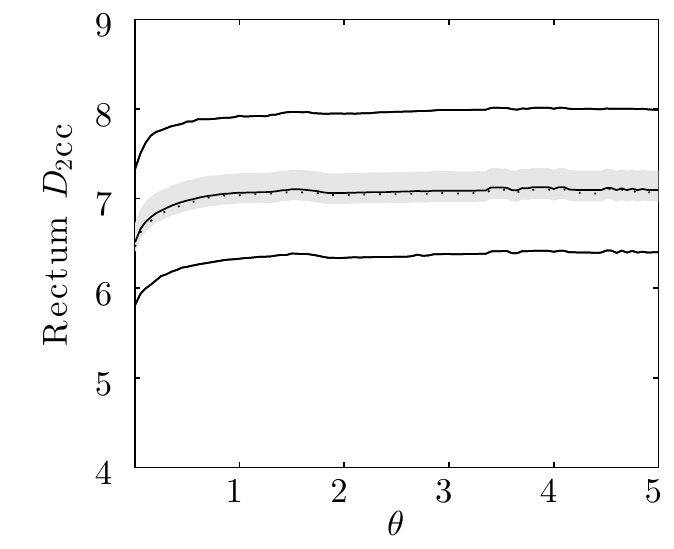}}
   \qquad\qquad
   \subfloat[]{\label{LPabsdevRD2ccpt3}\includegraphics[width=0.35\textwidth]{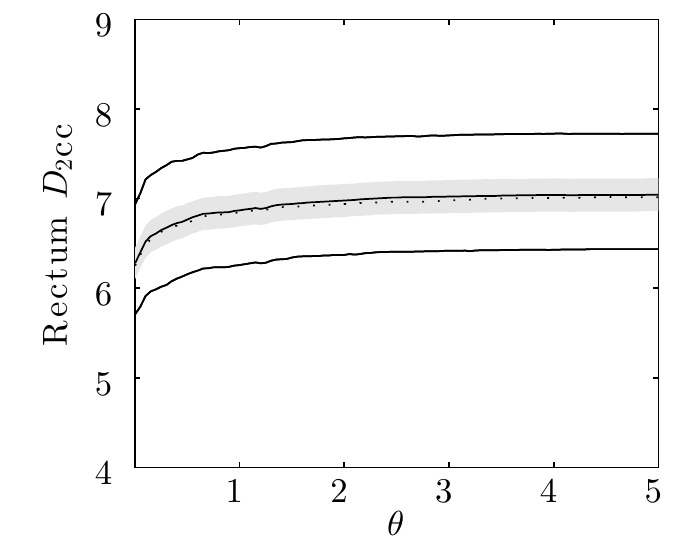}}
   \caption{$D_{2\textnormal{cc}}$(rectum) for all patients as function of the absolute dwell time difference $\theta$. The solid lines represent the minimum, mean and maximum values, and the dotted line is the pre-plan value. The grey area denotes values at most one standard deviation from the mean.}\label{LPabsdevRD2cc}
\end{figure}
\pagebreak
% Urethra D10
\begin{figure}[h]
   \centering
   \subfloat[]{\label{LPabsdevUD10pt1}\includegraphics[width=0.35\textwidth]{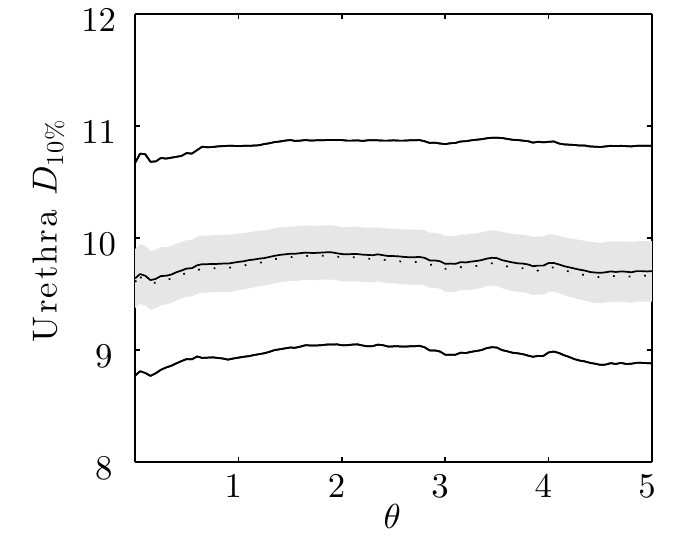}}
   \qquad\qquad
   \subfloat[]{\label{LPabsdevUD10pt2}\includegraphics[width=0.35\textwidth]{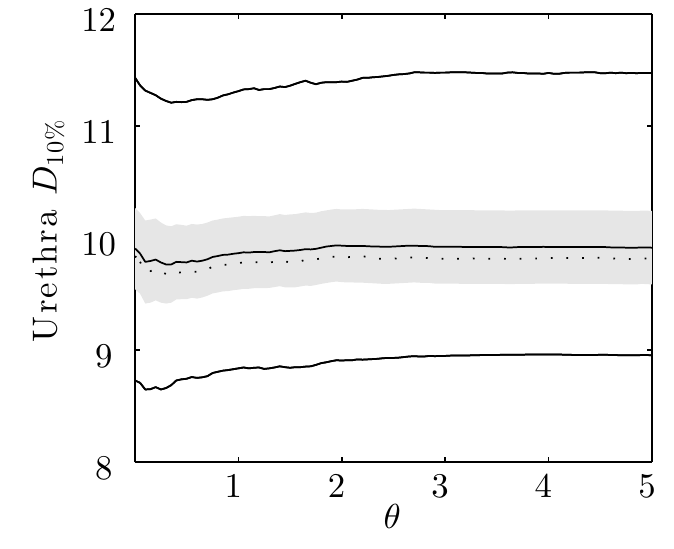}}
   \qquad\qquad
   \subfloat[]{\label{LPabsdevUD10pt3}\includegraphics[width=0.35\textwidth]{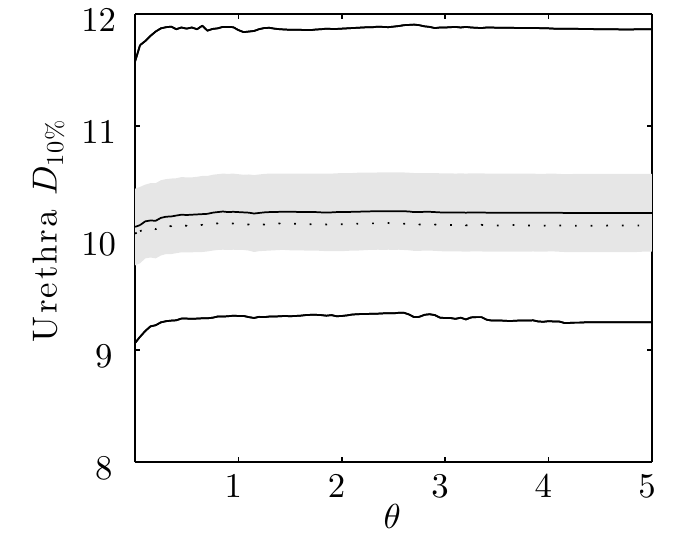}}
   \caption{$D_{10\%}$(urethra) for all patients as function of the absolute dwell time difference $\theta$. The solid lines represent the minimum, mean and maximum values, and the dotted line is the pre-plan value. The grey area denotes values at most one standard deviation from the mean.}\label{LPabsdevUD10}
\end{figure}
\nopagebreak[4]
% Urethra D01cc
\begin{figure}[h]
   \centering
   \subfloat[]{\label{LPabsdevUD0.1ccpt1}\includegraphics[width=0.35\textwidth]{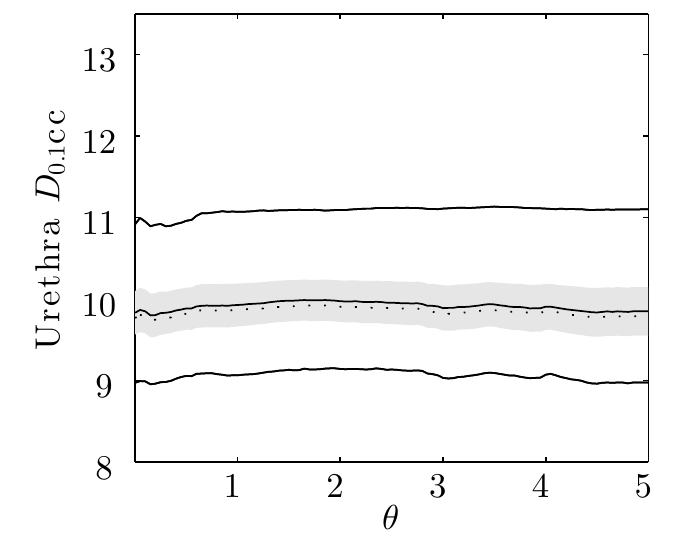}}
   \qquad\qquad
   \subfloat[]{\label{LPabsdevUD0.1ccpt2}\includegraphics[width=0.35\textwidth]{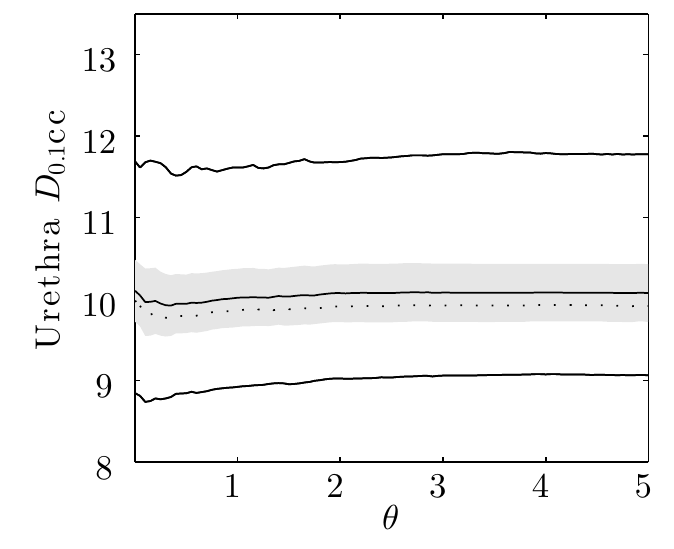}}
   \qquad\qquad
   \subfloat[]{\label{LPabsdevUD0.1ccpt3}\includegraphics[width=0.35\textwidth]{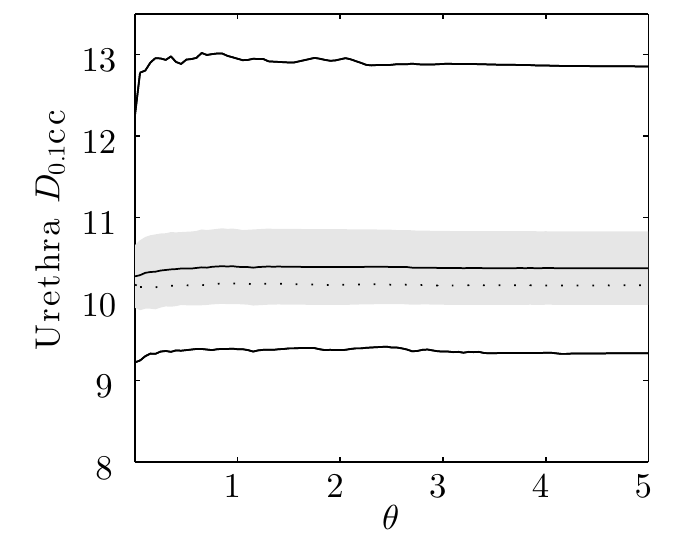}}
   \caption{$D_{0.1\textnormal{cc}}$(urethra) for all patients as function of the absolute dwell time difference $\theta$. The solid lines represent the minimum, mean and maximum values, and the dotted line is the pre-plan value. The grey area denotes values at most one standard deviation from the mean.}\label{LPabsdevUD0.1cc}
\end{figure}

\end{landscape}
\begin{landscape}
\subsection{(LDV) model}
%PTV V100
\begin{figure}[!h]
   \centering
   \subfloat[]{\label{DVHabsdevPTVV100pt1}\includegraphics[width=0.35\textwidth]{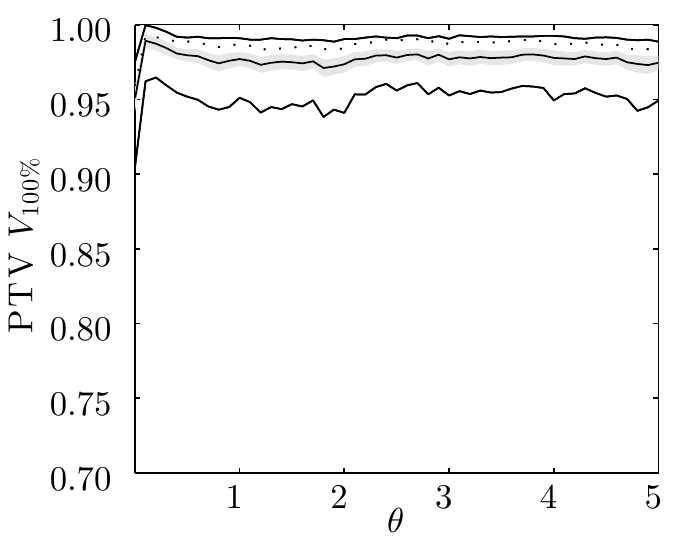}}
   \qquad\qquad
   \subfloat[]{\label{DVHabsdevPTVV100pt2}\includegraphics[width=0.35\textwidth]{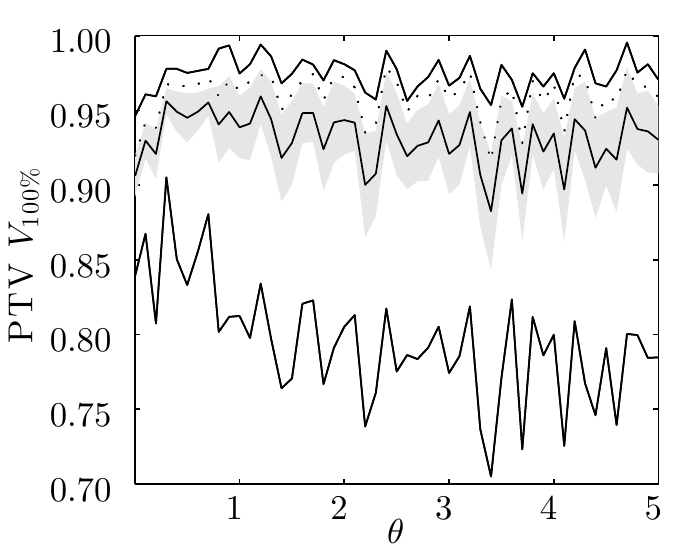}}
   \qquad\qquad
   \subfloat[]{\label{DVHabsdevPTVV100pt3}\includegraphics[width=0.35\textwidth]{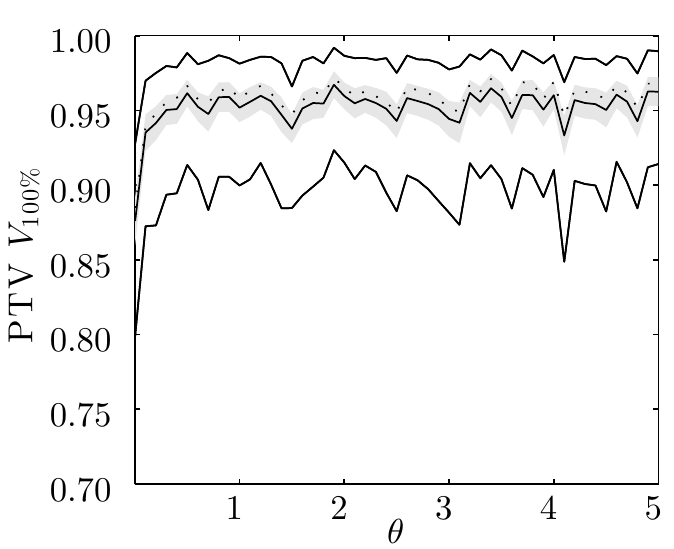}}
   \caption{$V_{100\%}$(PTV) for all patients as function of the absolute dwell time difference $\theta$. The solid lines represent the minimum, mean and maximum values, and the dotted line is the pre-plan value. The grey area denotes values at most one standard deviation from the mean.}\label{DVHabsdevPTVV100}
\end{figure}
%DHI
\begin{figure}[h]
   \centering
   \subfloat[]{\label{DVHabsdevDHIpt1}\includegraphics[width=0.35\textwidth]{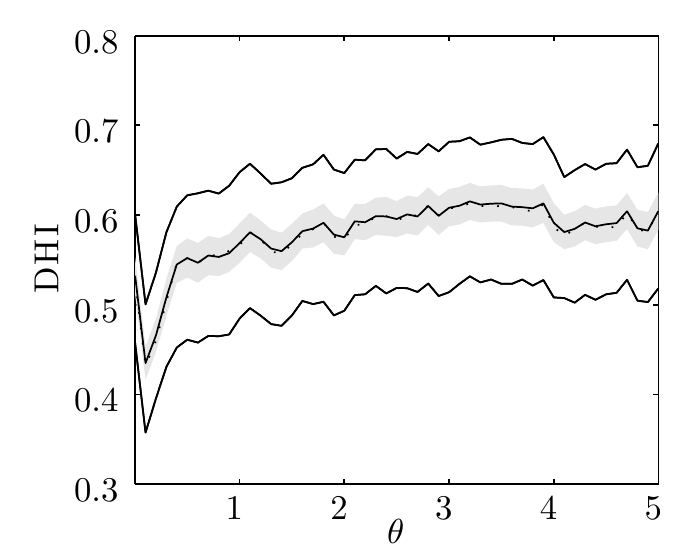}}
   \qquad\qquad
   \subfloat[]{\label{DVHabsdevDHIpt2}\includegraphics[width=0.35\textwidth]{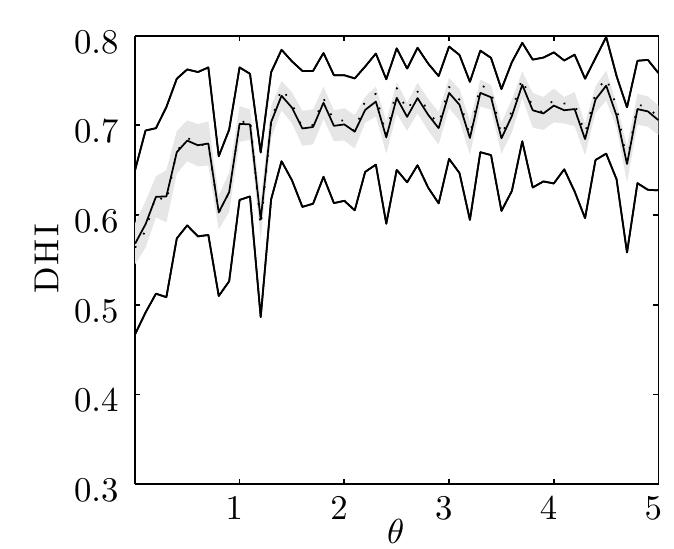}}
   \qquad\qquad
   \subfloat[]{\label{DVHabsdevDHIpt3}\includegraphics[width=0.35\textwidth]{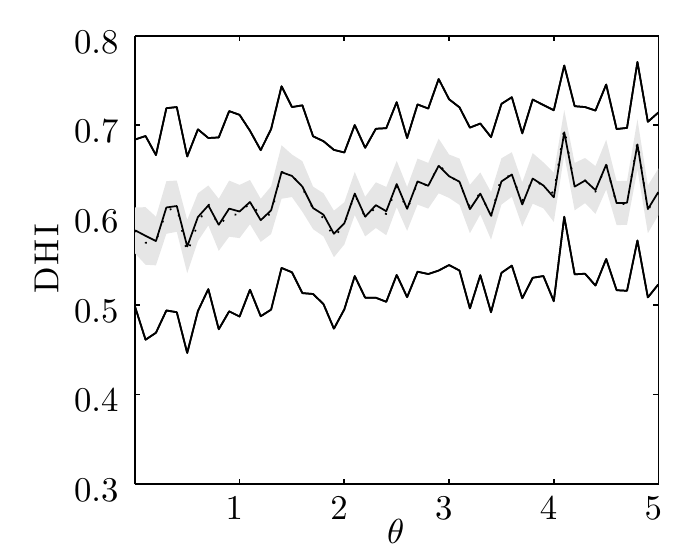}}
   \caption{DHI for all patients as function of the absolute dwell time difference $\theta$. The solid lines represent the minimum, mean and maximum values, and the dotted line is the pre-plan value. The grey area denotes values at most one standard deviation from the mean.}\label{DVHabsdevDHI}
\end{figure}
%DVHc
\begin{figure}[h]
   \centering
   \subfloat[]{\label{DVHabsdevDVHcpt1}\includegraphics[width=0.35\textwidth]{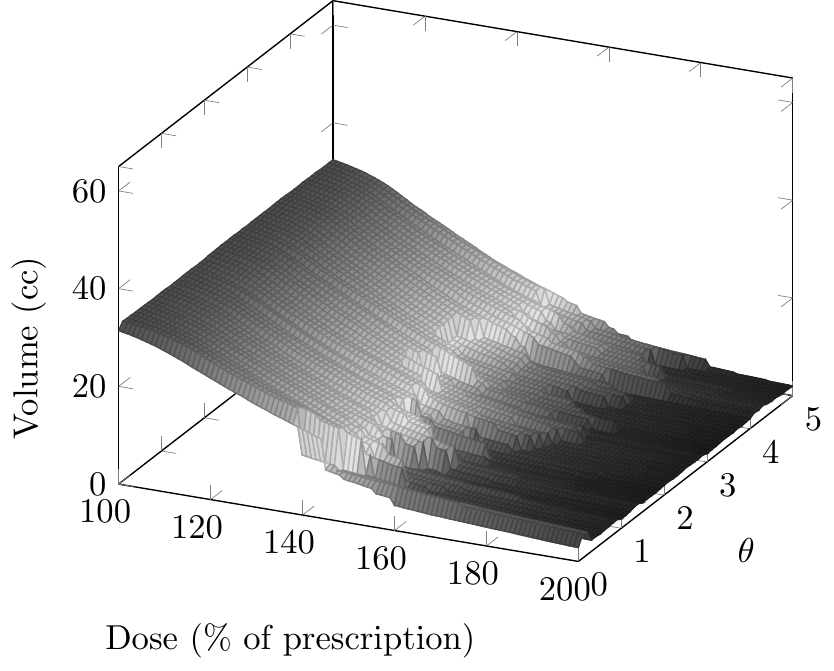}}
         \qquad
         \subfloat[]{\label{DVHabsdevDVHcpt2}\includegraphics[width=0.35\textwidth]{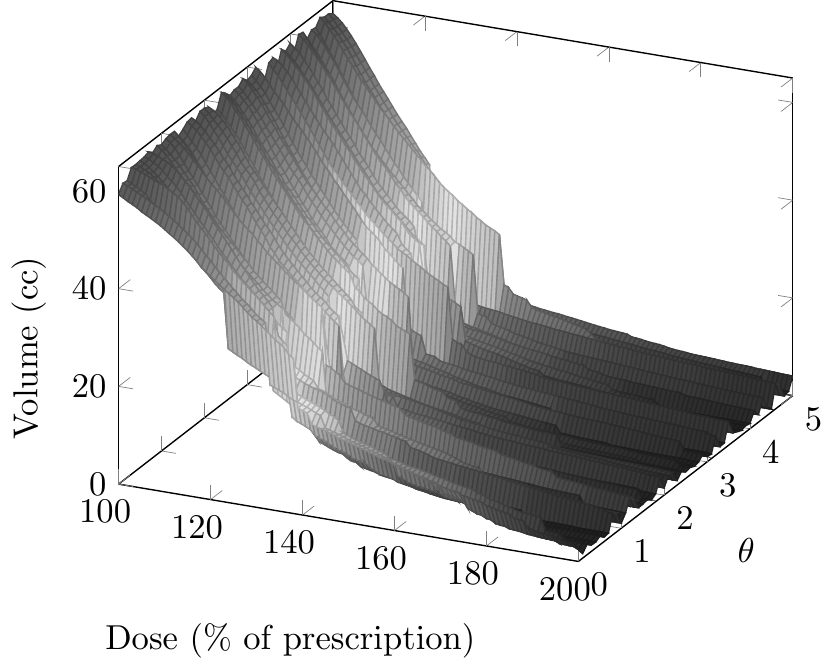}}
         \qquad
        \subfloat[]{\label{DVHabsdevDVHcpt3}\includegraphics[width=0.35\textwidth]{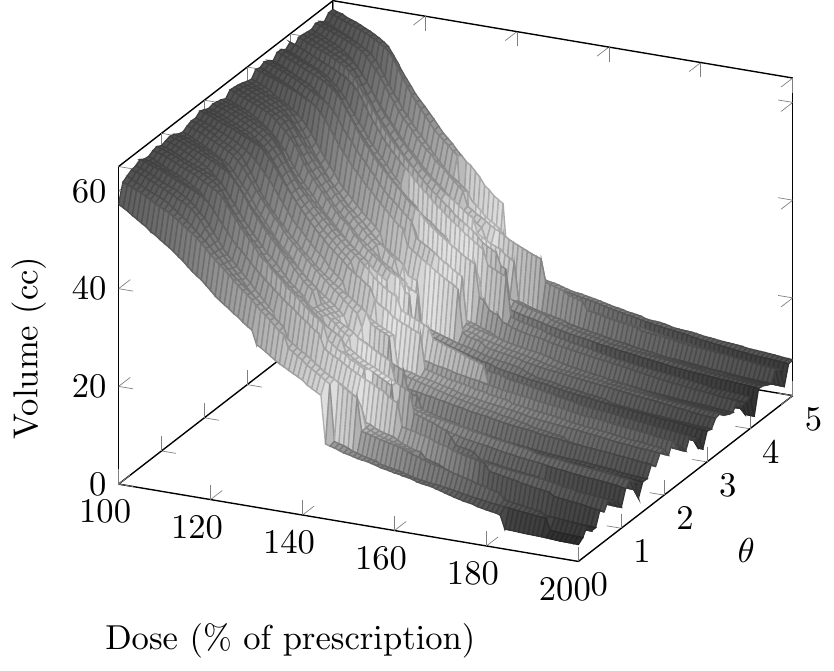}}
   \caption{PTV DVH$^c$ for all patients.}\label{DVHabsdevDVHc}
\end{figure}
\pagebreak[4]
%PTV D90
\begin{figure}[!h]
   \centering
   \subfloat[]{\label{DVHabsdevPTVd90pt1}\includegraphics[width=0.35\textwidth]{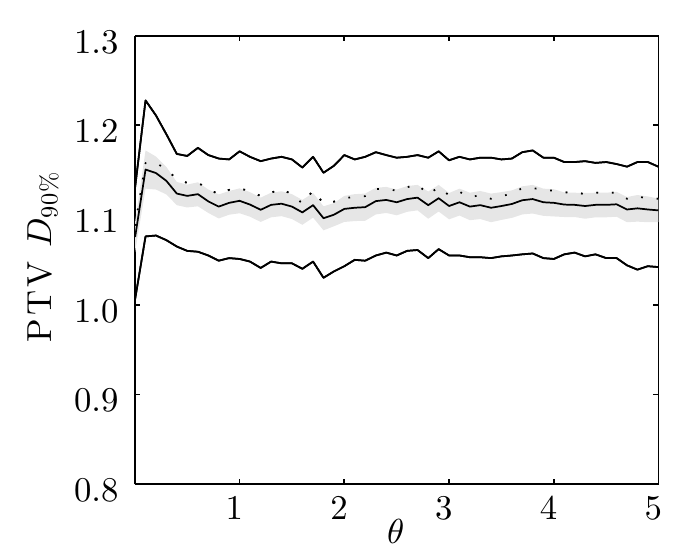}}
   \qquad\qquad
   \subfloat[]{\label{DVHabsdevPTVD90pt2}\includegraphics[width=0.35\textwidth]{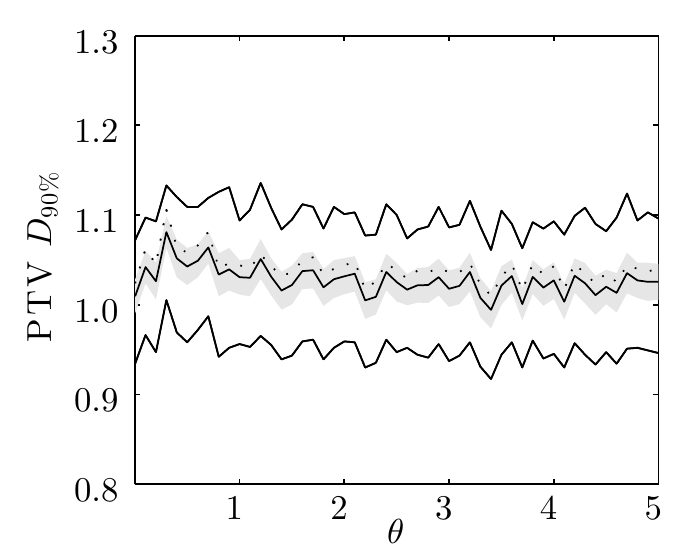}}
   \qquad\qquad
   \subfloat[]{\label{DVHabsdevPTVD90pt3}\includegraphics[width=0.35\textwidth]{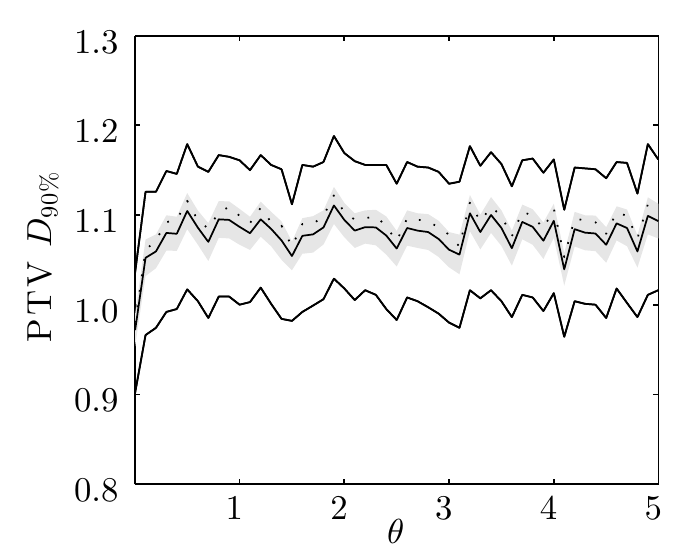}}
   \caption{$D_{90\%}$(PTV) for all patients as function of the absolute dwell time difference $\theta$. The solid lines represent the minimum, mean and maximum values, and the dotted line is the pre-plan value. The grey area denotes values at most one standard deviation from the mean.}\label{DVHabsdevPTVD90}
\end{figure}
%PTV V150
\begin{figure}[h]
   \centering
   \subfloat[]{\label{DVHabsdevPTVV150pt1}\includegraphics[width=0.35\textwidth]{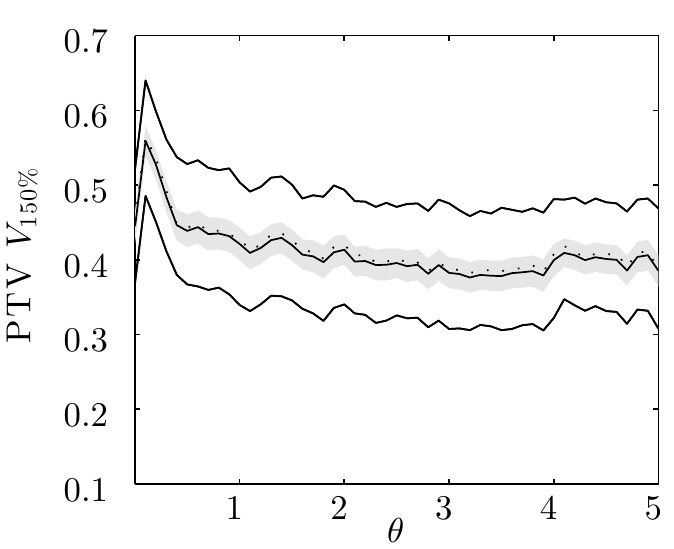}}
   \qquad\qquad
   \subfloat[]{\label{DVHabsdevPTVV150pt2}\includegraphics[width=0.35\textwidth]{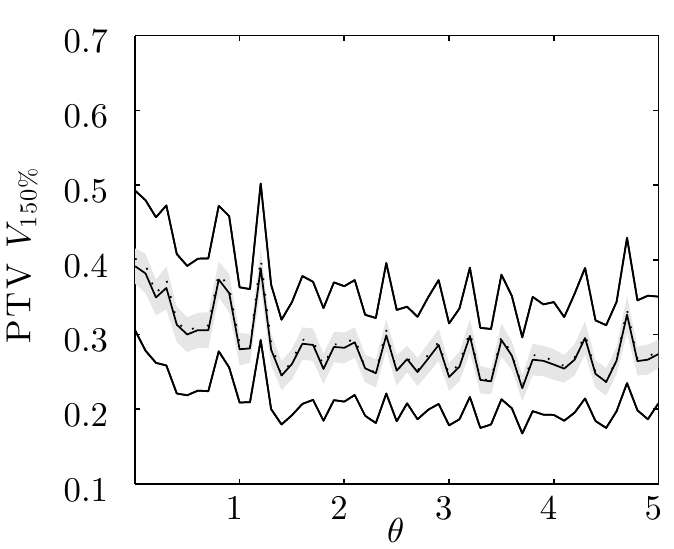}}
   \qquad\qquad
   \subfloat[]{\label{DVHabsdevPTVV150pt3}\includegraphics[width=0.35\textwidth]{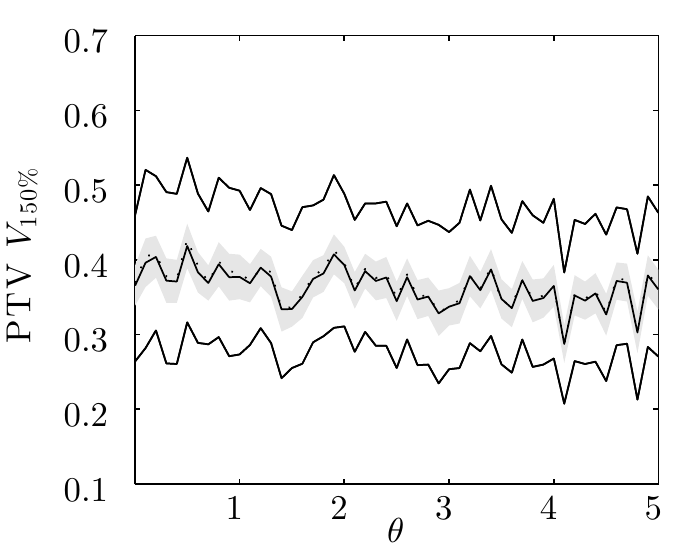}}
   \caption{$V_{150\%}$(PTV) for all patients as function of the absolute dwell time difference $\theta$. The solid lines represent the minimum, mean and maximum values, and the dotted line is the pre-plan value. The grey area denotes values at most one standard deviation from the mean.}\label{DVHabsdevPTVV150}
\end{figure}
%PTV V200
\begin{figure}[h]
   \centering
   \subfloat[]{\label{DVHabsdevPTVV200pt1}\includegraphics[width=0.35\textwidth]{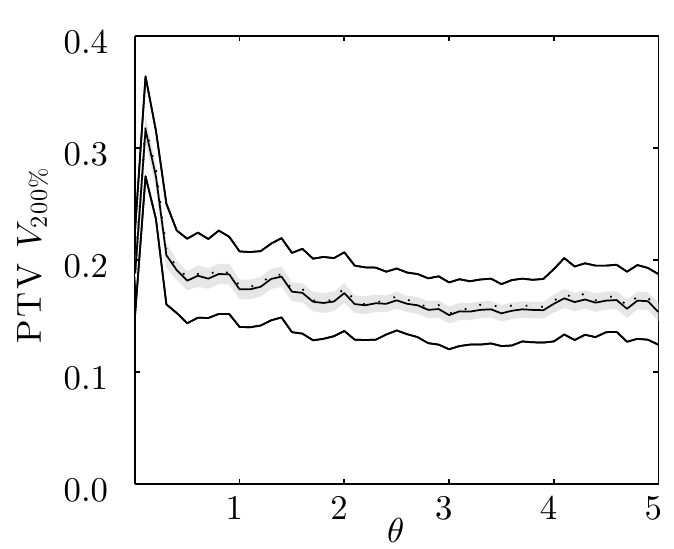}}
   \qquad\qquad
   \subfloat[]{\label{DVHabsdevPTVV200pt2}\includegraphics[width=0.35\textwidth]{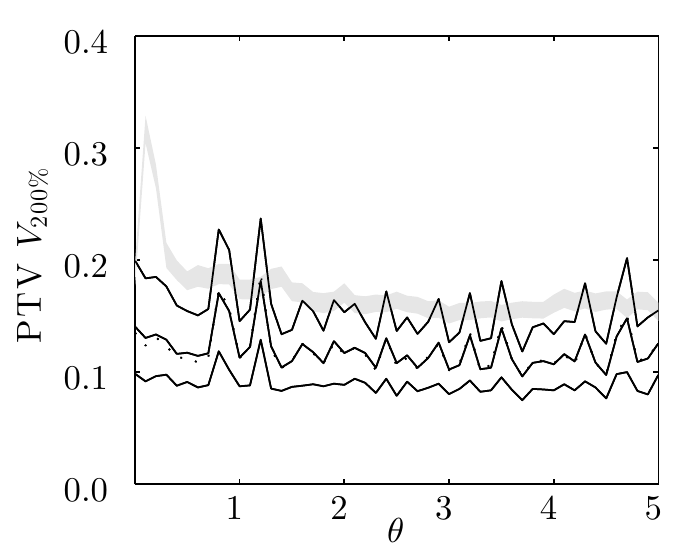}}
   \qquad\qquad
   \subfloat[]{\label{DVHabsdevPTVV200pt3}\includegraphics[width=0.35\textwidth]{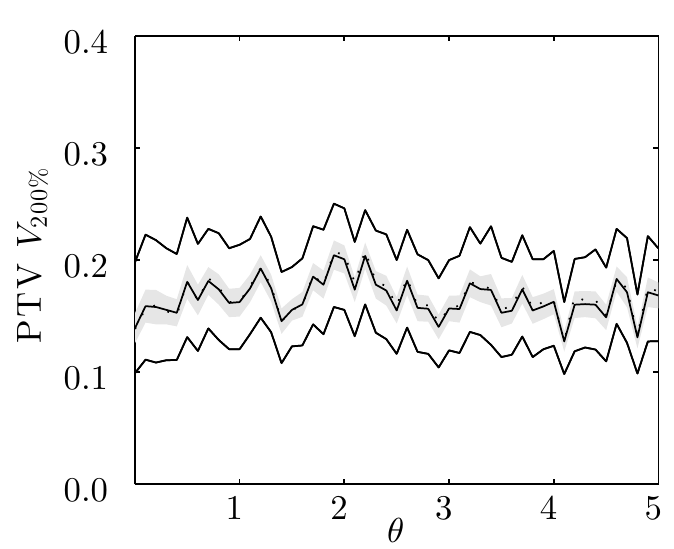}}
   \caption{$V_{200\%}$(PTV) for all patients as function of the absolute dwell time difference $\theta$. The solid lines represent the minimum, mean and maximum values, and the dotted line is the pre-plan value. The grey area denotes values at most one standard deviation from the mean.}\label{DVHabsdevPTVV200}
\end{figure}
% Rectum D10
\begin{figure}[h]
   \centering
   \subfloat[]{\label{DVHabsdevRD10pt1}\includegraphics[width=0.35\textwidth]{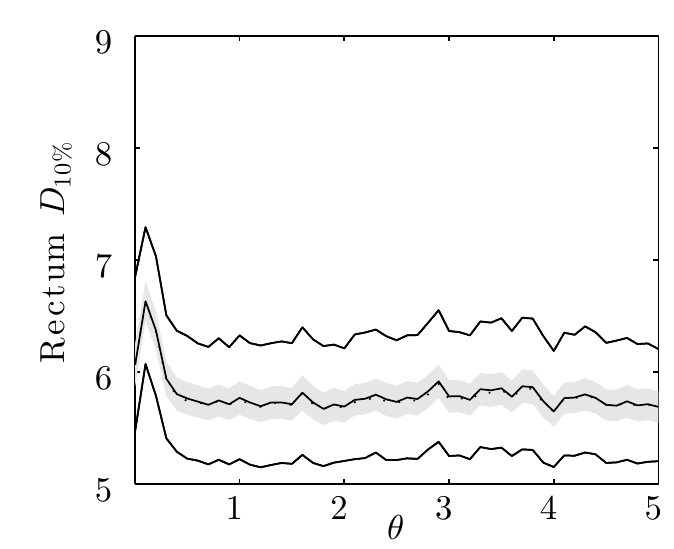}}
   \qquad\qquad
   \subfloat[]{\label{DVHabsdevRD10pt2}\includegraphics[width=0.35\textwidth]{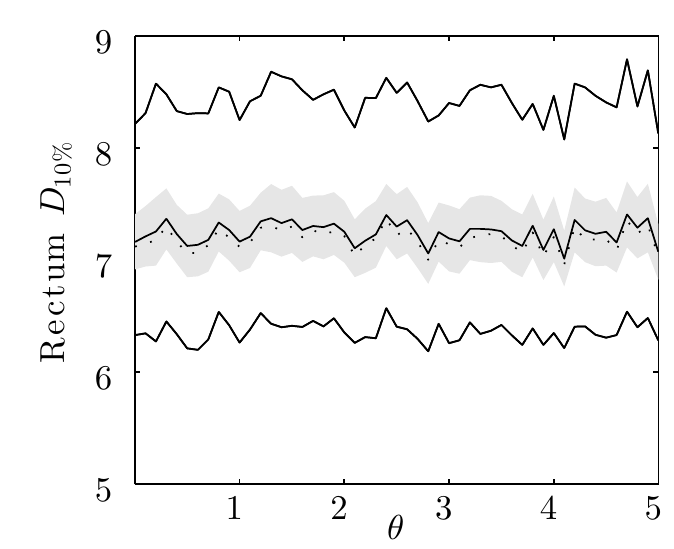}}
   \qquad\qquad
   \subfloat[]{\label{DVHabsdevRD10pt3}\includegraphics[width=0.35\textwidth]{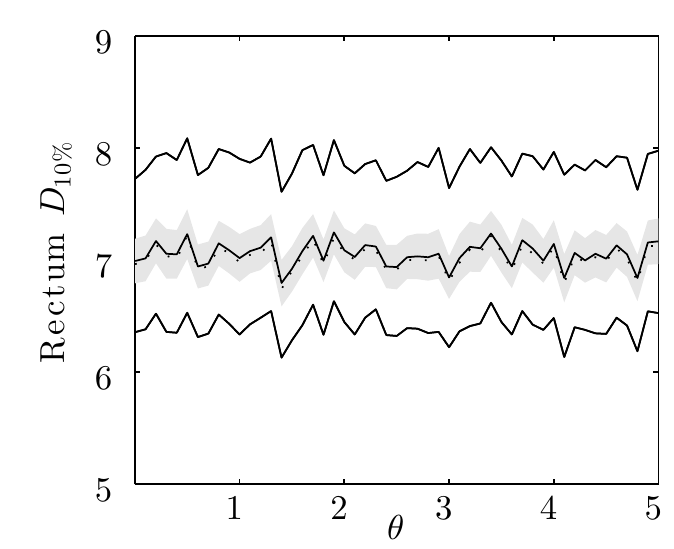}}
   \caption{$D_{10\%}$(rectum) for all patients as function of the absolute dwell time difference $\theta$. The solid lines represent the minimum, mean and maximum values, and the dotted line is the pre-plan value. The grey area denotes values at most one standard deviation from the mean.}\label{DVHabsdevRD10}
\end{figure}
% Rectum D2cc
\begin{figure}[h]
   \centering
   \subfloat[]{\label{DVHabsdevRD2ccpt1}\includegraphics[width=0.35\textwidth]{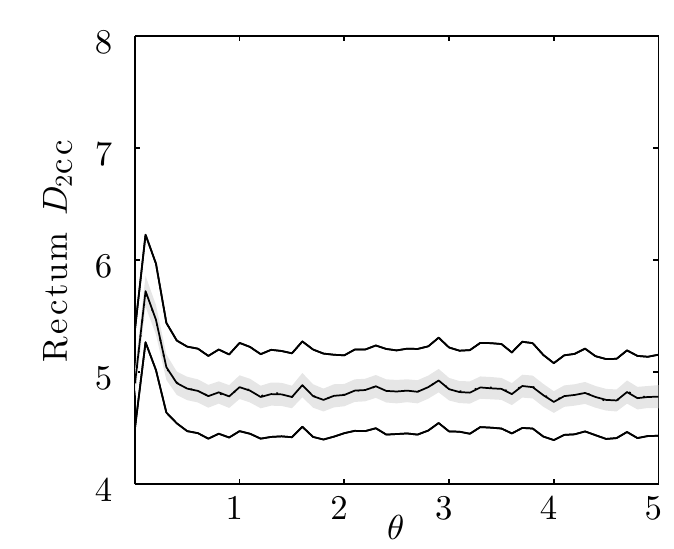}}
   \qquad\qquad
   \subfloat[]{\label{DVHabsdevRD2ccpt2}\includegraphics[width=0.35\textwidth]{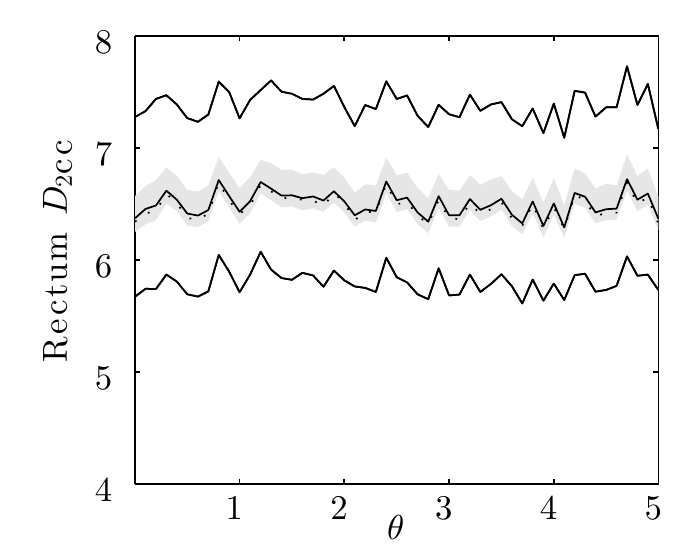}}
   \qquad\qquad
   \subfloat[]{\label{DVHabsdevRD2ccpt3}\includegraphics[width=0.35\textwidth]{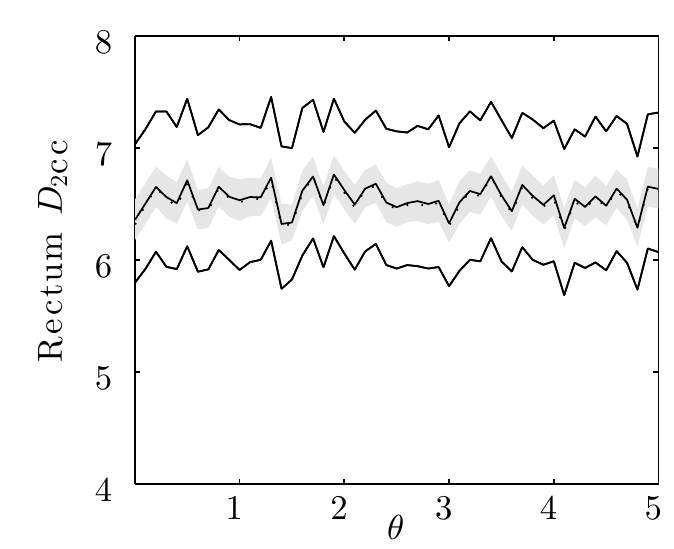}}
   \caption{$D_{2\textnormal{cc}}$(rectum) for all patients as function of the absolute dwell time difference $\theta$. The solid lines represent the minimum, mean and maximum values, and the dotted line is the pre-plan value. The grey area denotes values at most one standard deviation from the mean.}\label{DVHabsdevRD2cc}
\end{figure}
% Urethra D10
\begin{figure}[h]
   \centering
   \subfloat[]{\label{DVHabsdevUD10pt1}\includegraphics[width=0.35\textwidth]{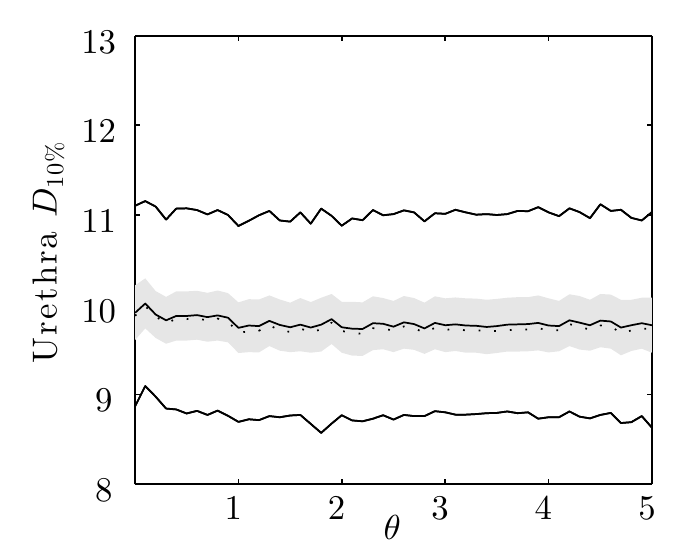}}
   \qquad\qquad
   \subfloat[]{\label{DVHabsdevUD10pt2}\includegraphics[width=0.35\textwidth]{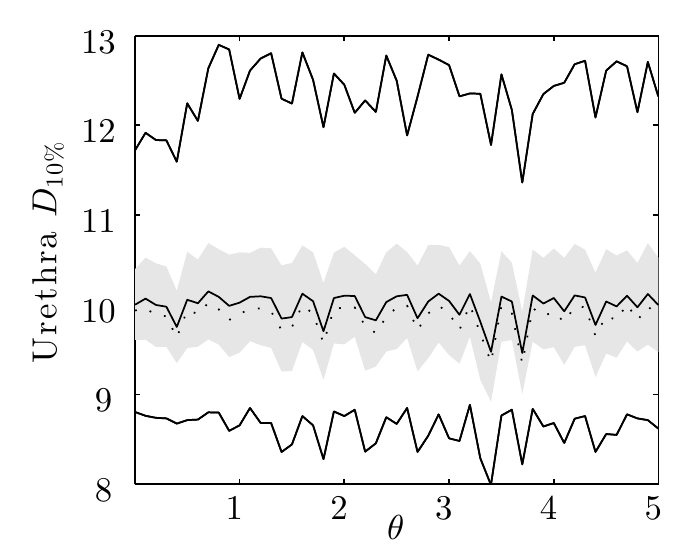}}
   \qquad\qquad
   \subfloat[]{\label{DVHabsdevUD10pt3}\includegraphics[width=0.35\textwidth]{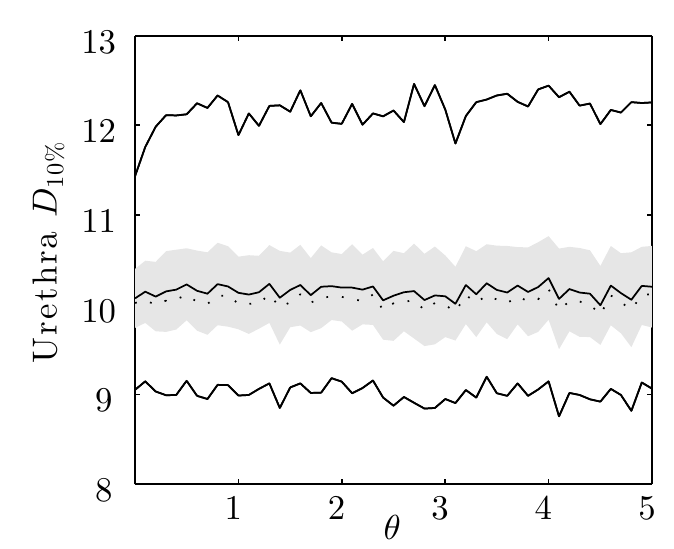}}
   \caption{$D_{10\%}$(urethra) for all patients as function of the absolute dwell time difference $\theta$. The solid lines represent the minimum, mean and maximum values, and the dotted line is the pre-plan value. The grey area denotes values at most one standard deviation from the mean.}\label{DVHabsdevUD10}
\end{figure}
\nopagebreak[4]
% Urethra D01cc
\begin{figure}[h]
   \centering
   \subfloat[]{\label{DVHabsdevUD0.1ccpt1}\includegraphics[width=0.35\textwidth]{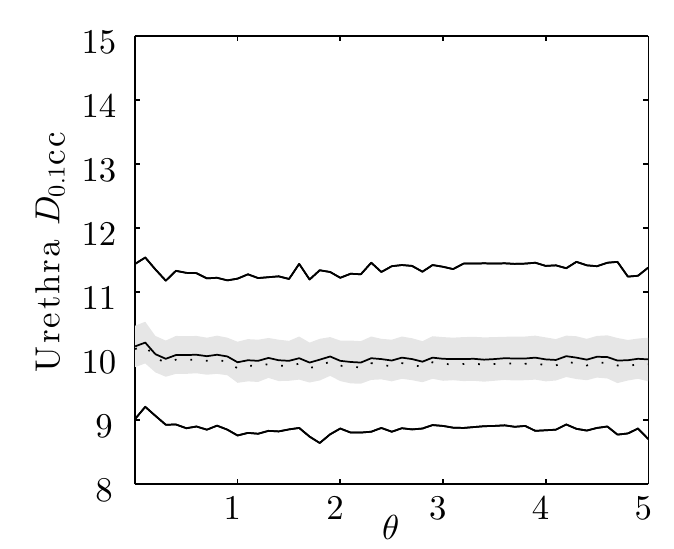}}
   \qquad\qquad
   \subfloat[]{\label{DVHabsdevUD0.1ccpt2}\includegraphics[width=0.35\textwidth]{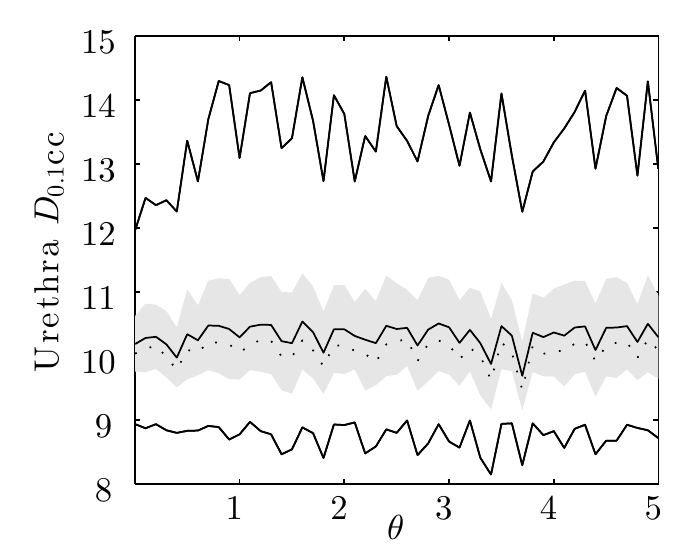}}
   \qquad\qquad
   \subfloat[]{\label{DVHabsdevUD0.1ccpt3}\includegraphics[width=0.35\textwidth]{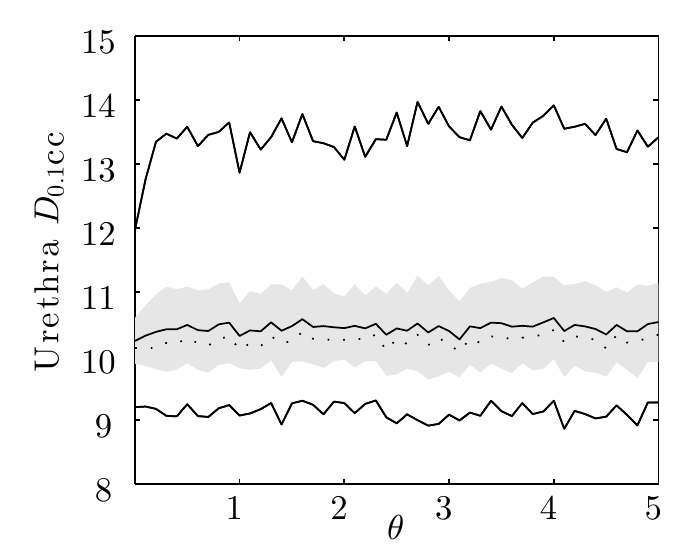}}
   \caption{$D_{0.1\textnormal{cc}}$(urethra) for all patients as function of the absolute dwell time difference $\theta$. The solid lines represent the minimum, mean and maximum values, and the dotted line is the pre-plan value. The grey area denotes values at most one standard deviation from the mean.}\label{DVHabsdevUD0.1cc}
\end{figure}

\end{landscape}
\begin{landscape}
\section{Quadratic dwell time difference restricted}\label{appmrrelQ}

\subsection{(LD) model}
\begin{figure}[h]
   \centering
   \subfloat[]{\label{LPDTGRLOVpt1}\includegraphics[width=0.35\textwidth]{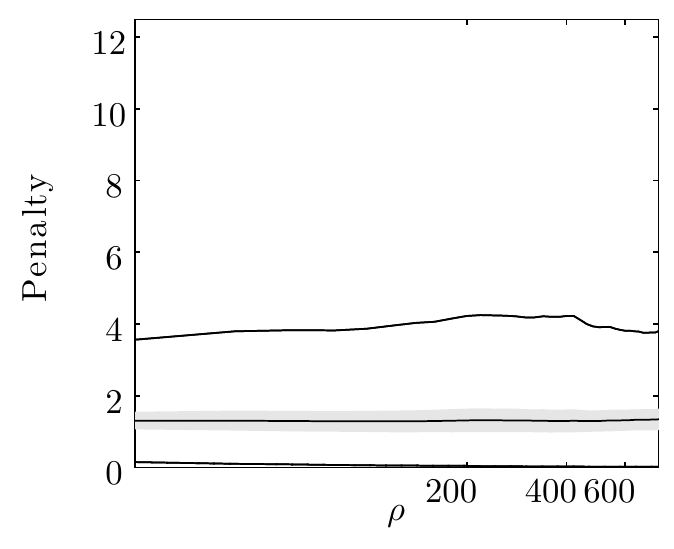}}
   \qquad\qquad
   \subfloat[]{\label{LPDTGRLOVpt2}\includegraphics[width=0.35\textwidth]{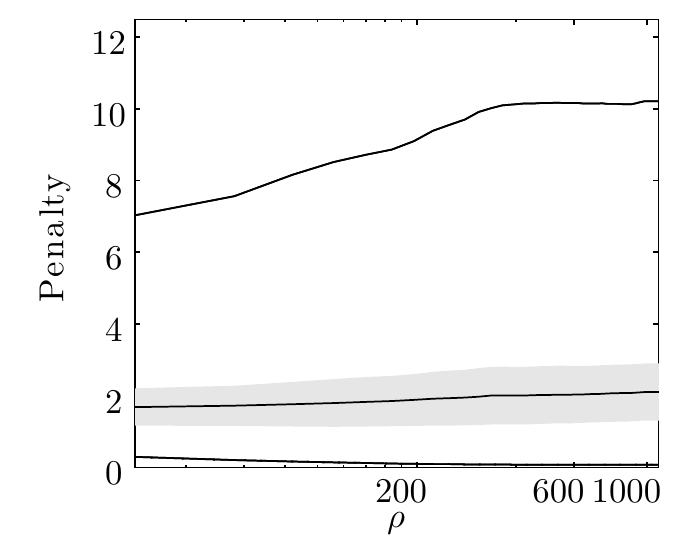}}
   \qquad\qquad
   \subfloat[]{\label{LPDTGRLOVpt3}\includegraphics[width=0.35\textwidth]{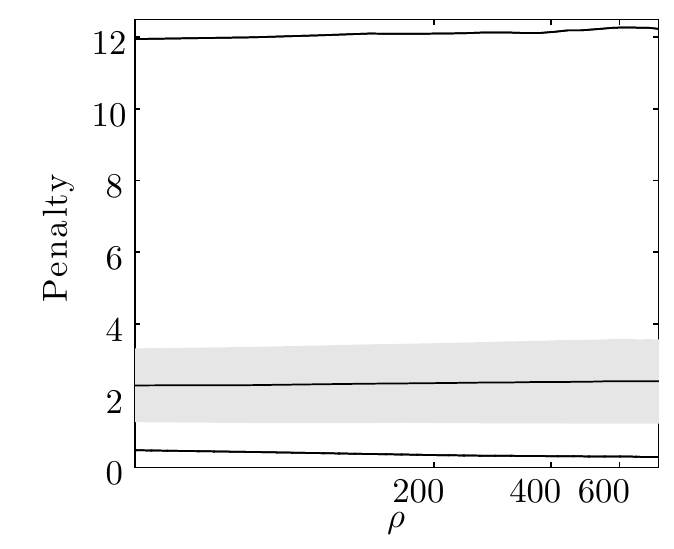}}
   \caption{Objective function value for all patients as function of the maximum sum of squared differences $\rho$. The solid lines represent the minimum, mean and maximum values, and the dotted line is the pre-plan value. The grey area denotes values at most one standard deviation from the mean.}\label{LPDTGRLOV}
\end{figure}
%DHI
\begin{figure}[h]
   \centering
   \subfloat[]{\label{LPDTGRDHIpt1}\includegraphics[width=0.35\textwidth]{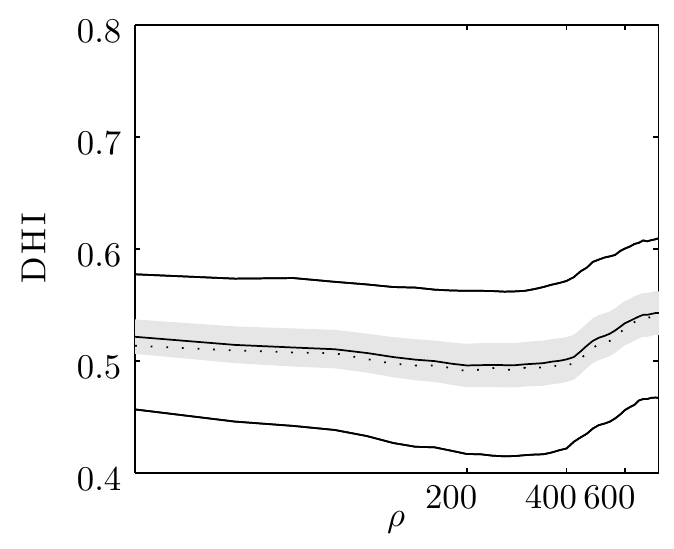}}
   \qquad\qquad
   \subfloat[]{\label{LPDTGRDHIpt2}\includegraphics[width=0.35\textwidth]{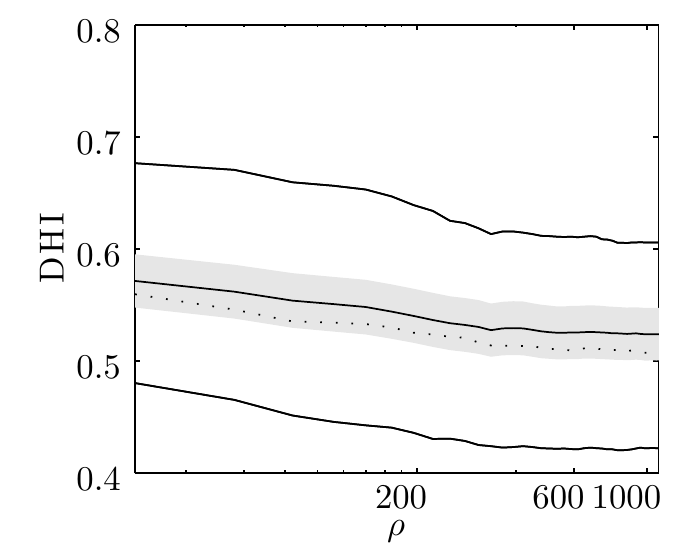}}
   \qquad\qquad
   \subfloat[]{\label{LPDTGRDHIpt3}\includegraphics[width=0.35\textwidth]{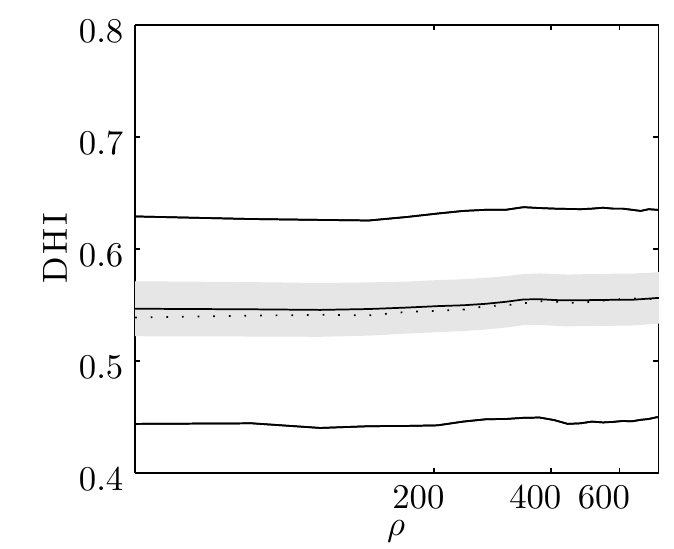}}
   \caption{DHI for all patients as function of the maximum sum of squared differences $\rho$. The solid lines represent the minimum, mean and maximum values, and the dotted line is the pre-plan value. The grey area denotes values at most one standard deviation from the mean.}\label{LPDTGRDHI}
\end{figure}
\pagebreak
%DVHc
\begin{figure}[h]
   \centering
   \subfloat[]{\label{LPDTGRDVHcpt1}\includegraphics[width=0.35\textwidth]{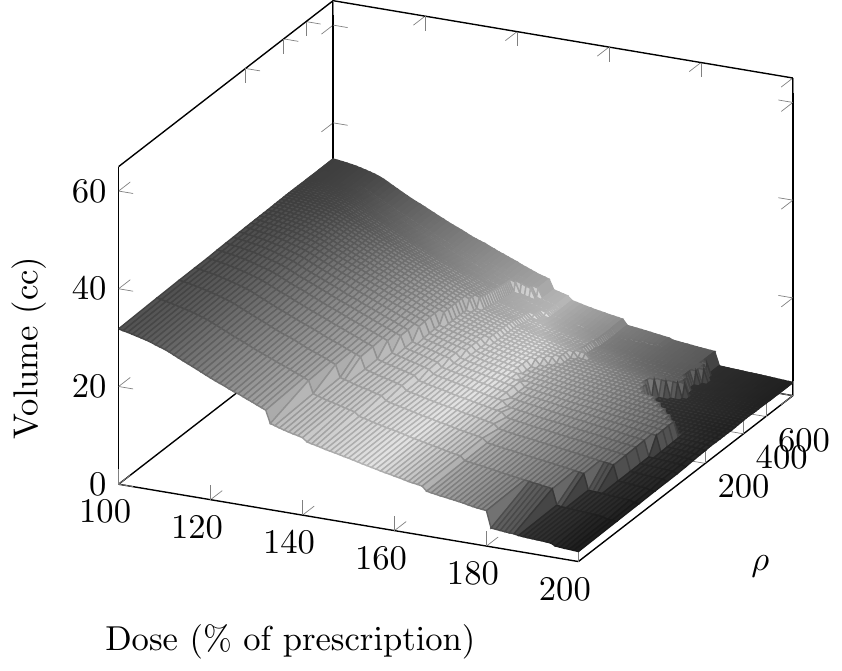}}
         \qquad
         \subfloat[]{\label{LPDTGRDVHcpt2}\includegraphics[width=0.35\textwidth]{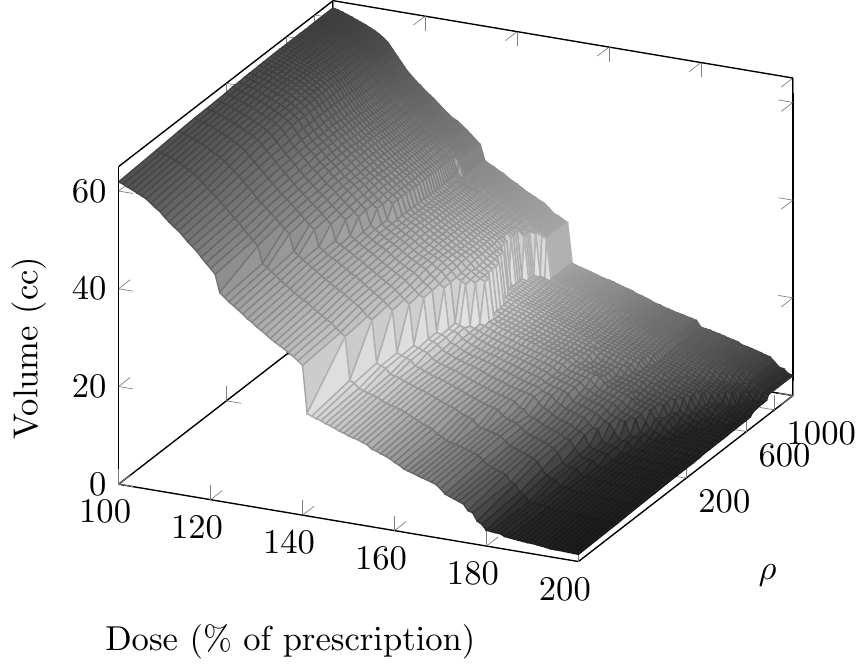}}
         \qquad
         \subfloat[]{\label{LPDTGRDVHcpt3}\includegraphics[width=0.35\textwidth]{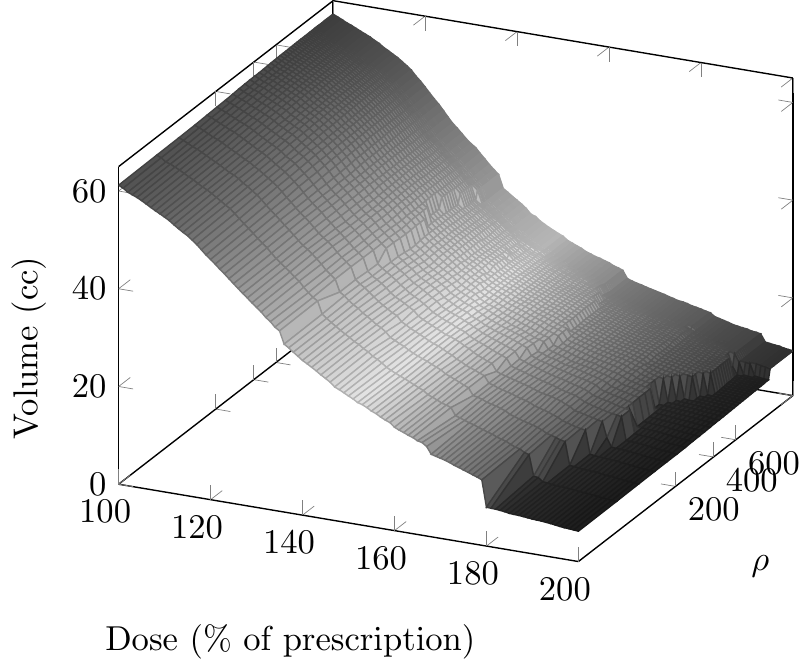}}
   \caption{PTV DVH$^c$ for all patients.}\label{LPDTGRDVHc}
\end{figure}
%PTV D90
\begin{figure}[h]
   \centering
   \subfloat[]{\label{LPDTGRPTVD90pt1}\includegraphics[width=0.35\textwidth]{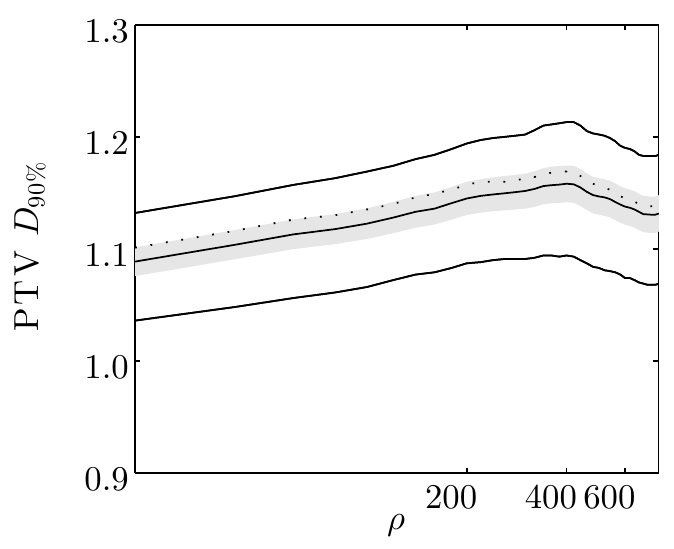}}
   \qquad\qquad
   \subfloat[]{\label{LPDTGRPTVD90pt2}\includegraphics[width=0.35\textwidth]{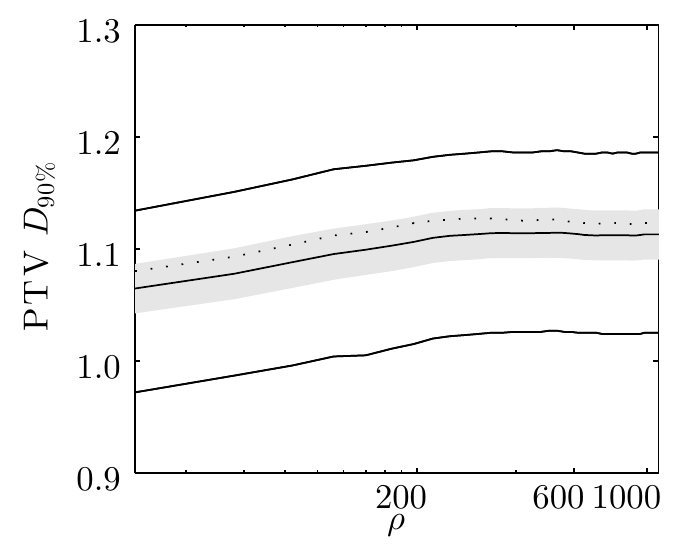}}
   \qquad\qquad
   \subfloat[]{\label{LPDTGRPTVD90pt3}\includegraphics[width=0.35\textwidth]{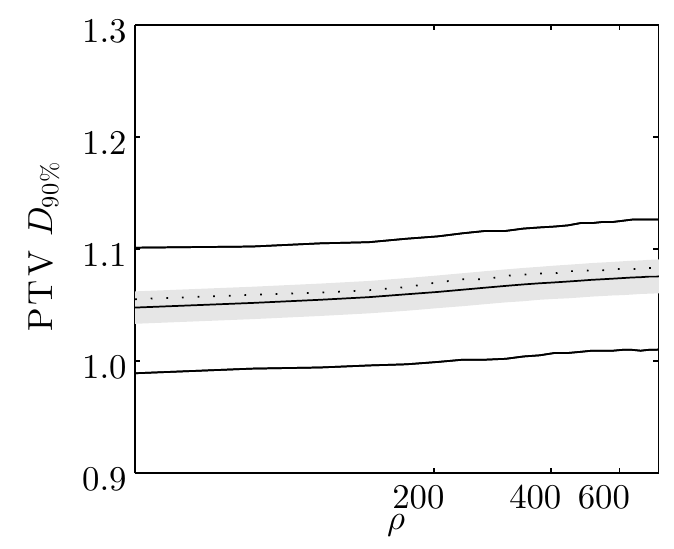}}
   \caption{$D_{90\%}$(PTV) for all patients as function of the maximum sum of squared differences $\rho$. The solid lines represent the minimum, mean and maximum values, and the dotted line is the pre-plan value. The grey area denotes values at most one standard deviation from the mean.}\label{LPDTGRPTVD90}
\end{figure}
%PTV V100
\begin{figure}[h]
   \centering
   \subfloat[]{\label{LPDTGRPTVV100pt1}\includegraphics[width=0.35\textwidth]{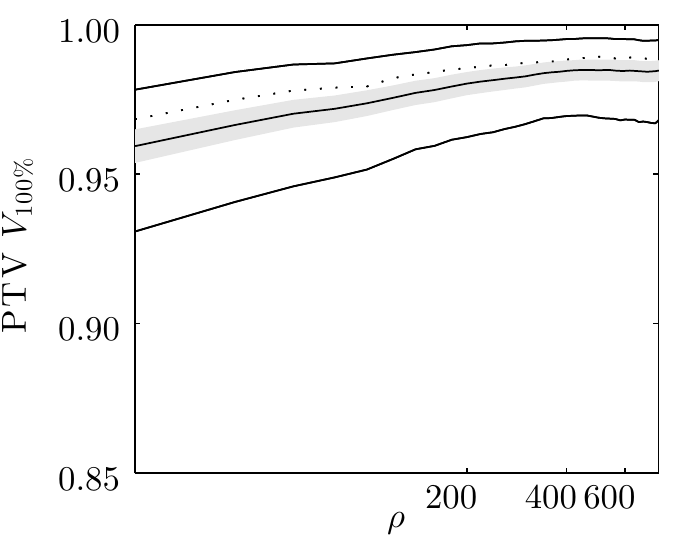}}
   \qquad\qquad
   \subfloat[]{\label{LPDTGRPTVV100pt2}\includegraphics[width=0.35\textwidth]{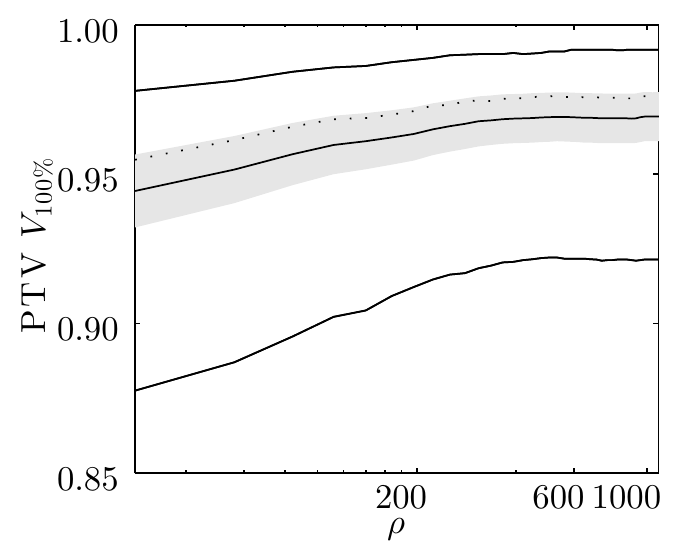}}
   \qquad\qquad
   \subfloat[]{\label{LPDTGRPTVV100pt3}\includegraphics[width=0.35\textwidth]{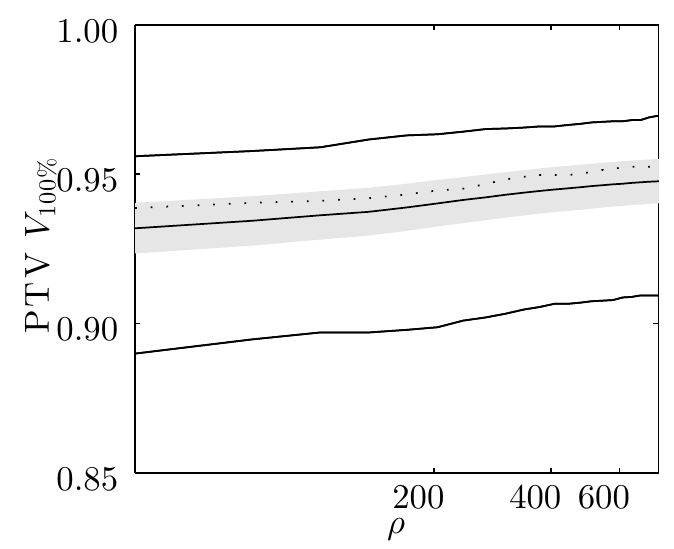}}
   \caption{$V_{100\%}$(PTV) for all patients as function of the maximum sum of squared differences $\rho$. The solid lines represent the minimum, mean and maximum values, and the dotted line is the pre-plan value. The grey area denotes values at most one standard deviation from the mean.}\label{LPDTGRPTVV100}
\end{figure}
%PTV V150
\begin{figure}[h]
   \centering
   \subfloat[]{\label{LPDTGRPTVV150pt1}\includegraphics[width=0.35\textwidth]{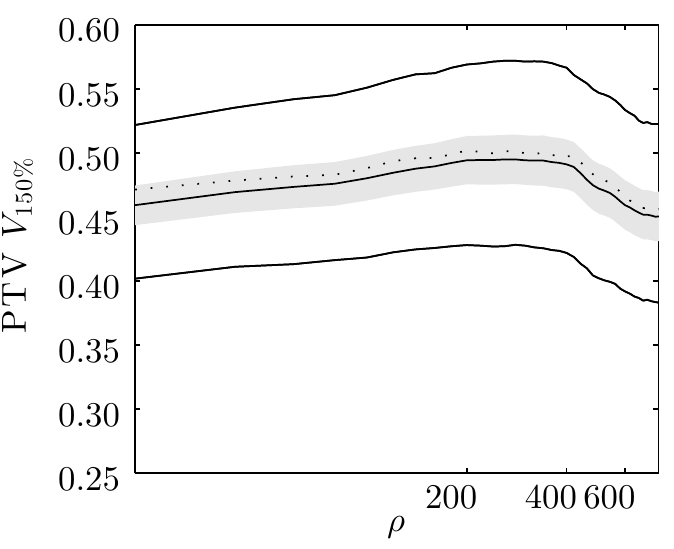}}
   \qquad\qquad
   \subfloat[]{\label{LPDTGRPTVV150pt2}\includegraphics[width=0.35\textwidth]{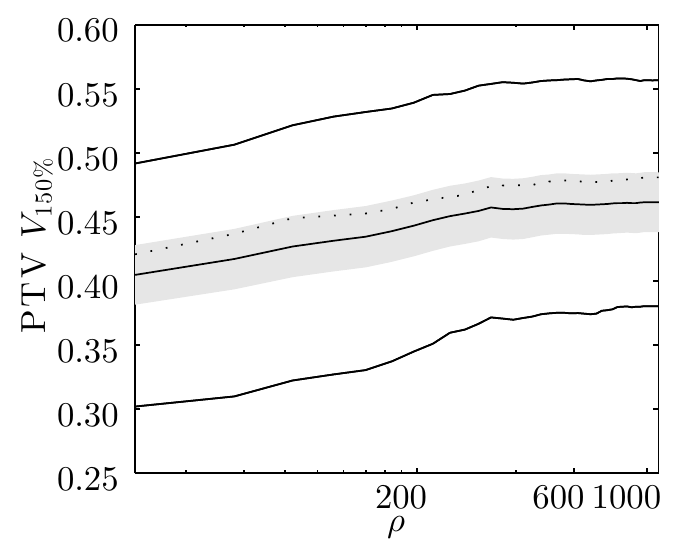}}
   \qquad\qquad
   \subfloat[]{\label{LPDTGRPTVV150pt3}\includegraphics[width=0.35\textwidth]{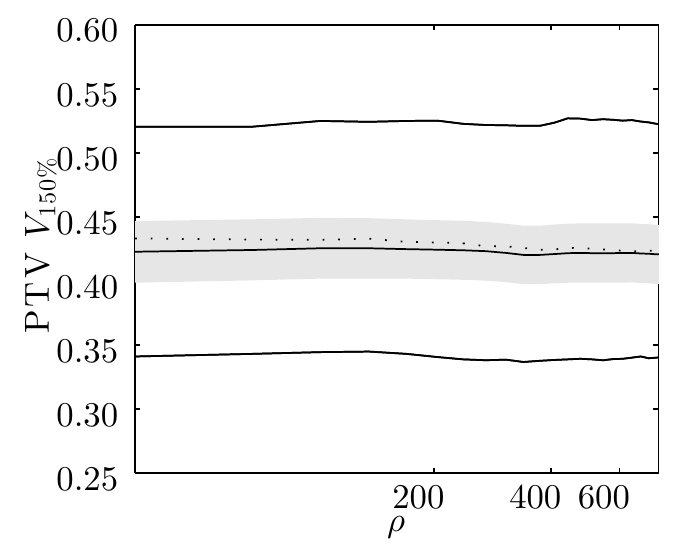}}
   \caption{$V_{150\%}$(PTV) for all patients as function of the maximum sum of squared differences $\rho$. The solid lines represent the minimum, mean and maximum values, and the dotted line is the pre-plan value. The grey area denotes values at most one standard deviation from the mean.}\label{LPDTGRPTVV150}
\end{figure}
%PTV V200
\begin{figure}[h]
   \centering
   \subfloat[]{\label{LPDTGRPTVV200pt1}\includegraphics[width=0.35\textwidth]{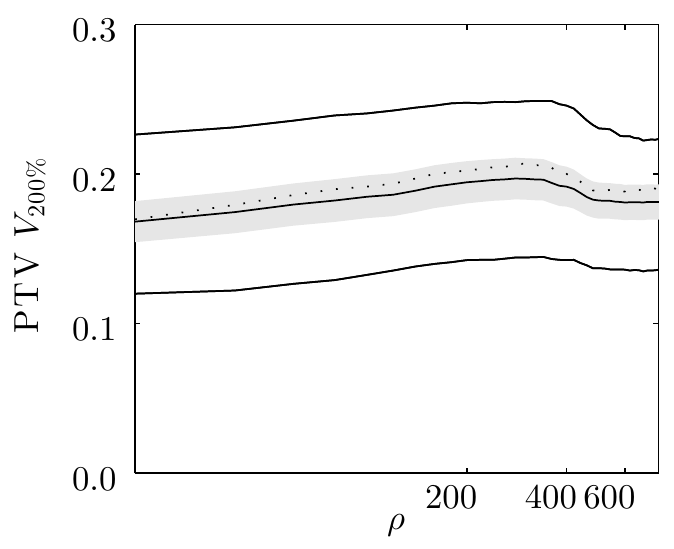}}
   \qquad\qquad
   \subfloat[]{\label{LPDTGRPTVV200pt2}\includegraphics[width=0.35\textwidth]{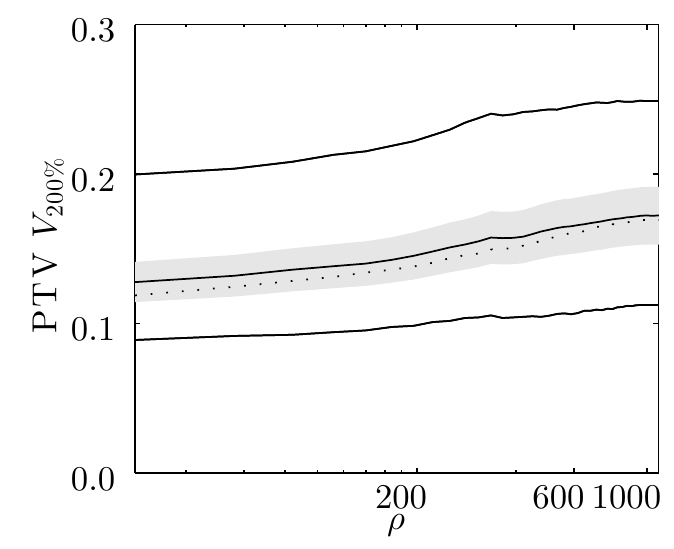}}
   \qquad\qquad
   \subfloat[]{\label{LPDTGRPTVV200pt3}\includegraphics[width=0.35\textwidth]{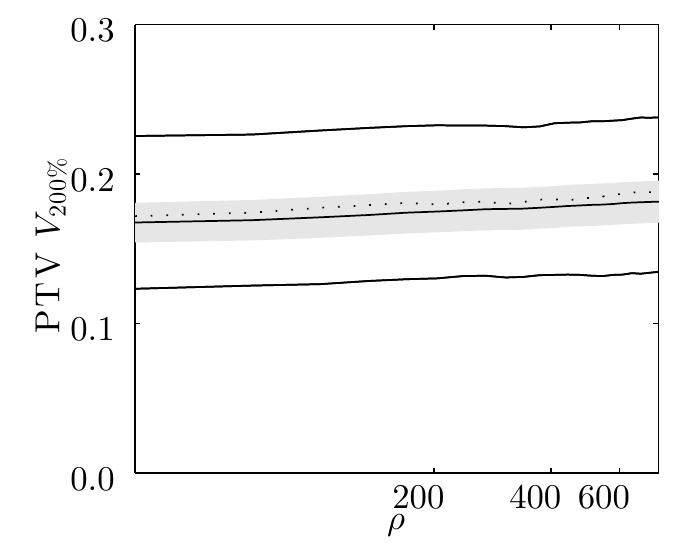}}
   \caption{$V_{200\%}$(PTV) for all patients as function of the maximum sum of squared differences $\rho$. The solid lines represent the minimum, mean and maximum values, and the dotted line is the pre-plan value. The grey area denotes values at most one standard deviation from the mean.}\label{LPDTGRPTVV200}
\end{figure}
\pagebreak
% Rectum D10
\begin{figure}[h]
   \centering
   \subfloat[]{\label{LPDTGRRD10pt1}\includegraphics[width=0.35\textwidth]{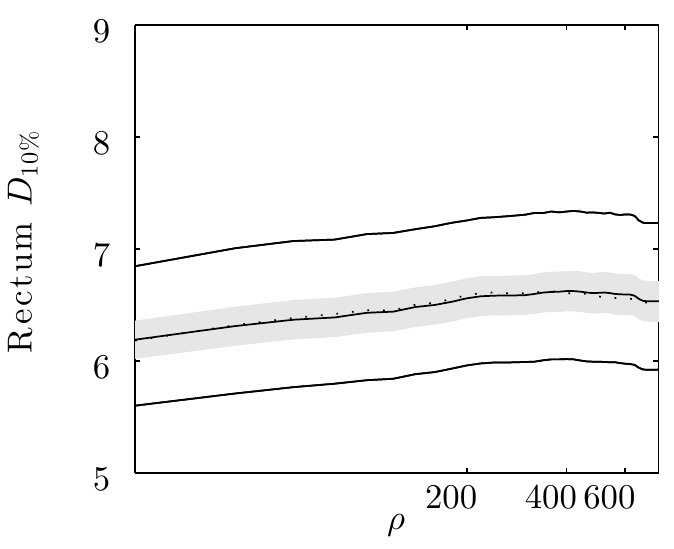}}
   \qquad\qquad
   \subfloat[]{\label{LPDTGRRD10pt2}\includegraphics[width=0.35\textwidth]{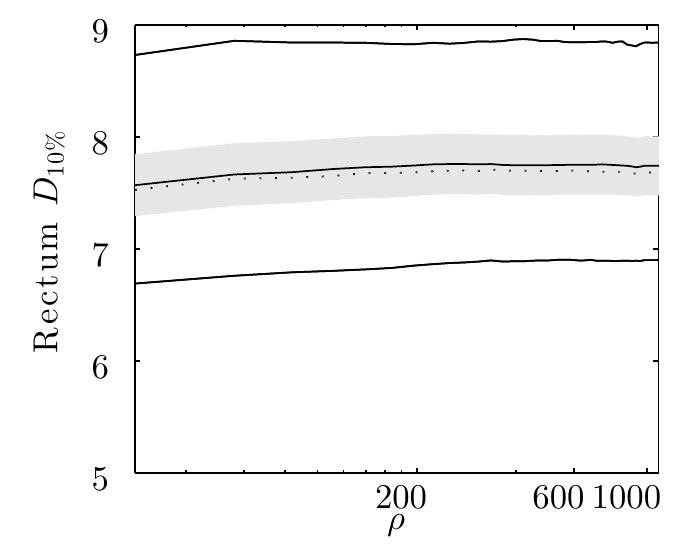}}
   \qquad\qquad
   \subfloat[]{\label{LPDTGRRD10pt3}\includegraphics[width=0.35\textwidth]{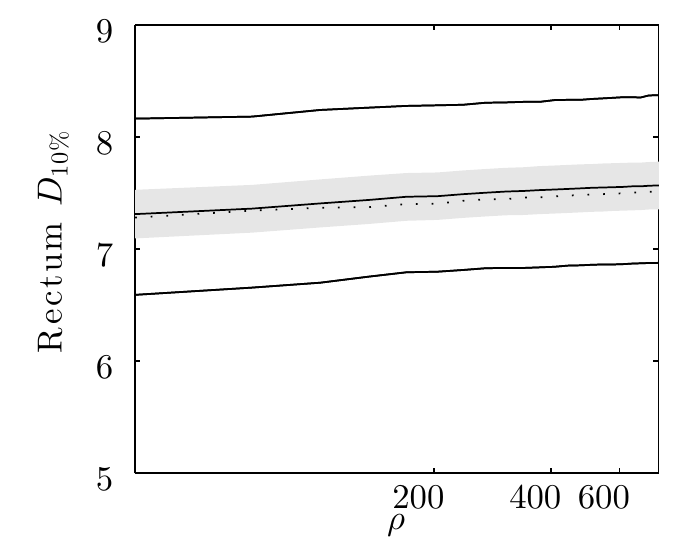}}
   \caption{$D_{10\%}$(rectum) for all patients as function of the maximum sum of squared differences $\rho$. The solid lines represent the minimum, mean and maximum values, and the dotted line is the pre-plan value. The grey area denotes values at most one standard deviation from the mean.}\label{LPDTGRRD10}
\end{figure}
\nopagebreak[4]
% Rectum D2cc
\begin{figure}[h]
   \centering
   \subfloat[]{\label{LPDTGRRD2ccpt1}\includegraphics[width=0.35\textwidth]{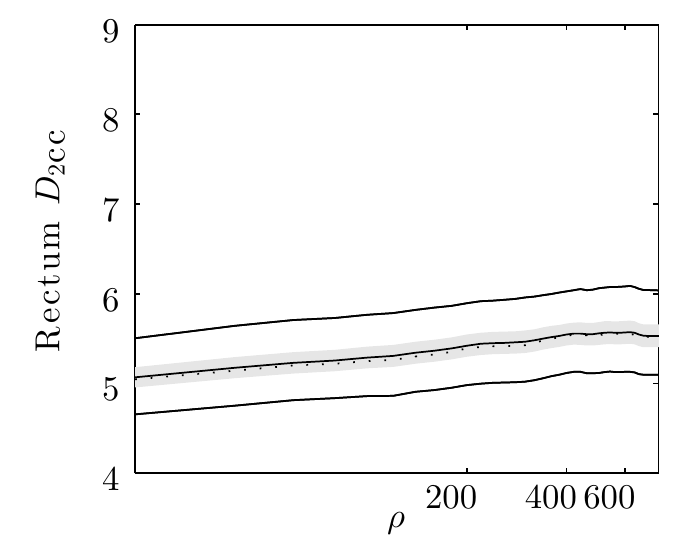}}
   \qquad\qquad
   \subfloat[]{\label{LPDTGRRD2ccpt2}\includegraphics[width=0.35\textwidth]{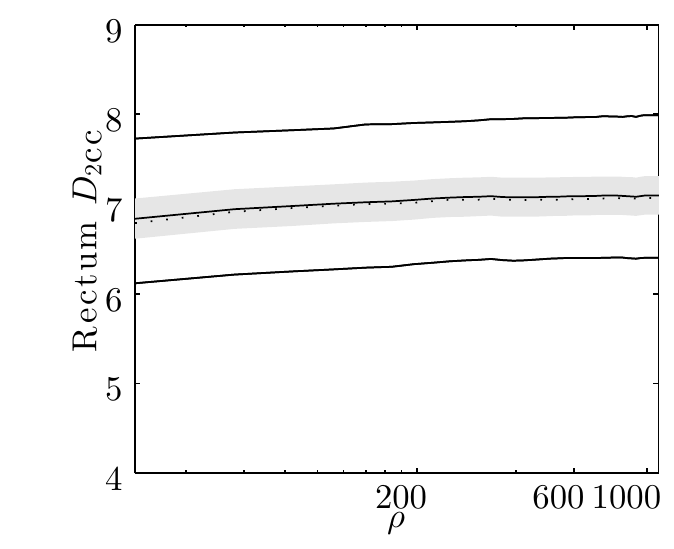}}
   \qquad\qquad
   \subfloat[]{\label{LPDTGRRD2ccpt3}\includegraphics[width=0.35\textwidth]{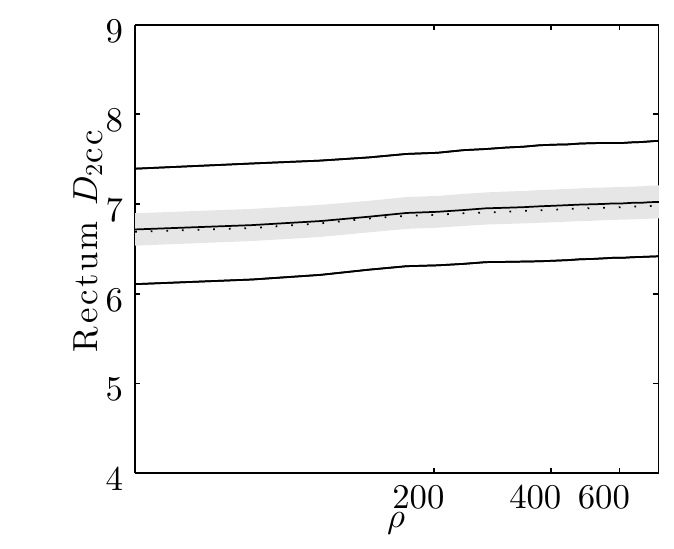}}
   \caption{$D_{2\textnormal{cc}}$(rectum) for all patients as function of the maximum sum of squared differences $\rho$. The solid lines represent the minimum, mean and maximum values, and the dotted line is the pre-plan value. The grey area denotes values at most one standard deviation from the mean.}\label{LPDTGRRD2cc}
\end{figure}
\pagebreak
% Urethra D10
\begin{figure}[h]
   \centering
   \subfloat[]{\label{LPDTGRUD10pt1}\includegraphics[width=0.35\textwidth]{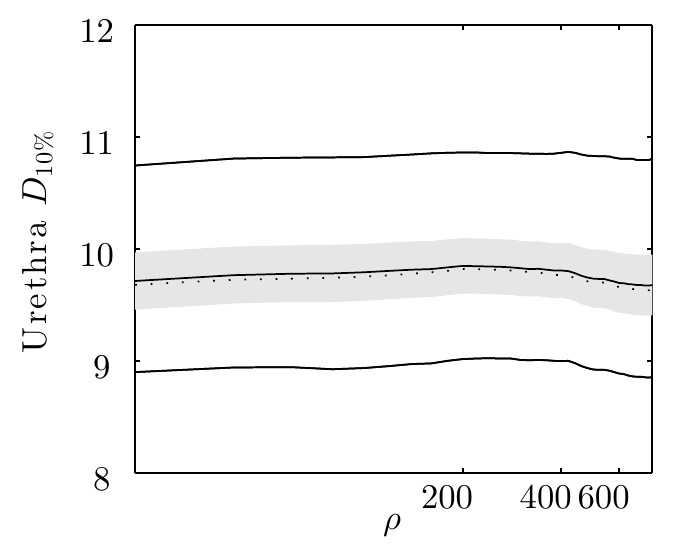}}
   \qquad\qquad
   \subfloat[]{\label{LPDTGRUD10pt2}\includegraphics[width=0.35\textwidth]{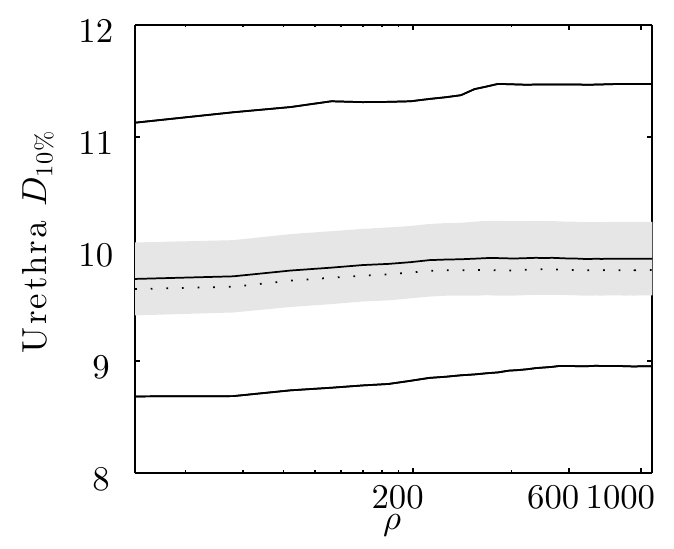}}
   \qquad\qquad
   \subfloat[]{\label{LPDTGRUD10pt3}\includegraphics[width=0.35\textwidth]{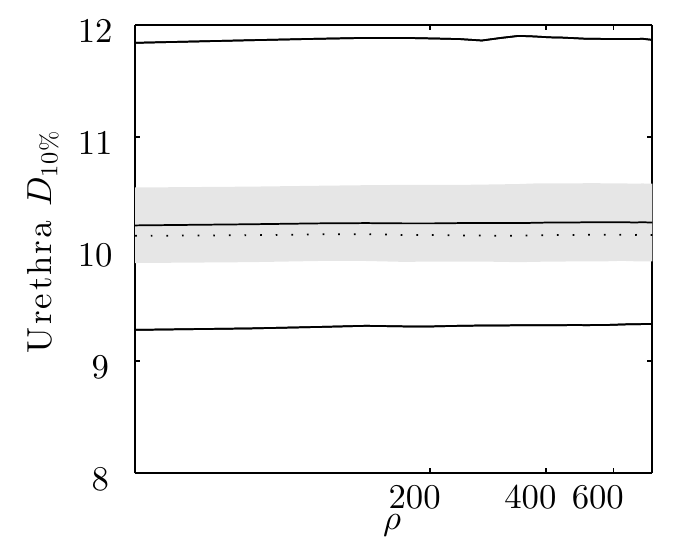}}
   \caption{$D_{10\%}$(urethra) for all patients as function of the maximum sum of squared differences $\rho$. The solid lines represent the minimum, mean and maximum values, and the dotted line is the pre-plan value. The grey area denotes values at most one standard deviation from the mean.}\label{LPDTGRUD10}
\end{figure}
\nopagebreak[4]
% Urethra D01cc
\begin{figure}[h]
   \centering
   \subfloat[]{\label{LPDTGRUD0.1ccpt1}\includegraphics[width=0.35\textwidth]{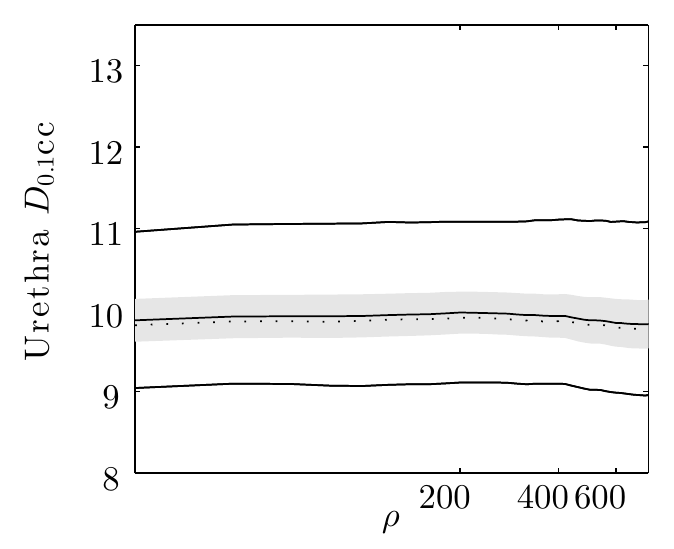}}
   \qquad\qquad
   \subfloat[]{\label{LPDTGRUD0.1ccpt2}\includegraphics[width=0.35\textwidth]{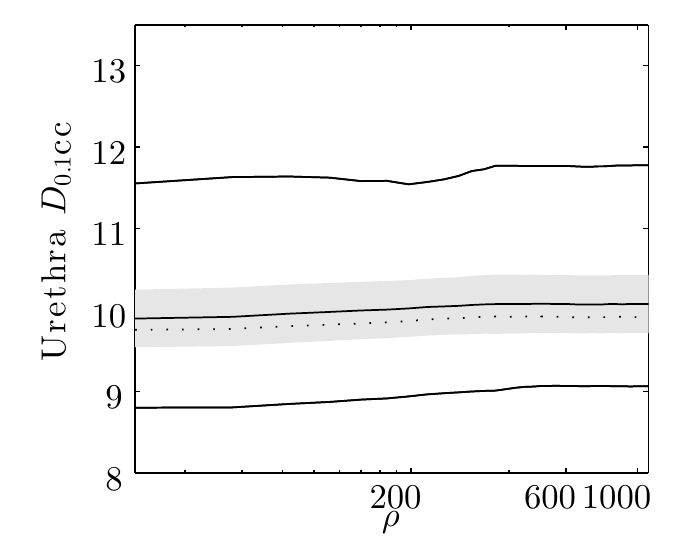}}
   \qquad\qquad
   \subfloat[]{\label{LPDTGRUD0.1ccpt3}\includegraphics[width=0.35\textwidth]{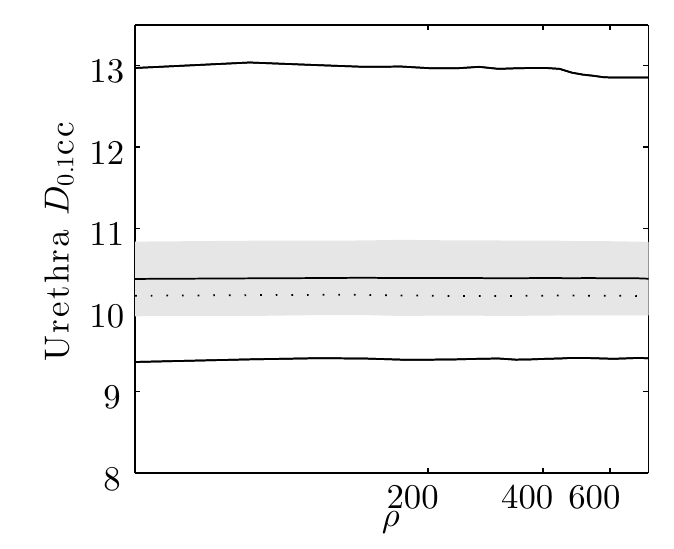}}
   \caption{$D_{0.1\textnormal{cc}}$(urethra) for all patients as function of the maximum sum of squared differences $\rho$. The solid lines represent the minimum, mean and maximum values, and the dotted line is the pre-plan value. The grey area denotes values at most one standard deviation from the mean.}\label{LPDTGRUD0.1cc}
\end{figure}

\end{landscape}

\end{document}